\documentclass[12pt]{article}
\pdfoutput=1
\usepackage{setspace}
\usepackage{color}
\usepackage{amsmath,mathtools,amssymb,slashed,url}
\usepackage{amsthm}
\usepackage{graphicx,epsfig}
\usepackage{array}
\usepackage{subfigure}
\usepackage{pifont}
\usepackage{empheq}
\usepackage{cancel}
\usepackage{amsfonts}
\numberwithin{equation}{section}
\usepackage{bm}
\usepackage{xcolor}
\usepackage[linktoc=all,breaklinks=true,colorlinks=true,linkcolor=purple,urlcolor=purple,citecolor=blue,%
  pdfsubject={hep-th},%
  pdfauthor={Massimo Porrati, Cedric Yu}]{hyperref}
\usepackage{dsfont}
\usepackage{soul}
\usepackage{pbox}
\usepackage{float}
\usepackage[nottoc,notlot,notlof]{tocbibind}

\numberwithin{equation}{section}

\def\n{\nu}
\def\p{\pi}

\def\slc{SL(2,\mathbb{C})}
\def\Tr{\mbox{Tr}\,}
\def\tr{\mbox{tr}\,}
\newcommand{\diag}{{\rm diag} \, }
\def\pd{\partial}

\def\beq{\begin{equation}}
\newcommand{\eeq}[1]{\label{#1}\end{equation}}
\def\bea{\begin{eqnarray}}
\newcommand{\eea}[1]{\label{#1}\end{eqnarray}}
\def\ba{\begin{array}}
\def\ea{\end{array}}
\renewcommand{\Im}{\mbox{Im}\,}
\renewcommand{\Re}{\mbox{Re}\,}
\date{}
\begin{document}
\def\draftnote#1{{\color{red}#1}}

\pagenumbering{roman}
\begin{titlepage}
\vspace{20pt}
\begin{center}
{\huge Kac-Moody and Virasoro Characters from the Perturbative Chern-Simons Path Integral}

\vspace{18pt}

{\large Massimo Porrati$^{a}$~\footnote{E-mail: \href{mailto:massimo.porrati@nyu.edu}{massimo.porrati@nyu.edu}} and Cedric Yu$^a$~\footnote{E-mail: \href{mailto:cedric.yu@nyu.edu}{cedric.yu@nyu.edu}}}

\vspace{12pt}

{$^a$ \em Center for Cosmology and Particle Physics, \\ Department of Physics, New York University, \\726 Broadway, New York, NY 10003, USA}

\vspace{12pt}

\end{center}

\abstract{We evaluate to one loop the functional integral that computes the partition functions of Chern-Simons theories based on compact groups, using the background field method and a covariant gauge fixing.  We compare our computation with the results of other, less direct methods. We find that our method correctly computes the characters of irreducible representations of Kac-Moody algebras. To extend the computation to non-compact groups we need to perform an appropriate analytic continuation of the partition function of the compact group. Non-vacuum characters are found by inserting a Wilson loop in the functional integral. We then extend our method to Euclidean Anti-de Sitter pure gravity in three dimensions. The explicit computation unveils several interesting features and lessons. The most important among them is that the very definition of gravity in the first-order Chern-Simons formalism requires non-trivial analytic continuations of the gauge fields outside their original domains of definition.}
\newpage
\end{titlepage}

\pagenumbering{arabic}

\begingroup
\hypersetup{linkcolor=black}
\tableofcontents
\vspace{0.8cm}
\hrule height 0.75pt
\endgroup

\section{Introduction}\label{intro}
Pure gravity in three-dimensional Anti-de Sitter space is a fascinating theory because it does not contain local degrees of 
freedom in the bulk but it possesses black hole solutions. So it may be simple enough to provide an example of a 
soluble yet
non-trivial model of quantum gravity. The most complete solution would be a holographic dual Conformal Field Theory (CFT), such as the 
extremal CFT proposed in~\cite{Witten:2007kt}. One of the most important quantities that can be computed once a dual 
CFT is known is the partition function $Z(\tau,\bar{\tau})$, which depends on the temperature and the angular-momentum chemical
potential through the complex parameter $\tau$. The partition function can also be computed in principle by a Euclidean functional integral on a solid 3D torus, whose boundary is identified with the 2-torus on which the CFT partition function is computed. The action of pure $AdS_3$ gravity is simple enough that the classical action of {\em all} its regular classical saddle points can be computed, together with the one-loop corrections around each saddle. This computation was performed in~\cite{Maloney:2007ud} but the result does not agree with the partition function found in~\cite{Witten:2007kt}, even in the semi-classical regime where the $AdS_3$ radius $l$ is much larger than the Planck length $G_N$ (in 3D the Newton constant $G_N$ is a length). In fact, when the perturbative Euclidean computation is interpreted as a partition function, it exhibits unphysical features incompatible with a  healthy CFT duals to gravity, 
such as negative-norm states and a continuous 
component in the spectrum. Since Euclidean quantum gravity is an educated guess rather than a complete theory, a possible cure for this unwanted result is to change the path of integration over 3D metrics to allow for saddle points with complex values of the metric~\cite{Maloney:2007ud}. This of course means that we would need to find another definition of the path of integration over metrics, which we do not have yet.

Classical $AdS_3$ gravity can be rewritten as a Chern-Simons theory~\cite{Achucarro:1987vz}\cite{Witten:1988hc}. The gauge group of  Euclidean  $AdS_3$ gravity in the Chern-Simons formulation is $SL(2,\mathbb{C})$ and the action can be written as $iI(A) - iI(A^*)$, where $A$ is an $SL(2,\mathbb{C})$ connection and $*$ denotes complex conjugation of the 
components of $A$ in the adjoint representation. Because the second-order metric formulation and the first-order Chern-Simons
 one are classically equivalent, one can {\em define} Euclidean quantum gravity in terms of the Chern-Simons theory, even if the
  equivalence ceases to hold at the quantum level. In fact any difference with the metric formulation may be welcome, in view of 
  the aforementioned problems found in~\cite{Maloney:2007ud}. 
A calculation of the partition function of  gravity in the $SL(2,\mathbb{C})$ does exist 
\cite{Honda:2015mel,Honda:2015hfa}. The calculation starts in $SL(2,\mathbb{C})$ Chern-Simons gravity but then sums 
the contributions of saddle points where $A^*$ is {\em not} the complex conjugate of $A$. This amounts to a redefinition of 
the path of integration that is left unexplained in \cite{Honda:2015mel,Honda:2015hfa}. Moreover, the result of the integration cannot be interpreted as a partition function of a unitary CFT, so~\cite{Honda:2015hfa} is forced to add boundary degrees of freedom to the action of pure gravity to fix this problem. Finally, the contributions from each saddle are not expressed in terms of characters of either the $SL(2,\mathbb{C})$ Kac-Moody or Virasoro algebras, so that their meaning in terms of semi-classical geometries and fluctuations around them is rather obscure.

Both ref.~\cite{Maloney:2007ud} and \cite{Honda:2015mel,Honda:2015hfa} make clear that an important ingredient in the definition
of Euclidean quantum gravity is the definition of the path of integration. In a semi-classical calculation this is the same as
finding the classical saddle points that contribute to the path integral and computing the contribution of each one of them.
These saddle points need not correspond to real Euclidean metrics. Complex metrics --or saddle points where $A^*$ is not the complex conjugate of $A$-- can appear because boundary conditions imposed by turning on certain chemical potentials may require it, or because real boundary fields develop a complex part inside Euclidean $AdS_3$. 

Even when the set of relevant saddle points is known and even when working in the semi-classical approximation, one must 
know the result of a perturbative calculation around each saddle point. For Chern-Simons theory, this is not as obvious as it 
sounds, first of all because the gauge groups relevant to 3D gravity are non-compact~\footnote{The group for Euclidean 
$AdS_3$ gravity is $SL(2,\mathbb{C})$ while it is $SL(2,\mathbb{R})\times SL(2, \mathbb{R})$ for Lorentzian gravity.}. Another 
source of complexity is that a perturbative bulk calculation requires to fix a space-time metric. If we want to maintain a relation 
between this ancillary metric and the metric defined by the connections $A,A^*$ through the formulas given later in the paper, we
 should use as background metric that of thermal Euclidean $AdS_3$. This metric has some nice aspects such as maximal 
 symmetry and metric completeness, but it makes the underlying space non-compact. This implies that the fluctuations of the 
 gauge fields around a classical background cannot be standard Dirichlet or Neumann boundary conditions but rather asymptotic
  conditions determined by requiring self-adjointness of the wave operators of the fluctuations.

In fact, a direct calculation of the one-loop functional integral of Chern-Simons theory in an open manifold with a conformal boundary and with asymptotic boundary conditions is still lacking in the literature. So, as a modest first step toward the program of computing the partition function of pure gravity and eventually also indices of $AdS_3$ extended pure supergravity theories, in this paper we will perform a perturbative bulk calculation of the partition function of Chern-Simons theories of compact gauge group ($U(1)$ and $SU(2)$), noncompact gauge group $SL(2,\mathbb{R})$, and Euclidean $AdS_3$ gravity. Each will present  challenges and each will be tested against known results. Specifically, for $U(1)$, $SU(2)$ and $SL(2,\mathbb{R})$ Chern-Simons theories, the one-loop computation will be tested against known formulas for Kac-Moody vacuum characters,  while the calculation for gravity will be tested against the results of refs.~\cite{Maloney:2007ud}\cite{Giombi:2008vd}. Further checks will be provided by computing partition functions in the presence of a Wilson loop, which will be compared with characters of irreducible Kac-Moody representations for $SU(2)$ and $SL(2,\mathbb{R})$. A Wilson loop insertion is also a necessary ingredient of the Chern-Simons gravity calculations, because the $SL(2,\mathbb{C})$ gauge field corresponding to the regular negative-curvature metric on the solid 3D torus has non-trivial holonomy even around a contractible circle. In all cases we will find a quite non-trivial agreement between our perturbative calculations and the results obtained from the theory of representations of Kac-Moody algebras. 

We organize the paper as follows. In \hyperref[sec:CSWZW]{Section \ref*{sec:CSWZW}} we perform the reduction of the partition function of Chern-Simons theory on a solid 3D torus $T$ for an arbitrary compact gauge group and for Euclidean $AdS_3$ gravity to that of a chiral 2D Wess-Zumino-Witten (WZW) model defined on the boundary of the torus. For compact gauge groups the calculation was done in~\cite{Elitzur:1989nr}. Our calculation differs from theirs only in the choice of the boundary conditions for the ``time'' component of the gauge field. Yet, already here we find the need for an analytic continuation of the boundary value of the gauge field. The calculation for gravity requires a more drastic analytic continuation of the gauge fields in the bulk, which concretely shows that the definition of the path of integration of Chern-Simons gravity is far from obvious. While the reduction to a boundary WZW model is an efficient way of computing the partition function and obtaining characters of representations of Kac-Moody algebras, it obscures the meaning of a classical saddle point, in particular its relations to a semi-classical background metric. To find this relation we need a bulk calculation that we perform in the rest of the paper, starting in \hyperref[sec:resonance]{Section \ref*{sec:resonance}}. There, we define a method for computing functional determinants using
 the singularities of an appropriate scattering operator.  \hyperref[sec:resonance]{Section \ref*{sec:resonance}} also clarifies the basis and validity of the calculation of functional determinants by the quasinormal-modes method used in~\cite{Denef:2009kn}. \hyperref[sec:gfcom]{Section \ref*{sec:gfcom}} defines the covariant gauge fixing that we will use throughout our paper as well as the steepest-descent procedure that renders the one-loop contribution to the functional integral well-defined. Though apparently simple, the continuation can nevertheless give wrong results if not properly defined. Sections \hyperref[u1vac]{\ref*{u1vac}}, \hyperref[su2vac]{\ref*{su2vac}} and \hyperref[sec:wilson]{\ref*{sec:wilson}} compute progressively more complex functional integrals and verify that the analytic continuation correctly reproduces the expected quantities in various settings. \hyperref[u1vac]{Section \ref*{u1vac}} works out the $U(1)$ Chern-Simons theory. It computes the functional determinants with the method of scattering poles introduced in \hyperref[sec:resonance]{Section \ref*{sec:resonance}} and also defines a key ingredient in the computation: the eta-invariant \cite{Atiyah:1975jf,Atiyah:1976jg,Atiyah:1980jh}, which is necessary for computing the phase of the functional determinants appearing at one loop. \hyperref[su2vac]{Section \ref*{su2vac}} computes the partition function of $SU(2)$ Chern-Simons theory and shows that it reproduces the vacuum character of the corresponding Kac-Moody algebra in the presence of a chemical potential. A key technique introduced in the section is the computation of twisted functional determinants. \hyperref[sec:wilson]{Section \ref*{sec:wilson}} computes the partition function in the presence of an $SU(2)$ Wilson loop and compares it to non-vacuum characters. \hyperref[sl2rcs]{Section \ref*{sl2rcs}} studies in some detail the analytic continuation from $SU(2)$ Chern-Simons theory needed to compute the vacuum and non-vacuum characters of the $\widehat{\mathfrak{sl}}(2,\mathbb{R})_{k}$ Kac-Moody algebra. The section also discusses the relation between such continuation from the 
$SU(2)$ Chern-Simons theory and the gauge-fixing for non-compact groups introduced in ref.~\cite{BarNatan:1991rn}. \hyperref[sec:grav]{Section \ref*{sec:grav}} finally tackles the partition function of quantum gravity. It re-examines the issue of analytic continuation of the partition function from the steepest descent path integral defined for $SU(2)$ Chern-Simons. It also shows how to impose the extra conditions on the asymptotic value of the gauge fields that are necessary to guarantee the existence of an asymptotically Anti-de Sitter metric. We impose these {\em Brown-Henneaux}~\cite{Brown:1986nw} conditions as extra gauge invariances and extra gauge fixings, both  defined only on the restriction of bulk fields to the conformal boundary. Several important but more technical calculations are given in various appendices.

\subsection{Notation and Convention}\label{notation}
Throughout this paper we work in Euclidean signature. The Euclidean $AdS_3$ global metric is
\begin{subequations}
\begin{align}
ds^2&=\frac{dr^2}{1+r^2}+(1+r^2)dt^2+r^2d\phi^2,\;r\in[0,\infty)\\
&=d\xi^2+\sinh^2{\xi}d\phi^2+\cosh^2{\xi}dt^2,\;r=\sinh{\xi},\;\xi\in[0,\infty)\\
&=\frac{dx^2}{4x(1-x)^2}+\frac{x}{1-x}d\phi^2+\frac{1}{1-x}dt^2,\;x=\tanh^2{\xi},\;x\in[0,1]\label{eq:xmetric}\\
&=\frac{d{r'}^{2}}{{r'}^{2}}+\left({r'}^{2}+\frac{1}{16{r'}^{2}} \right)(d\phi^2+dt^2)-\frac{1}{4}(d\phi+idt)^2-\frac{1}{4}(d\phi-idt)^2,\;r'=\frac{r+\sqrt{1+r^2}}{2}.
\end{align}
\end{subequations}
Here $\xi=0$ is the origin where $x\sim \xi^2$, and $x=1$ is infinity where $(1-x)\sim e^{-2\xi}\sim r^{-2}$. $r'\in[\frac{1}{2},\infty)$ is the Fefferman-Graham radial coordinate. We set the $AdS$ length $l=1$ throughout. 

Thermal $AdS_3$ is a non-compact, in fact open, manifold. It has the periodicities $\phi\sim\phi+2\pi$ and $(\phi,t)\sim(\phi+2\pi\tau_1,t+2\pi\tau_2)$ and the complex structure of the \textit{conformal} torus boundary is $\tau=\tau_1+i\tau_2$, $q=\exp{(2\pi i \tau)}$, where $\tau_1$ is the angular potential and $\tau_2$ is the inverse temperature. Define the holomorphic cylindrical coordinates $w=\phi+it$ and $\bar{w}=\phi-it$, $w\sim w+2\pi\sim w+2\pi\tau$ such that Fourier modes have arguments
\begin{equation}
\begin{aligned}
\omega t-k\phi&=\underbrace{\left( \frac{-i\omega-k}{2}\right)}_{k_w}w+\underbrace{\left( \frac{i\omega-k}{2}\right)}_{k_{\bar{w}}}{\bar{w}},\\
k&=-(k_w+k_{\bar{w}}),\;\omega=i(k_w-k_{\bar{w}}).
\end{aligned}
\end{equation}
For any one-form $A$,
\begin{equation}
\begin{aligned}
A_\phi&=A_w+A_{\bar{w}},\;A_t=i(A_w-A_{\bar{w}}),\;\\
A_{w}&=\frac{A_\phi- iA_{t}}{2},\;A_{\bar{w}}=\frac{A_\phi+ iA_{t}}{2}.
\end{aligned} 
\end{equation}
Our manifold is oriented with the Levi-Civita \textit{symbol}
\begin{align}
\varepsilon^{tr\phi}=\varepsilon_{tr\phi}=+1.
\end{align}
The generators $\{T^a\}$ of Lie algebra $\mathfrak{g}$ of $G$ are such that a group element of $g\in G$ connected to the identity has the form $g=e^{\theta^aT^a}$, where for $G$ compact $\theta^a$ are real. For $SU(2)$ the generators in the fundamental representation are anti-Hermitian,
\begin{align}
T^a&=-\frac{i\sigma^a}{2},\;[T^a,T^b]=+\delta^{cd}\epsilon_{abc}T^d,
\end{align}
where $\epsilon_{123}=\epsilon^{123}=+1$, and the Killing metric $\Tr$ is negative-definite, $\Tr[T^aT^b]=-\frac{1}{2}\delta^{ab}$. Define the raising and lowering operators,
\begin{equation}
\begin{aligned}
T^\pm\equiv-\frac{T^1\mp iT^2}{\sqrt{2}}&,\;\Tr[T^+ T^-]=-\frac{1}{2},\\
[T^3,T^\pm]=\pm i T^\pm&,\;[T^+,T^-]=iT^3.
\end{aligned}
\end{equation}
An $\mathfrak{su}(2)$-valued field $\psi$ has expansion
\begin{align}
\psi&=\psi^3(x)T^3+\psi^1(x)T^1+\psi^2(x)T^2=\psi^3(x)T^3+\psi^+(x)T^++\psi^-(x)T^-,
\end{align}
with the components $\psi^{1,2,3}$ real functions and
\begin{align}
\psi^\pm=-\frac{1}{\sqrt{2}}(\psi^1\pm i\psi^2),\;(\psi^\pm)^*&=\psi^\mp.\label{eq:reality}
\end{align}

\section{Chern-Simons Theory and Chiral WZW Model}\label{sec:CSWZW}
\subsection{The Chern-Simons Path Integral}
The Chern-Simons action of a Lie group $G$ on a 3-manifold $M$ is
\begin{align}
W[A]\equiv \frac{1}{4\pi}\int_M\Tr[AdA+\frac{2}{3}A\wedge A\wedge A] ,
\end{align}
where $A$ is a $\mathfrak{g}$-valued connection in the fundamental representation and $\Tr$ is the trace (Killing metric) therein. Since $\Tr$ is negative-definite, for $G=U(1)$ we define $W[A]=-\frac{1}{4\pi}\int_M AdA$.

The Chern-Simons path integral $Z_G(k,\zeta)$ is a function of the level $k$ and other parameters (such as the boundary condition on $A$) collectively denoted as $\zeta$:
\begin{align}
Z_G(k,\zeta)&=\int_{\mathcal{U}_G} DA\; e^{-I[A]}=\int_{\mathcal{U}_G} DA\;e^{-(-ikW[A]+I_{bt}[A]) },\label{eq:pathint}
\end{align}
where $\mathcal{U}_G$ is the space of connections in $G$ modulo gauge redundancy, which we may assume to be trivial
while $k$ is the level; in this work we need not discuss whether it is quantized or why. We have included possible boundary terms
 $I_{bt}[A]$ that are necessary to have a well-defined classical variational problem associated to given boundary conditions. 

\subsection{Reduction to Boundary Dynamics}\label{WZWred}
It is well known that the canonical quantization of Chern-Simons Theory on a space $D\times \mathbb{R}$ ($D$=disk) is equivalent to the quantization of a chiral WZW model~\cite{Elitzur:1989nr}. Here we will show that the Euclidean partition function of a Chern-Simons theory on the 3D torus $T=D\times S^1$ reduces to the partition function of a chiral WZW model. Using appropriate analytic continuations for the path of integration used in the definition of the theory, the equivalence extends to the non-compact $\slc$ theory which is classically equivalent to pure 3D gravity in Anti-de Sitter space~\cite{Achucarro:1987vz}\cite{Witten:1988hc}. We will also compute explicitly the partition function of the WZW model for compact gauge groups in Appendix A.\footnote{We thank Blagoje Oblak for sharing with us his unpublished, detailed calculations of the partition function of a free chiral boson WZW\cite{bob}.}

To compute the partition function of a Chern-Simons Theory with compact gauge group $G$ on the solid torus $T$ (which is a compact manifold with a boundary), we parametrize the disk with angular coordinate $\phi\in[0,2\pi)$ and radial coordinate $r\in[0, R]$, and the thermal circle $S^1$ with $t\in[0, 2\pi\beta)$. Next, we decompose the connection into the sum of its pullback to the disk $\hat{A}=A_\phi d\phi + A_r dr$ and its pullback to $S^1$,  $A_0=A_t dt$. $A=\hat{A} + A_0$ is a one-form valued in the real Lie algebra $\mathfrak{g}$. The partition function is
  \bea
  Z(\tau) &=& \int_{\mathcal{U}_G} DA \exp\Big( +\frac{ik}{4\pi}\int_T \Tr[AdA+\frac{2}{3} A\wedge A\wedge A]\Big)  \nonumber \\
   &=& \int_{\mathcal{U}_G} D\hat{A}DA_0 \exp{\left(+{ik\over 4\pi} \int_T \Tr[2A_0  \hat{F}+ dt \pd_t \hat{A}  \hat{A}] -{ik\over 4\pi} \int_{\pd T} \Tr[A_t dt  A_\phi d\phi] \right)}, \nonumber \\
          \hat{F} &=&\hat{d} \hat{A} + \hat{A}  \hat{A}.
          \eea{m1}
   The integral over $A_t|_{\mathring{T}}$, i.e. the restriction of $A_t$ to the interior of $T$, produces a functional Dirac delta that enforces the constraint $\hat{F}=0$ {\em when $A_t|_{\mathring{T}}$ takes value over real Lie-algebra valued fields}. On a disk, flat connections are {\em globally} pure gauge so $\hat{A} = U^{-1}\hat{d} U $ for some $U:T\rightarrow G$. The second term in the second line of~\eqref{m1} contributes to the action a term
   \bea
   &&{ik\over 4\pi} \int_T dt \Tr[\pd_t (U^{-1}\hat{d} UU^{-1}\hat{d} U)]\nonumber\\
  &=& {ik\over 4\pi}\left( \int_{\pd T} d\phi dt  \Tr[U^{-1}\pd_t U  U^{-1}  \partial_\phi U] -{1\over 3} \int_T \Tr[U^{-1}d U U^{-1}  dU U^{-1} d U] \right) .
   \eea{m2}
   The last contribution to the action is the term proportional to 
   $\int_{\pd T}\Tr[ A_t dt  A_\phi d\phi]$. It is well defined also when we fix as boundary condition $A_{\bar{w}}|_{\partial T}=0$, in which case $A_t|_{\pd T}$ takes values in the {\em complex} Lie algebra $\mathfrak{g}$
   \beq
   A_t|_{\pd T}=i A_\phi |_{\pd T}\Leftrightarrow A_{\bar{w}}|_{\partial T}=0 .
   \eeq{m3}
   This condition  reduces the Chern-Simons action to that of a Euclidean-signature chiral WZW model
    \beq
    I[U]= -{k\over 4\pi}\left(2\int_{\pd T} dt d\phi \Tr[U^{-1}\partial_\phi U U^{-1}\partial_{\bar{w}}U] -{i\over 3} \int_T \Tr[U^{-1}d U U^{-1}  dU U^{-1} d U] \right).
    \eeq{m4}
    
To complete the evaluation of the functional integral in~\eqref{m1}  we need to compute the 
 change in the integration measure
 \beq
 \delta[\hat{F}]D\hat{A} = JDU.
 \eeq{m5}
 The Jacobian $J$ is computed from the metric on the $\hat{A}$ space ($i=t,\phi$) 
 \bea
 ds^2 &=& \int_D drd\phi \sqrt{g}\delta A_i \delta A_j g^{ij}=\int_D drd\phi \sqrt{g} (D_i u D_j u + D_i t  D_j t)g^{ij} , 
\;\; \delta A_i = D_i u + \epsilon_{ij} D^j t, \nonumber \\
 A_i &=& U ^{-1}\pd_i U ,\;\; D_i  =\pd_i  + [A_i, \; ]  .
 \eea{m6}
 Here $g_{ij}$ is an arbitrary regular metric on the disk $D$. For a metric $ds^2= G_{ab} dy^a dy^b$ the volume factor is
 $\sqrt{\det G}$, so the change of variables~\eqref{m6} contributes a $\det{D_iD^i}$ to the Jacobian. This factor cancels 
 against the delta function, giving $J=1$ and making the partition function~\eqref{m1} exactly equal to a chiral WZW partition function
\beq
Z(\tau)=\int_{U\in G} DU e^{-I[U]}.
\eeq{m7}

The partition function of Euclidean quantum gravity in  the first-order, Chern-Simons formulation~\cite{Achucarro:1987vz}\cite{Witten:1988hc} is formally
defined by the functional integral
\beq
Z(\tau)= \int_{A\in \mathcal{U}_{\slc}} DA DA^* \exp\left( 2{k\over 4\pi}\Im \int_T \Tr [A d A +{2\over 3} A\wedge A \wedge A]\right) .
\eeq{m8}
By decomposing the $\slc$ gauge connection as in the previous case ($A=\hat{A} + A_0$) and following manipulations 
analogous to the compact gauge group case we recast~\eqref{m8} into
\beq
Z(\tau)=\int_{A\in \mathcal{U}_{\slc}} D\hat{A}D\hat{A}^* DA_0 DA_0^* \exp\left[2{k\over 4\pi}\Im \left( \int_T \Tr[2A_0  \hat{F}+ dt \pd_t \hat{A}  \hat{A}]-\int_{\partial T}\Tr[A_tdt A_\phi d\phi] \right)\right].
 \eeq{m9}
The term $\Im\Tr \int_T dt A_t  \hat{F}= \Tr \int_T dt (A_t^R  \hat{F}^R - A_t^I  \hat{F}^I)$ 
($A=A^R+iA^I$ etc.) is real and unbounded below, so the integral of its exponential is not even formally convergent (see also Appendix E of \cite{Mikhaylov:2014aoa}).  
Clearly, to make sense of~\eqref{m9} even in the loosest sense of the word we need to change the path of integration.
One simple choice that does make sense is to continue both the real and imaginary parts of $A_t$ to imaginary values
$A^R_t\rightarrow iA^R_t$, $A^I_t\rightarrow iA_t^I$. After this redefinition the integral over $A_t$ produces a delta
function that constrains the $\slc$ connection to be flat. The boundary condition on $A_t$ is as in the compact case
$A_t=iA_\phi$, but this time it does not require an analytic continuation outside the original space of connections.
Following step by step the procedure we used for compact groups we arrive at 
\beq
Z(\tau)= \int_{U\in \slc}  DU DU^* \exp \left[ 2{k\over 4\pi}\Im \left(-2i\int_{\pd T} dt d\phi \Tr[U^{-1}\partial_\phi U U^{-1}\partial_{\bar{w}}U] -{1\over 3} \int_T \Tr[(U^{-1}d U)^3]\right) \right].
\eeq{m10}
We have again a problem here, because the integral is real and unbounded below. A possible solution is to write the partition function as 
\bea
Z(\tau)&=&\int_{V=U^*} DU DV \exp \left[ {k\over 4\pi}\left( -2\int_{\pd T} dt d\phi \Tr[U^{-1}\partial_\phi U^{-1}\partial_{\bar{w}}U]+{i\over 3} \int_T \Tr[(U^{-1}d U)^3 ]\right)+\right. \nonumber \\
&&\left.{k\over 4\pi}\left( -2\int_{\pd T} dt d\phi \,  \Tr[V^{-1}\partial_\phi VV^{-1}\pd_{w} V] -{i\over 3} \int_T \Tr[(V^{-1}d V)^3 ]\right)\right],
\eea{m11}
evaluate it on a {\em different} path of integration and analytically continue the results. Unless the computation is done 
exactly this procedure is fraught with problems. In particular, the naive analytic continuation of the semi-classical saddle-point computation may suffer from ambiguities due to Stokes phenomena. A possible choice that reproduces the results of the second-order formulation of gravity is to continue from the contour $U\in SU(2)$, $V\in SU(2)$, $V$ independent of $U$. The computation for $SU(2)$ is done using Hamiltonian reduction in \hyperref[sec:chWZW]{Appendix \ref*{sec:chWZW}}; it gives the vacuum character of the level-$k$ Kac-Moody algebra
of $SU(2)$. In the main part of this work we make use of a very different method to compute Chern-Simons path integrals, namely a covariant gauge-fixing in thermal $AdS_3$, which allows for a perturbative saddle-point computation at one-loop. We first compute the $U(1)$ and $SU(2)$ Chern-Simons path integrals at one-loop and confirm that they give results
 identical to the WZW reduction; namely, they  both give vacuum Kac-Moody characters. This is a useful check of
  the validity of our approach. Then we move on to define the Euclidean gravity partition function in covariant gauge-fixing by analytic continuations of parameters in the path integrals.

\section{Functional Determinants and Resonance Poles}\label{sec:resonance}
In what follows we will make use of resonance/scattering poles to compute functional determinants of various elliptic (first- and second-order) operators in thermal $AdS_3$. Such method is well-known in the mathematical literature and was introduced to the context of (A)dS/CFT in \cite{Denef:2009kn}. We now review it in the case of the Klein-Gordon operator. The purpose here is to give a physicist's intuition without providing mathematical rigor, in part to clear up some confusion 
that is found in the physics literature.

Consider the functional integral of a complex massive scalar field $\varphi$ on Euclidean $AdS_{d+1}$,
\begin{align}
Z(s)=\int D\varphi\; \exp{\left(-\int_M d^{d+1}x\sqrt{g} \varphi^* (\Delta_0+s(s-d)) \varphi\right)},
\end{align}
where $\Delta_0=-\nabla^2$ is the positive-definite scalar Laplacian while the mass term is related to the conformal dimension $s$ by $m^2=s(s-d)$. Unitarity requires $s\geq\frac{d-2}{2}$ \cite{Breitenlohner:1982jf}. One must specify the domain of the Laplacian, i.e. the functional space of $\varphi$ to be integrated over, as part of the definition of $Z(s)$. In order that $Z(s)$ be expressible in terms of the functional determinant $\det{(\Delta_0+m^2)}$, we must require that the Laplacian be strictly self-adjoint.

Define the positive-definite $L^2$-norm for any complex scalars $\psi$ and $\varphi$ by
\begin{align}
(\psi,\varphi)=\int_M d^{d+1}x\sqrt{g} \psi^*(x) \varphi(x),
\end{align}
where ${}^*$ means complex conjugation. The requirement that the Laplacian be strictly self-adjoint means, by definition,
\begin{align}
(\Delta_0\psi,\varphi)\overset{!}{=}(\psi,\Delta_0\varphi)
\end{align}
\textit{and} that the domain of the Laplacian coincide with the domain of its adjoint. $EAdS_{d+1}$ is an open non-compact manifold without a boundary. We do not compactify it so we cannot impose any boundary conditions on $\varphi$. The domain is instead the space of $L^2$-normalizable functions. This is neither relative (Dirichlet) nor absolute (Neumann) boundary condition, as can be explicitly seen from \cite{Camporesi:1994ga} or from heat kernel expansions. Such domain is  formally spanned by the smooth, delta-function orthonormal eigenfunctions of $\Delta_0$ which belong a continuous spectrum $[\frac{d}{2}+s(s-d),\infty)$.

Defining the path integral by integrating on such space, we have 
\begin{align}
Z(s)=\int D\varphi\; \exp{\left(-\int_M d^{d+1}x\sqrt{g} \varphi^* (\Delta_0+s(s-d)) \varphi\right)}=\det\nolimits^{-1}{(\Delta_0+s(s-d))}.
\end{align}
In the last equality we have normalized the path integral to absorb the delta-function norms of the eigenfunctions which span the functional space. Instead of explicitly computing the spectral function of the Laplacian and evaluate the determinant using zeta function regularization, as was done in \cite{Camporesi:1994ga} for the hyperbolic space $\mathbb{H}^{d+1}$, we follow \cite{Denef:2009kn} and consider $\det{(\Delta_0+s(s-d))}$ as a meromorphically continued function of $s\in \mathbb{C}$.

\subsection{Functional Determinants from Scattering Poles}
\subsubsection*{Functional Determinant and the Green's Function}
First we write the logarithm of the Klein-Gordon determinant as a trace over a basis of smooth, delta-function normalizable functions,
\begin{align}
\log{\det{(\Delta_0+s(s-d))}}=\tr\log{(\Delta_0+s(s-d))}.
\end{align}
Differentiating both sides wrt. $s$, we have
\begin{align}
\frac{d}{ds}\log{\det{(\Delta_0+s(s-d))}}=(2s-d)\tr{(\Delta_0+s(s-d))^{-1}}\equiv (2s-d)\tr\hat{\mathcal{R}}(s),
\end{align}
where we have defined the resolvent of the Klein-Gordon operator, $\hat{\mathcal{R}}(s)\equiv(\Delta_0+s(s-d))^{-1}$. The trace of the resolvent can be expressed as an integral of a Green's function $G_s(x;x')$ over $EAdS_{d+1}$,
\begin{align}
\tr\hat{\mathcal{R}}(s)=\int_M d^{d+1}x\sqrt{g}G_s(x;x'=x),
\end{align}
where $G_s(x;x')$ satisfies
\begin{align}
\left(\Delta_{0x}+s(s-d)\right)G_{s}(x;x')=\delta^{d+1}(x;x').
\end{align} 
The determinant is then related to the Green's function as
\begin{align}
\frac{d}{ds}\log{\det{(\Delta_0+s(s-d))}}=(2s-d)\tr\hat{\mathcal{R}}(s)=(2s-d)\int_M d^{d+1}x\sqrt{g}G_s(x;x'=x).\label{eq:detgreens}
\end{align}

Now we meromorphically continue both the determinant and the resolvent/Green's function to the complex plane of $s$. We choose to continue the resolvent that is originally meromorphic on the half-plane $\Re{(s)}>d/2$ containing physical values $s>d/2$ (i.e. $s_+$), with only simple poles  at the location of $L^2$-normalizable bound states. In our case of thermal $AdS_3$ there are no such bound states. Next, one constructs a meromorphic continuation of the resolvent to the whole complex plane $s\in\mathbb{C}$ \cite{MAZZEO1987260}.\footnote{In general, the meromorphic continuation is only to $\mathbb{C}\backslash \frac{1}{2}(d-\mathbb{N})$, with possible essential singularities at $\frac{1}{2}(d-\mathbb{N})$ \cite{MAZZEO1987260}. But as pointed out in \cite{10.2307/2693876}\cite{guillarmou2005}, $\frac{1}{2}(d-\mathbb{N})$ contains at most poles of finite order for manifolds with constant curvature near infinity, such as thermal $AdS_3$. The poles at $\frac{1}{2}(d-\mathbb{N})$ are independent of the structure of the conformal boundary and can be absorbed into a polynomial $\text{Pol}(s)$ as was done in \cite{Denef:2009kn}.} The meromorphically continued resolvent is analytic on the half-plane $\Re{(s)}\geq d/2$, including the line $\Re{(s)}=d/2$, and contains finite-order poles on the $\Re{(s)}<d/2$ half-plane. By definition, these poles (that appear only when $x'=x$)(excluding $\frac{d}{2}(d-\mathbb{N})$) are the resonance poles $\{s_r\}$ with $\Re{(s_r)}<d/2$.

Given $G_s(x;x')$ meromorphic in $\mathbb{C}$ with only resonance poles $\{s_r\}$ and no bound states, the trace of the resolvent can be written as partial fractions,
\begin{align}
(2s-d)\int_M d^{d+1}x\sqrt{g}G_s(x;x'=x)=\sum_r \frac{m_r}{s-s_r}+\text{Pol}(s),
\end{align}
where $\text{Pol}(s)$ is a polynomial in $s$ and $m_r$ is a positive integer which counts the multiplicity (rank) of the resonance poles. Then, as meromorphic functions of $s$,
\begin{align}
\frac{d}{ds}\log{\det{(\Delta_0+s(s-d))}}&=\sum_r \frac{m_r}{s-s_r}+\text{Pol}(s),\;\text{or}\label{eq:logderi}\\
\det{(\Delta_0+s(s-d))}&=e^{\text{Pol}(s)}\prod_r (s-s_r)^{m_r}.
\end{align}

We have managed to express the determinant in terms of a product of resonance poles. To recover the prescription of \cite{Denef:2009kn}, we need to establish their correspondence with poles of the scattering operator, with multiplicities.

\subsubsection*{The Green's Function and the Scattering Operator}
The Poincar\'e metric for $EAdS_{d+1}$ space reads
\begin{align}
ds^2=\frac{dy^2+dx^idx_i}{y^2},\;y>0.
\end{align}
The Green's function we used is meromorphically continued from $\Re(s)>d/2$, in which case it is simply the well-known bulk-to-bulk propagator \cite{DHoker:1999mqo}\cite{Klebanov:1999tb}, which for $y\rightarrow 0$ \cite{Klebanov:1999tb}\cite{Giddings:1999qu} is
\begin{align}
G_{s}(y,\vec{x};y',\vec{x'})\xrightarrow{y\rightarrow 0}y^s \frac{K_{s}(\vec{x};y',\vec{x'})}{2s-d},
\end{align}
where $K_{s}(\vec{x};y',\vec{x'})$ is the bulk-to-boundary propagator,
\begin{align}
K_{s}(\vec{x};y',\vec{x'})={y'}^{s}\pi^{-d/2}\frac{\Gamma(s)}{\Gamma(s-d/2)}\frac{1}{({y'}^{2}+|\vec{x}-\vec{x'}|^2)^s}\xrightarrow{y'\rightarrow 0}{y'}^{d-s} \delta^{d}(\vec{x},\vec{x'}).
\end{align}
A smooth bulk free field $\varphi$ satisfying the Klein-Gordon equation is completely determined by the source $\varphi_0(\vec{x})$ on the boundary,
\begin{align}
\varphi(y,\vec{x})&=\int_{\partial M}d^dx'K_{s}(\vec{x};y'=y,\vec{x'})\varphi_0(\vec{x'})\\
&=y^{d-s}\left(\varphi_0(\vec{x})+O(y^2)\right)+y^{s}\left(A(\vec{x})+O(y^2)\right).
\end{align}
Expanding $K_{s}(\vec{x};y',\vec{x'})$ at small $y'$ away from $\vec{x}=\vec{x'}$ such that the delta function can be ignored, one deduces
\begin{align}
A(\vec{x})&=\int_{\partial M}d^d x' \pi^{-d/2}\frac{\Gamma(s)}{\Gamma(s-d/2)}\frac{1}{(y^{2}+|\vec{x}-\vec{x'}|^2)^s}\varphi_0(\vec{x'})\\
&=(2s-d)\int_{\partial M}d^d x' \left.\left((yy')^{-s}G_{s}(y,\vec{x};y',\vec{x'}) \right)\right|_{y,y'\rightarrow 0} \varphi_0(\vec{x'}).\label{eq:Greensfunction}
\end{align}
This relation is to be understood to be meromorphic in $s\in \mathbb{C}$, including the line $\Re{(s)}=d/2$.

On the other hand, when we consider a scattering problem in the spacetime $EAdS_{d+1}\times \mathbb{R}$ where $\mathbb{R}$ is an auxiliary time direction parametrized by $T$, we take $s=\frac{d}{2}-i\alpha$, $\alpha>0$, with an $e^{-i\Omega T}$ time dependence. Near the boundary $y\rightarrow 0$, the smooth spatial wavefunction $\varphi(y,\vec{x})$ becomes a superposition of the ingoing and outgoing wavefunctions
\begin{align}
\varphi(y,\vec{x})&\xrightarrow{y\rightarrow 0} y^{d-s}f_{in}(\vec{x})+y^{s}f_{out}(\vec{x})\overset{s=\frac{d}{2}-i\alpha}{=} y^{\frac{d}{2}+i\alpha}f_{in}(\vec{x})+y^{\frac{d}{2}-i\alpha}f_{out}(\vec{x}).
\end{align}
The scattering kernel, meromorphically continued to $s\in\mathbb{C}$, is defined to be the mapping of the ingoing wavefunction $f_{in}(\vec{x})$ to the outgoing wavefunction $f_{out}(\vec{x})$ of the smooth $\varphi(y,\vec{x})$ on the boundary,
\begin{align}
S(s)(\vec{x};\vec{x'})&:C^\infty(\partial M)\rightarrow C^\infty(\partial M)\text{, such that}\\
f_{out}(\vec{x})=&\int_{\partial M}d^dx' S(s)(\vec{x};\vec{x'}) f_{in}(\vec{x'}).\label{eq:Smatrix}
\end{align}
By comparing~\eqref{eq:Greensfunction} and~\eqref{eq:Smatrix}, we see that the source term $\varphi_0(\vec{x})$, being the most divergent at physical $s>d/2$, corresponds to the ingoing wave $f_{in}(\vec{x})$ and $A(\vec{x})$ corresponds to the outgoing one. The relation between the scattering kernel and the Green's function, both meromorphically continued, is:
\begin{equation}
S(s)(\vec{x};\vec{x'})=(2s-d)\left.(yy')^{-s}G_{s}(y,\vec{x};y',\vec{x'}) \right|_{y,y'\rightarrow 0},\label{eq:ressca}
\end{equation}
which means that the meromorphically continued (from $\Re{(s)}>d/2$) Green's function restricted on the conformal boundary, coincides with the meromorphically continued scattering kernel. This result is more rigorously derived in \cite{joshi2000}, \cite{10.2307/2951846} (for $d=1$) and reviewed in \cite{10.2307/2693876}. $S(s)(\vec{x},\vec{x'})$ defined under the decomposition of in- and out-wavefunctions is not well-defined when $2s=\mathbb{Z}$ or $s\in (-\infty,0]$; the meromorphically continued scattering kernel may have essential singularities there. But one can use~\eqref{eq:ressca} to define the continuation of $S(s)$ on the whole $\mathbb{C}$ using the Green's function. In effect one should consider the renormalized scattering kernel $S_R(s)$,
\begin{align}
S_R(s)&\equiv \frac{\Gamma(s-d/2)}{\Gamma(d/2-s)}S(s)\label{eq:Sren}, 
\end{align}
which satisfies $S_R(s)S_R(d-s)=S(s)S(d-s)=\mathds{1}$ and is finite-meromorphic on $\mathbb{C}$. From now on, we only consider $S_R(s)$ and drop the subscript $R$.

Define the scattering poles $\{s_\star\}$ as the poles of the renormalized scattering operator $S_R(s)$ on $\Re{(s)}<d/2$ (excluding $\frac{1}{2}(d-\mathbb{N})$). From~\eqref{eq:ressca}, \cite{10.2307/2693876}\cite{guillarmou2005} have proved that the resonance poles and the scattering poles coincide and have the same multiplicities: $\{s_r\}=\{s_\star\}$ and $m_r=\nu_\star$ (except when there exist bound states). The proof therein applies to any convex co-compact hyperbolic manifolds, i.e. orbifolds $M=\mathbb{H}^{d+1}/\Gamma$ where $\Gamma$ is a discrete subgroup of the isometry group of $\mathbb{H}^{n+1}$ with no parabolic elements, such that $M$ is of infinite volume and has no cusps. In three dimensions $\Gamma$ is a classical Schottky group freely generated by loxodromic generators. For thermal $AdS_3$ or BTZ black holes, $\Gamma\cong \mathbb{Z}$.

Thus we can expand the trace of the scattering operator $\hat{S}(s)$ as a partial fraction,
\begin{align}
\tr\hat{S}(s)&=\sum_{k,\omega}\langle k,\omega|\hat{S}(s)|k,\omega\rangle=\sum_{k,\omega}\sum_{\star_{k,\omega}}(s-s_{\star_{k,\omega}})^{-1}+\text{Pol}(s)\\
&=\sum_{\star}\frac{\nu_{\star}}{s-s_{\star}}+\text{Pol}(s)=\sum_{r}\frac{m_r}{s-s_{r}}+\text{Pol}(s),
\end{align}
where $k$ and $\omega$ are the Fourier mode labels in $EAdS_{d+1}$, and $\text{Pol}(s)$ is a polynomial in $s$. The prescription for finding the scattering/resonance poles is as follows. Solve the Klein-Gordon equation with $s\in\mathbb{C}$ for each Fourier mode. In the \textit{smooth} classical solution, the coefficients of the most divergent term when $s>d/2$ and that generating the other series are respectively identified as the ingoing and outgoing wavefunctions. The diagonal scattering matrix elements in Fourier basis are then the ratio of the latter to the former and the scattering poles $s_\star$ can be found accordingly. The corresponding smooth scattering pole solutions that satisfy the Klein-Gordon equation are by definition non-normalizable, in fact not even delta-function normalizable since they diverge exponentially at infinity. 

The resonance poles in the $s$-plane can be recast as poles in the complex plane of frequency $\omega$, with $s$ fixed and physical. When continued to Lorentzian signature, the solutions become quasinormal modes. However, we are reminded that for our purpose they are only a proxy for computing functional determinants at one-loop and have nothing to do here with conditions imposed at the classical level, such as those in~\cite{Brown:1986nw}. The above construction also proves the result in \cite{Bytsenko:2006ph}\cite{Bytsenko:2008wj}, where the heat kernel in Euclidean BTZ (/thermal $AdS_3$) is related to the Selberg Zeta function~\cite{Perry:2003}, whose zeros coincide with the scattering poles $\{s_\star\}$ with exactly half the multiplicity. 

Although we have only heuristically demonstrated the relation between the resonance and scattering 
poles for the scalar Laplacian, analogous relations are expected to hold for higher-spin first- and second-order elliptic operators. As concrete examples, ref. \cite{Datta:2011za} demonstrated, in terms of quasinormal modes in BTZ background, that the higher-spin Laplace determinants thus obtained agree with the heat kernel results \cite{David:2009xg} up to local polynomial terms. A mathematical proof is also given in \cite{GUILLARMOU20102464} for the Dirac operator in odd-dimensional convex co-compact hyperbolic manifolds.

\subsubsection*{The Local Contributions \texorpdfstring{$\exp(\text{Pol}(s))$}{exp(Pol(s))}}
The determinant as a meromorphic function of $s$ is determined completely by the zeros (there are none here), poles which we have just discussed, and its behavior at $s\rightarrow \infty$. The factor $\exp{(\text{Pol}(s))}$ needs to be determined independently from the large-$s$ behavior of the determinant, which can be done using the heat kernel expansion in small-$t$. In this limit $s\rightarrow \infty$, the field becomes very massive. As demonstrated by \cite{Denef:2009kn} in various examples, $\text{Pol}(s)$ is a polynomial in $s$ whose coefficients are local, i.e. spacetime integrals of local curvature invariants and the background fields. 

In thermal $AdS_3$, these coefficients cannot depend on the complex structures $q$ and $\bar{q}$ of the \textit{conformal} boundary since these are non-local quantities, so in particular they cannot take the form $\log{(1-q^a\bar{q}^b)}$ and modify the infinite products which we will obtain from the resonance poles. However, they \textit{can} depend on $\tau_2$ (but not $\tau_1$) since $\tau_2$ characterizes the (holographically renormalized) volume of the manifold. In what follows, instead of doing heat kernel expansions, we will simply obtain $\text{Pol}(s)$ by matching with existing exact determinants from heat kernel calculations \cite{Giombi:2008vd}.

This establishes 
\begin{align}
\det{(\Delta_0+s(s-d))}=e^{\text{Pol}(s)}\prod_r (s-s_r)^{m_r}=e^{\text{Pol}(s)}\prod_\star (s-s_\star)^{\nu_\star},
\end{align}
with $\text{Pol}(s)$ a polynomial of $s$ with local coefficients.

\subsection{Warm-up: Scalar Scattering Poles in Thermal \texorpdfstring{$AdS_3$}{AdS3}}\label{sec:scares}
The calculation here follows closely ref.~\cite{Castro:2017mfj}; we work in the metric~\eqref{eq:xmetric}. To find scalar resonances modes in thermal $AdS_3$ regular at the origin, we Fourier-expand the scalar $\varphi$ with the correct periodicity,
\begin{align}
\varphi&=R(x)e^{-i\omega t+ik\phi}=R(x)e^{-ik_w w-ik_{\bar{w}} {\bar{w}}},\;k\in \mathbb{Z},\;\omega=\frac{-n+k\tau_1}{\tau_2},\;n\in \mathbb{Z}.
\end{align}

A scattering pole solution obeys the Klein-Gordon equation
\begin{align}
(\Delta_0+s(s&-2))\varphi=0\Rightarrow x(1-x)R''+(1-x)R'+\left(-\frac{k^2}{4x}-\frac{\omega^2}{4}-\frac{s(s-2)}{4(1-x)} \right)R=0.
\end{align}
The solutions regular at the origin are \cite{Zwillinger}
\begin{align}
R(x)=(1-x)^{\frac{s}{2}}x^{\frac{|k|}{2}}F(\frac{s+|k|+i\omega}{2},\frac{s+|k|-i\omega}{2};1+|k|;x),\;&k\in \mathbb{Z},
\end{align}
where $F(a,b;c;x)={}_2F_1(a,b;c;x)$ is the hypergeometric function. Expanding around infinity ($x=1$) and normalizing the coefficient of $(1-x)^{\frac{2-s}{2}}$, we find
\begin{align}
\hspace*{-1cm}R\rightarrow &(1-x)^{\frac{s}{2}}\left(\frac{\Gamma(1-s)}{\Gamma(s-1)}\frac{\Gamma(\frac{s+|k|+i\omega}{2})\Gamma(\frac{s+|k|-i\omega}{2})}{\Gamma(\frac{2-s+|k|+i\omega}{2})\Gamma(\frac{2-s+|k|-i\omega}{2})} +O(1-x)\right)+(1-x)^{\frac{2-s}{2}}\left(1 +O(1-x)\right).\label{eq:scalarquasinormal}
\end{align}

A diagonal scattering matrix element is identified as the ratio of the coefficients of the $(1-x)^\frac{s}{2}\sim e^{-s \xi}$ term to that of the $(1-x)^\frac{2-s}{2}\sim e^{-(2-s) \xi}$ term, the latter being most divergent at $s>d/2=1$. In other words,
\begin{align}
\langle k\geq 0,\omega|\hat{S}(s)|k\geq 0,\omega\rangle&=\frac{\Gamma(\frac{s+k+i\omega}{2})\Gamma(\frac{s+k-i\omega}{2})}{\Gamma(\frac{2-s+k+i\omega}{2})\Gamma(\frac{2-s+k-i\omega}{2})}  ,\\
\langle k<0,\omega|\hat{S}(s)|k<0,\omega\rangle&=\frac{\Gamma(\frac{s-k+i\omega}{2})\Gamma(\frac{s-k-i\omega}{2})}{\Gamma(\frac{2-s-k+i\omega}{2})\Gamma(\frac{2-s-k-i\omega}{2})}.
\end{align}
These matrix elements have poles $s_\star$ at
\begin{equation}
\begin{aligned}
\begin{dcases}
\frac{s_\star+|k|+i\omega}{2}&=-p\text{ or }\\
\frac{s_\star+|k|-i\omega}{2}&=-p
\end{dcases},\;p=0,1,\ldots.
\end{aligned}
\end{equation}
All $s_\star$ indeed have $\Re{(s_\star)}<d/2$, which is consistent with the fact that they are also poles of the resolvent that is meromorphically continued from the half-plane $\Re{(s)}>d/2$, where it is analytic.

Therefore, substituting $\omega=\frac{-n+k\tau_1}{\tau_2}$ we have the determinant
\begin{align}
&\quad\det{(\Delta_0+s(s-2))}\propto\prod_{s_\star}(s-s_\star)^{\nu_\star}\nonumber\\
&\;\nonumber\\
&\propto \prod_{\substack{k>0\\p\geq0\\n\in \mathbb{Z}}}\left( n^2+\left[\tau_2(s+2p+k)-ik\tau_1\right]^2 \right)\left( n^2+\left[\tau_2(s+2p+k)+ik\tau_1\right]^2 \right)\times \prod_{\substack{p\geq0\\n\in \mathbb{Z}}} \left( n^2+\tau_2^2(s+2p)^2 \right)\nonumber\\
&\propto \prod_{\substack{k>0\\p\geq0}}\left(1-e^{-2\pi [\tau_2(s+2p+k)-ik\tau_1]}\right)^2\left(1-e^{-2\pi [\tau_2(s+2p+k)+ik\tau_1]}\right)^2\left(1-e^{-2\pi \tau_2(s+2p)}\right)^2\nonumber\\
&=\prod_{\substack{k>0\\p\geq0}}\left(1-q^k (q\bar{q})^{p+\frac{s}{2}}\right)^2\left(1-(q\bar{q})^{p+\frac{s}{2}}\bar{q}^k\right)^2\left(1-(q\bar{q})^{p+\frac{s}{2}}\right)^2=\prod^\infty_{l,l'=0}\left(1-q^{l+\frac{s}{2}}\bar{q}^{l'+\frac{s}{2}}\right)^2.
\end{align}
We have discarded local terms absorbed in $\exp{(\text{Pol}(s))}$. From this we can match with the heat kernel computation,~eq.\eqref{eq:hksca}, to obtain the local term $\exp{(\text{Pol}(s))}$, which is proportional to the holographically renormalized volume factor \cite{Karch:2005ms} $\text{Vol}_{ren}(\mathbb{H}^3/\Gamma)=-\pi^2 \tau_2$.
We finally obtain
\begin{equation}
\det{(\Delta_0+s(s-2))}=q^{-\frac{(s-1)^3}{24}}\bar{q}^{-\frac{(s-1)^3}{24}}\prod^\infty_{l,l'=0}\left(1-q^{l+\frac{s}{2}}\bar{q}^{l'+\frac{s}{2}}\right)^2.\label{eq:Delta0}
\end{equation}
To this result one substitutes the physical value $s_+>d/2$, since the resolvent we used was continued from the $\Re{(s)}>d/2$ half-plane where it is analytic.

\section{Covariant Gauge Fixing in Chern-Simons Theory of a Compact Group}\label{sec:gfcom}
\subsection{Gauge Fixing for Compact Gauge Groups}
We will use a covariant gauge-fixing for compact gauge groups around a background connection, following \cite{Witten:1988hf}\cite{BarNatan:1991rn}. Consider a compact gauge group $H$ and a level $k$ chosen to be \underline{positive}. We will compute the path integral as a sum of the steepest descent integral around each saddle point $\bar{A}_\sigma$ whose contribution is counted by $\mathfrak{n}_\sigma$:
\begin{align}
Z_H(k,\zeta)&=\int_{\mathcal{U}_H} DA\exp{\left( -(-ikW[A]+I_{bt}[A]) \right)}=\sum_{\sigma}\mathfrak{n}_\sigma Z_{H,\mathcal{J}_\sigma}(k,\zeta).\label{eq:expsadd}
\end{align}
The $\mathfrak{n}_\sigma$ are determined by the defining path of integration $\mathcal{U}_H$, as explained in \cite{Witten:2010cx}.

Recalling that the Killing metric $\Tr$ is negative-definite we define the $L^2$-norm for $\mathfrak{h}$-valued scalars and vectors as
\begin{align}
(\varphi,\varphi')=-\int_M \sqrt{g} {\rm Tr}[\varphi\varphi'],\;(A,A')=-\int_M \sqrt{g} g^{\mu\nu}{\rm Tr}[A_\mu A'_\nu].\label{eq:L2norm}
\end{align}
Since $H$ is compact, an $\mathfrak{h}$-valued field $\psi=\psi^aT^a$ has $\psi^a$ real so the $L^2$-norm is a positive-definite inner product without complex conjugation; this is no longer true when the field is complex-valued or when the group is non-compact.

Working in the background field gauge, we first expand $A$ around a flat connection $\bar{A}$ and consider the fluctuation $A'$ (dropping the prime from now on). The Chern-Simons action is invariant under a proper gauge transformation that vanishes at infinity. For $H$ compact, we use the Landau gauge in the BRST formalism. With the adjoint-valued ghosts $c$ and $\bar{c}$ and the auxiliary scalar $\varphi$, the BRST transformations ($D=\nabla +[\bar{A},\cdot\;]$) are
\begin{equation}
\begin{aligned}
\delta_B A&=-Dc,\qquad&\delta_B \bar{c}&&=i\varphi,\\
\delta_B c^a&=\frac{1}{2}f^{a}_{\;\;bc}c^bc^c,\qquad&\delta_B \varphi&&=0,
\end{aligned}
\end{equation}
where $f^a_{\;\;bc}$ are the structure constants of $H$.
Adding the gauge-fixing term 
\begin{align}
V=\frac{k}{2\pi}\int_M \sqrt{g} {\rm Tr}[\bar{c}D^{\mu}A_\mu]
\end{align}
the total gauge-fixed action is
\begin{align}
I[A]-\delta_B V&=I[\bar{A}]-\frac{ik}{4\pi}\int_M {\rm Tr}[ADA+2\sqrt{g}\varphi D^\mu A_\mu]+\frac{k}{2\pi}\int_M \sqrt{g}{\rm Tr}[\bar{c}D^\mu D_\mu c]\\
&=I[\bar{A}]+\left( \frac{k}{2\pi}\right)\frac{i}{2}(H,L_- H)+(\bar{c},\Delta_0 c).\label{eq:Igf}
\end{align}
In the second line we have canonically normalized $c$ and $\bar{c}$, and discarded the boundary terms involving the fluctuations, which we can do because $D^\mu$ is self-adjoint under the $L^2$-norm~\eqref{eq:L2norm}. In the last line we have denoted
\begin{equation}
\begin{aligned}
H=\begin{pmatrix}
A\\ \varphi
\end{pmatrix},\;L_-=\begin{pmatrix}
\frac{\varepsilon^{\mu\nu\rho}}{\sqrt{g}}D_\nu & -D^\mu \\
D^\rho & 0
\end{pmatrix}
\end{aligned}
\end{equation}
and $\Delta_0\equiv -D^2$, $\Delta_1\equiv L_-^2$. When the background connection is trivial both are positive-definite.

\subsection{The Steepest Descent Path}\label{sec:SDH}
Since $L_-$ is a first-order, self-adjoint operator and has a real continuous spectrum, the Gaussian integral involving $H$ is oscillatory. In other words, regardless of what each background $\bar{A}_\sigma$ is, the initial path $\mathcal{U}_H$ is never the steepest descent path $\mathcal{J}_{\sigma}$; the latter is some middle-dimensional space in $\mathcal{U}_{H_\mathbb{C}}$. As pointed out in \cite{Witten:2010cx}, the steepest descent path $\mathcal{J}_{\sigma}$ is uniquely determined by the Morse function chosen to be the real part of the action (at least when the boundary terms drops out, as in the present case for the one-loop fluctuations). In a Gaussian integral it varies continuously with the parameters $k$ and $\zeta$ defining the path integral $Z_H(k,\zeta)$. In particular, it is independent of the choice of $H$ which defines the initial integration cycle $\mathcal{U}_H$, so $Z_{H,\mathcal{J}_\sigma}(k,\zeta)=Z_{\mathcal{J}_\sigma}(k,\zeta)$. One can in fact consider any real forms $H'$ of $H_{\mathbb{C}}$ defining the initial path $\mathcal{U}_{H'}$-- for $H_\mathbb{C}=SL(2,\mathbb{C})$, $H'$ can be $SU(2)$, $SL(2,\mathbb{R})$ or $SU(1,1)$-- the steepest descent path defined by $\bar{A}_\sigma$ is always 
the same. The intersection number $\mathfrak{n}_{\sigma}$ may be different but their determination is beyond the scope of discussion of this work.

We will not solve the steepest descent equation in \cite{Witten:2010cx} for general complex background connections. Instead we will consider a real $\bar{A}^a(x)$ that is parametrized by some real parameters such as the chemical potential. When
 $H=SU(2)$ we can use the gauge fixing given in the previous section to find the corresponding unique steepest descent path unambiguously. After that, we continue the final answer to complex values of the parameters defining the path integral, which corresponds to assuming that the steepest descent path varies continuously with the parameters. As we will see, 
this procedure gives a result that agrees with the Kac-Moody characters, which we take as a suggestion that our assumption is valid.

$L_-$ is self-adjoint under the $L^2$-norm~\eqref{eq:L2norm}, so when the background $\bar{A}^a(x)$ is real, the spectrum of $L_-$ is continuous on $\mathbb{R}$ and unbounded. Denote the eigenfunctions of $L_-$ as $H_i=H_i^a T^a$, normalized to $(H_i,H_j)=+\delta_{ij}$ and $H_i^a$ real,
\begin{align}
L_-H_i=\lambda_iH_i,\;\lambda_i\in \mathbb{R}.
\end{align}
This eigenvalue problem is independent of $k$. Expand $H=\sum_i c_i H_i$, $c_i\in \mathbb{R}$, and evaluate the one-loop path integral on $H$, (recall $Z\sim e^{-I}$)
\begin{align}
Z(k>0)\sim\int DH\exp{\left(-\left( \frac{k}{2\pi}\right)\frac{i}{2}(H,L_-H) \right)}&=\int DH\exp{\left(-\frac{i}{2}(H,L_-H) \right)}\nonumber\\
&=\prod_i\int_{-\infty}^\infty dc_i \exp{\left(-\frac{i}{2}\lambda_i (c_i)^2 \right)},\;\lambda_i\in\mathbb{R}.\label{eq:SD1eqn}
\end{align}
We have canonically normalized the field $H$ by $H\rightarrow H/\sqrt{2\pi k}$, which is real and positive when $k>0$. The Gaussian integral for each eigenvalue $\lambda_i$ is oscillatory and needs to be regularized. Equivalently, the path of integration needs to be deformed to a steepest descent path \cite{Witten:1988hf}\cite{BarNatan:1991rn}\cite{Witten:2010cx}. To do this we deform $L_-$,
\begin{align}
\int DH\exp{\left(-\frac{i}{2}(H,L_-H) \right)}&\rightarrow\int DH\exp{\left(-\frac{i}{2}(H,L_-H) -\frac{1}{2}\varepsilon (H,H)\right)}\label{eq:phaseL}\\
&=\prod_i\int_{-\infty}^\infty dc_i \exp{\left(-\frac{i}{2}\left(\lambda_i-i\varepsilon\right) (c_i)^2\right)}\\
&\propto\prod_i\frac{1}{\sqrt{\varepsilon+i \lambda_i}}=\prod_i\frac{1}{\sqrt{|\lambda_i|}}\exp{\left(-\frac{i\pi}{4}\text{sign}(\lambda_i)\right)}.
\end{align}
The second line clearly indicates that, for a real background connection and given $k>0$, the overall sign of the action uniquely determines the deformation of $L_-$, i.e. the ``$i\varepsilon$" prescription. In the present case, we must consider the operator $(L_--im)$, $m>0$ and take $m=0^+$ in the end. Crucially, there is a branch cut associated to the square root and in the last line we have chosen the principal branch $\rm{Arg}{(z)}\in(-\pi,\pi]$. This branch cut will become important later.

\begin{figure}[ht]
\begin{center}
\hspace*{1.6cm}\includegraphics[width=0.55\linewidth,keepaspectratio]{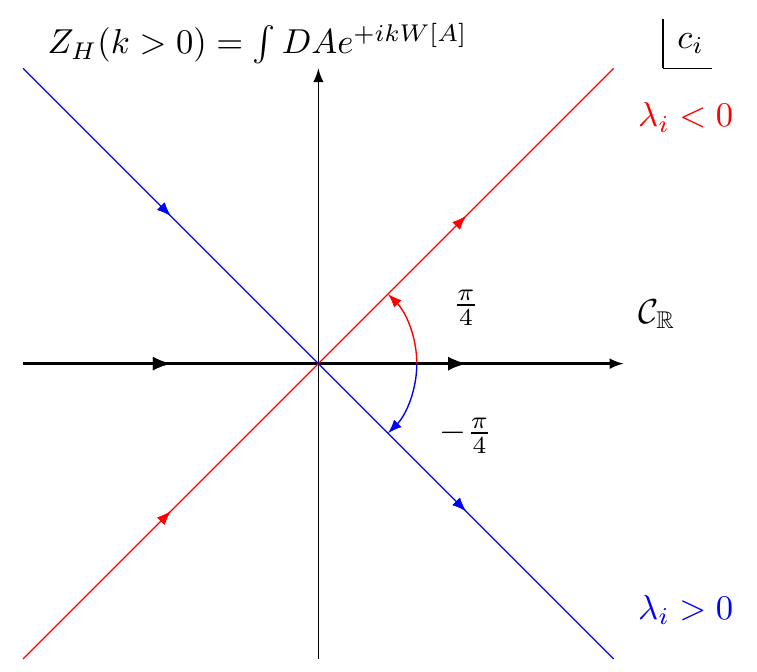} 
\caption{Steepest Descent Paths of the $L_-$ path integral~\eqref{eq:SD1eqn}.}\label{fig:SD1a}
\end{center}
\end{figure}

Equivalently, one analytically continues each Gaussian integral into a line integral in the complex $c_i$-plane along the real axis. One then deforms the real contour $\mathcal{C}_{\mathbb{R}}$ to the steepest descent path, $c_i=\exp{(-i\text{sign}(\lambda_i)\pi/4)}\mathbb{R}$, which renders the integral convergent; see \hyperref[fig:SD1a]{Fig.\ref*{fig:SD1a}}. Furthermore, the orientation of this contour is uniquely determined by closing a Cauchy contour in such a way that the segments at infinity do not contribute to the integral. Here we have chosen the same principal branch as above. Hence the steepest descent path is \textit{unique} and equivalent to the $(L_--i\varepsilon)$ deformation. The Jacobians of the integration measures then give rise to a total phase of $\exp{(-\sum_i\text{sign}(\lambda_i)i\pi/4)}$. Since there are infinitely many eigenvalues $\lambda_i$, we regularize this phase by first defining
\begin{align}
\eta(L_-,s)\equiv \sum_i \text{sign}(\lambda_i)|\lambda_i|^{-s}.
\end{align}
The eta-invariant is then defined by analytically continuing 
$\eta(L_-,s)$ to $\eta(L_-,0)$, $\eta(L_-)\equiv\eta(L_-,0)$. Hence we have
\begin{align}
\frac{1}{\sqrt{\det{(L_--i\varepsilon)}}}=\frac{1}{\sqrt{|\det{(L_--i\varepsilon)}|}}e^{-\frac{i\pi}{4}\eta(L_-)}.\label{eq:etainv}
\end{align}

To evaluate $\det{(L_--im)}$ for $m>0$, consider the resolvent
\begin{align}
\hat{\mathcal{R}}_-(m)\equiv (L_--im)^{-1}=\begin{pmatrix}
\frac{\varepsilon^{\mu\nu\rho}}{\sqrt{g}}D_\nu\bm{-}img^{\mu\rho} & -D^\mu \\
D^\rho & \bm{-}im
\end{pmatrix}^{-1},
\end{align}
initially analytic on the $\Re{(m)}>0$ half-plane containing the physical values $m>0$. The continuous spectrum of $L_-$ is $\mathbb{R}$, which in terms of $m$ becomes the line $\Re{(m)}=0$. The longitudinal mode of the vector gets mixed with the scalar mode $\varphi$, hence it is natural to factorize $\tr\hat{\mathcal{R}}_-(m)$ into the scalar/longitudinal part as a meromorphic function of $s$, with $m^2=s(s-2)$, and a transverse part as a meromorphic function of $m$. The former is insensitive to the $i\varepsilon$ prescription and we can simply apply the scalar result computed in \hyperref[sec:scares]{Section \ref*{sec:scares}}. 

For the transverse part of the resolvent $\hat{\mathcal{R}}^{(T)}_-(m)$, we assume the following to be true in thermal $AdS_3$ and we expect the proof to be similar to that given in~\cite{GUILLARMOU20102464} for the Dirac operator. The meromorphic continuation of $\hat{\mathcal{R}}^{(T)}_-(m)$ is analytic on the ${\rm Re}(m)\geq 0$ half-plane including $\Re{(m)}=0$, and only has finite-order poles on the ${\rm Re}(m)<0$ half-plane, which by definition are the resonance poles. Furthermore a
relation  analogous to~\eqref{eq:ressca} holds between $\hat{\mathcal{R}}_-(m)$ and the scattering operator $\hat{S}(m)$, which implies that the resonance poles counted with their multiplicities coincide with the scattering poles.

We remark that the resonance solutions are in general complex; therefore, their $L^2$-norm~\eqref{eq:L2norm} is not positive-definite. This is not a problem because the resonance solutions never (need to) satisfy any normalizability condition other than not being bound states. 

\section{The \texorpdfstring{$\widehat{\mathfrak{u}}(1)$}{u(1)-hat} Vacuum Character}\label{u1vac}
Let us begin with the prototypical example of the $U(1)$ Chern-Simons Theory with a trivial background $\bar{A}=0$. To match with the convention used in the non-Abelian case, where the Killing metric $\Tr$ is negative-definite, we use the path integral
\begin{align}
Z(k)=\int_{\mathcal{U}_{U(1)}}DA\exp{\left(-\frac{ik}{4\pi}\int_M AdA\right)},\;k>0
\end{align}
and the positive-definite $L^2$-norm $(u,v)=+\int_Md\mu \;u\star v$. In this case the $i\varepsilon$ prescription for $L_-$ is again $(L_--i\varepsilon)$.

\subsection{\texorpdfstring{$\det{L_-}$}{det(L)}: The Scattering Poles}\label{sec:vectorpole}
\subsubsection*{The Scattering Solutions}
The computation done here is similar to that in \cite{Datta:2011za}\cite{Castro:2017mfj}, but in Euclidean signature. To compute $\det\nolimits^{(T)}{(L_--i\varepsilon)}$, where the superscript $T$ denotes the transverse part, we look for transverse resonance modes regular at the origin. They satisfy $(L_--im)H=0$, i.e.
\begin{empheq}{align}
\frac{\varepsilon^{\mu\nu\rho}}{\sqrt{g}}\nabla_\nu A_\rho&=\nabla^\mu \varphi+im A^\mu\label{eq:Lminus1st1} ,\\
\nabla^\mu A_\mu&=+im\varphi .\label{eq:Lminus1st2}
\end{empheq}
Contracting~\eqref{eq:Lminus1st1} with $\sqrt{g}\varepsilon_{\alpha\beta\mu}\nabla^\beta$ and $\nabla_\mu$ respectively, we get the vector and scalar Laplace equations
\begin{align}
(\Delta_1+m^2)A_\mu=0,\;(\Delta_0+s(s-2))\varphi=0,
\end{align}
where $\Delta_1=-(\nabla^\rho \nabla_\rho+2)$ is the positive-definite vector Laplacian. Fourier-expanding\footnote{The angular quantum number $k$ should not be confused with the Chern-Simons level $k$; which is which should be clear from the context.}
\begin{align}
(A_\mu,\varphi)&=(R_\mu(x),R(x))e^{-i\omega t+ik \phi},\;k\in \mathbb{Z},\;\omega=\frac{-n+k\tau_1}{\tau_2},\;n\in\mathbb{Z},
\end{align}
the equations of motion read ($\varepsilon_{x\phi t}=\varepsilon^{x \phi t}=+1$)
\begin{subequations}
\begin{empheq}{align}
2x (R'+imR_x)&=ik R_t+i\omega R_\phi\label{eq:m1stvec3}\\
-ik R-imR_\phi&=2x(1-x)(R'_t+i\omega R_x)\label{eq:m1stvec4}\\
-i\omega R+imR_t&=2(1-x)(R'_\phi-ik R_x)\label{eq:m1stvec5}\\
4x(1-x)(xR_x)'&=-ik R_\phi+ix\omega R_t+\frac{imx}{1-x}R.\label{eq:m1stvec6}
\end{empheq}
\end{subequations}
They imply the following inhomogeneous second-order Laplace equations, in terms of $R_{w,\bar{w}}=\frac{1}{2}(R_\phi\mp iR_t)$ and $k_{w,\bar{w}}=\frac{\mp i\omega-k}{2}$,
\begin{equation}
\begin{aligned}
\begin{cases}
x(1-x)R_w''+(1-x)R'_w+\left( -\frac{k^2}{4x}-\frac{\omega^2}{4}-\frac{(m-1)^2-1}{4(1-x)}\right)R_w&=\frac{+k_{w}}{2(1-x)}R\\
x(1-x)R_{\bar{w}}''+(1-x)R'_{\bar{w}}+\left( -\frac{k^2}{4x}-\frac{\omega^2}{4}-\frac{(m+1)^2-1}{4(1-x)}\right)R_{\bar{w}}&=\frac{+k_{\bar{w}}}{2(1-x)}R\label{eq:m2ndvec}
\end{cases}.
\end{aligned}
\end{equation}
The regularity of $R$ and $R_{w,\bar{w}}$ implies their general solutions (for any $m\neq0$; $m=0$ is not a resonance pole)
\begin{align}
R&=(1-x)^{\frac{s}{2}}x^{\frac{|k|}{2}}F(\frac{s+|k|+i\omega}{2},\frac{s+|k|-i\omega}{2};1+|k|;x),\\
R_w&=e_w(1-x)^{\frac{m}{2}}x^{\frac{|k|}{2}}F(\frac{m+|k|+i\omega}{2},\frac{m+|k|-i\omega}{2};1+|k|;x)+\frac{k_{w}}{m}R,\\
R_{\bar{w}}&=e_{\bar{w}}(1-x)^{-\frac{m}{2}}x^{\frac{|k|}{2}}F(\frac{-m+|k|+i\omega}{2},\frac{-m+|k|-i\omega}{2};1+|k|;x)+\frac{k_{\bar{w}}}{m}R.
\end{align}
Regularity also requires that $R_{\phi}=R_w+R_{\bar{w}}$ vanish at the origin $x=0$ because the $\phi$-coordinate parametrizes the contractible cycle. All solutions with $k\neq 0$ already have $R_{\phi}(x=0)=0$. Imposing this requirement on the $k=0$ solutions implies that $e_w+e_{\bar{w}}=0$. From~\eqref{eq:m1stvec3} then we automatically have $R_\xi$ regular at the origin for all $k\in\mathbb{Z}$ (we use $\xi$ instead of $x$ since $g^{\xi\xi}=1$).

Moreover, the equations of motion are first-order: using~\eqref{eq:m1stvec4} and~\eqref{eq:m1stvec5} we have a relation between the polarization constants $e_{w,\bar{w}}$ for $k\neq 0$, but not $k=0$. Together with the condition $e_w+e_{\bar{w}}=0$ for the $k=0$ solutions, we have the relations between $e_{w,\bar{w}}$,
\begin{equation}
\left\{\begin{alignedat}{3}
e_w&=-e_{\bar{w}}\quad&&,\quad&&k=-(k_w+k_{\bar{w}})=0,\\
(m+k-i\omega)e_w&=(-m+k+i\omega)e_{\bar{w}}\quad&&,\quad&&k=-(k_w+k_{\bar{w}})>0,\label{eq:polcons}\\
(m+|k|+i\omega)e_w&=(-m+|k|-i\omega)e_{\bar{w}}\quad&&,\quad&&k=-(k_w+k_{\bar{w}})<0.
\end{alignedat}\right.
\end{equation}
The scalar mode $R$ enters into $R_{w,\bar{w}}$ (and $R_x$) as a particular solution and longitudinal mode, while the complementary solutions are the transverse vector modes. This justifies our factorization of the resolvent into the longitudinal/scalar one as a meromorphic function of $s$, and the transverse one as a meromorphic function of $m$. The scattering poles of the longitudinal/scalar modes were already computed in \hyperref[sec:scares]{Section \ref*{sec:scares}}. In the rest of this subsection we only consider the transverse part of the solutions and drop the particular solution.

\subsubsection*{Defining the Scattering Matrix}\label{sec:defLscmatrix}
Expanding the transverse part of $R_{w,\bar{w}}$ near infinity (i.e. around $x=1$), we obtain
\begin{align}
\begin{split}
R_{w}&\rightarrow e_w\left[(1-x)^{\frac{m}{2}}\frac{\Gamma(1+|k|)\Gamma(-m+1)}{\Gamma(\frac{2-m+|k|+i\omega}{2})\Gamma(\frac{2-m+|k|-i\omega}{2})}+O(1-x)\right]\\
&\qquad+e_w\left[(1-x)^{\frac{2-m}{2}}\frac{\Gamma(1+|k|)\Gamma(m-1)}{\Gamma(\frac{m+|k|+i\omega}{2})\Gamma(\frac{m+|k|-i\omega}{2})}+O(1-x)\right],
\end{split}\\
\begin{split}
R_{\bar{w}}&\rightarrow e_{\bar{w}}\left[(1-x)^{-\frac{m}{2}}\frac{\Gamma(1+|k|)\Gamma(m+1)}{\Gamma(\frac{2+m+|k|+i\omega}{2})\Gamma(\frac{2+m+|k|-i\omega}{2})}+O(1-x)\right]\\
&\qquad +e_{\bar{w}}\left[(1-x)^{\frac{2+m}{2}}\frac{\Gamma(1+|k|)\Gamma(-m-1)}{\Gamma(\frac{-m+|k|+i\omega}{2})\Gamma(\frac{-m+|k|-i\omega}{2})}+O(1-x)\right]
\end{split}
\end{align}
for all $k$. As with scalar, in a solution regular at the origin we identify the most divergent term (when $m>0$) as the ingoing wave and the leading term of the other series as the outgoing wave.  For the transverse modes, we take the $(1-x)^{-\frac{m}{2}}$ in $R_{\bar{w}}$ as incoming and the $(1-x)^{+\frac{m}{2}}$ in $R_{w}$ as outgoing. The diagonal scattering matrix elements are the ratio of the latter's coefficient to the former's.

\subsubsection*{The Transverse Scattering Poles}

For the poles of the transverse modes, we first normalize the $(1-x)^{-\frac{m}{2}}$ term in $R_{\bar{w}}$ then plug in the polarization relations~\eqref{eq:polcons}.

For $k=0$, since $e_w+e_{\bar{w}}=0$ by regularity, we have
\begin{align}
R_{w}&\rightarrow -(1-x)^{\frac{m}{2}}\frac{\Gamma(-m+1)}{\Gamma(m+1)}\underbrace{\frac{\Gamma(\frac{2+m+i\omega}{2})\Gamma(\frac{2+m-i\omega}{2})}{\Gamma(\frac{2-m+i\omega}{2})\Gamma(\frac{2-m-i\omega}{2})}}_{\equiv \langle k= 0,\omega|\hat{S}(m)|k= 0,\omega\rangle}+\ldots,\;R_{\bar{w}}\rightarrow (1-x)^{-\frac{m}{2}}+\ldots.
\end{align}
For $k>0$, 
\begin{align}
R_{w}&\rightarrow (1-x)^{\frac{m}{2}}\frac{\Gamma(-m+1)}{\Gamma(m+1)}\underbrace{\frac{\Gamma(\frac{2+m+k+i\omega}{2})\Gamma(\frac{m+k-i\omega}{2})}{\Gamma(\frac{-m+k+i\omega}{2})\Gamma(\frac{2-m+k-i\omega}{2})}}_{\equiv \langle k> 0,\omega|\hat{S}(m)|k> 0,\omega\rangle}+\ldots,\;R_{\bar{w}}\rightarrow (1-x)^{-\frac{m}{2}}+\ldots,
\end{align}
while for $k<0$, 
\begin{align}
R_{w}\rightarrow (1-x)^{\frac{m}{2}}\frac{\Gamma(-m+1)}{\Gamma(m+1)}\underbrace{\frac{\Gamma(\frac{m+|k|+i\omega}{2})\Gamma(\frac{2+m+|k|-i\omega}{2})}{\Gamma(\frac{2-m+|k|+i\omega}{2})\Gamma(\frac{-m+|k|-i\omega}{2})}}_{\equiv \langle k<0,\omega|\hat{S}(m)|k<0,\omega\rangle}+\ldots,\;R_{\bar{w}}\rightarrow (1-x)^{-\frac{m}{2}}+\ldots.
\end{align}

The scattering poles are

\begin{equation}
\begin{aligned}
 k= 0:&\begin{cases}
\frac{2+m_\star+i\omega}{2}\quad\;&=-p,\;\text{or}\\
\frac{2+m_\star-i\omega}{2}&=-p
\end{cases}\\
k> 0:&\begin{cases}
\frac{2+m_\star+k+i\omega}{2}&=-p,\;\text{or}\\
\frac{m_\star+k-i\omega}{2}&=-p
\end{cases}\\
k<0:&\begin{cases}
\frac{m_\star-k+i\omega}{2}&=-p,\;\text{or}\\
\frac{2+m_\star-k-i\omega}{2}&=-p
\end{cases}\label{eq:transversepoles}
\end{aligned}\quad,\quad p=0,1,\ldots.
\end{equation}
All poles indeed have $\Re{(m_\star)}<0$, which is consistent with that they are also poles of the resolvent $\hat{\mathcal{R}}_-(m)$ of $(L_--im)$ that is meromorphically continued from $\Re{(m)}>0$ where it is analytic.

\subsubsection*{The determinant \texorpdfstring{$\det{(L_--im)}$}{det(L)}}
Substituting $\omega=\frac{-n+k\tau_1}{\tau_2}$, the transverse $(T)$ part of $\det{(L_--im)}$ becomes
\begin{align}
&\quad\det\nolimits{^{(T)}}{(L_--im)}\propto \prod_{\star}(m-m_\star)^{\nu_\star}\nonumber\\
&\propto \prod_{\substack{k>0\\p\geq0\\n\in \mathbb{Z}}}%
    \begin{array}{c}
      \left(n^2+\left[\tau_2(2(p+1)+k+m)+ik\tau_1\right]^2\right)\\
      \left(n^2+\left[\tau_2(2p+k+m)-ik\tau_1\right]^2\right)
    \end{array}\times\prod_{\substack{p\geq0\\n\in \mathbb{Z}}}\left(n^2+\left[\tau_2(2(p+1)+m)\right]^2\right)\nonumber
\end{align}
\begin{align}
&\propto \prod_{\substack{k>0\\p\geq0}}\left(1-e^{-2\pi\left[\tau_2 (2(p+1)+k+m)+ik\tau_1\right]}\right)^2\left(1-e^{-2\pi\left[\tau_2 (2p+k+m)-ik\tau_1\right]}\right)^2\times\prod_{p\geq0}\left(1-e^{-2\pi\left[\tau_2(2(p+1)+m)\right]}\right)^2\nonumber\\
&=\prod_{l,l'=0}^\infty \left(1-q^{l+1+\frac{m}{2}}\bar{q}^{l'+\frac{m}{2}}\right)^2.\label{eq:transverse0}
\end{align}
We have again discarded some local terms which are absorbed in $\exp{(\text{Pol}(m))}$. As discussed earlier, the undetermined factor $\exp{(\text{Pol}(m))}$ where $\text{Pol}(m)$ is a polynomial with local coefficients, so it
 cannot modify the $q$-dependent part of the infinite product from the resonance poles. Taking $m=0^+$, we match our determinant with the heat kernel result. First, we have the operator relation
\begin{align}
(L_--im)(L_-+im)=(\Delta_1+m^2).
\end{align}
For a Dirac operator on a manifold without boundary in any dimension, a multiplicative anomaly is absent in the massless limit but not in general \cite{Cognola:1999xv}. Assuming this to be true for our case, by using the determinant for the vector Laplacian \cite{Giombi:2008vd}~\eqref{eq:hkvec} we have
\begin{align}
\det\nolimits{^{(T)}}{(\Delta_1)}=\prod_{l,l'=0}^\infty \left(1-q^{l+1}\bar{q}^{l'}\right)^2\left(1-q^{l}\bar{q}^{l'+1}\right)^2=\det\nolimits{^{(T)}}{(L_--i0^+)}\det\nolimits{^{(T)}}{(L_-+i0^+)},
\end{align}
where we recall that $\det\nolimits{^{(T)}}{(\Delta_1)}$ contains no volume factor in the massless limit $m\rightarrow 0$. Since $\bar{A}=0$,  
\begin{align}
\det\nolimits{^{(T)}}{(L_-+i0^+)}=(\det\nolimits{^{(T)}}{(L_--i0^+)})^*,
\end{align}
implying that~\eqref{eq:transverse0} is at most off from the correct answer by a pure phase:
\begin{equation}
\det\nolimits{^{(T)}}{(L_--i0^+)}=(\text{pure phase})\times\prod_{l,l'=0}^\infty \left(1-q^{l+1}\bar{q}^{l'}\right)^2.\label{eq:transverse1}
\end{equation}

\subsection{The Eta-Invariant \texorpdfstring{$\eta(L_-)$}{eta(L)}\label{sec:4.3}}
The eta-invariant $\eta(L_-)$, i.e. the phase of $\det{(L_-)}$ only entered when we regularized the operator in~\eqref{eq:phaseL} with an $i\varepsilon$ deformation. Since the scalar/longitudinal contributions are independent of the $i\varepsilon$ prescription, the only contributions to the phase are from the transverse modes. The infinite product from the resonance poles that we just found gives non-local contributions which depend on the (conformal) boundary complex structure. The remaining contribution, coming from the undetermined phase, is entirely local. In particular it cannot depend on $\tau_1$ so it cannot
 give a contribution of the form $(q/\bar{q})^{N}=e^{4\pi i \tau_1 \cdot N}$ to the determinant, with $N$ a number not determined by the resonance-poles method.

Since in the present case $\bar{A}=0$ and thermal $AdS_3$ is homogeneous, the pure phase may only be proportional to the renormalized volume $\text{Vol}(\mathbb{H}^3/\Gamma)_{ren}=-\pi^2\tau_2$. However, were such contribution present in the zero temperature limit $\tau_2\rightarrow \infty$ where we recover the hyperbolic space $\mathbb{H}^3$, it would lead to an infinite oscillatory phase in the path integral. This is a contradiction since this was precisely avoided by integrating along the steepest descent path.

Hence for $U(1)$ with trivial background connection, including the longitudinal/scalar mode \eqref{eq:Delta0} with $s=2$ we have
\begin{align}
\det{(L_--i0^+)}&=\det\nolimits^{(T)}{(L_--i0^+)}\times\det{\Delta_0}\nonumber\\
&=\left(\prod_{l,l'=0}^\infty \left(1-q^{l+1}\bar{q}^{l'}\right)^2\right)\times \left(q^{-\frac{1}{24}}\bar{q}^{-\frac{1}{24}}\prod_{l,l'=0}^\infty\left(1-q^{l+1}\bar{q}^{l'+1}\right)^2\right).
\end{align}
The ratio of determinants is
\begin{equation}
\frac{\det{\Delta_0}}{\sqrt{\det{(L_--i0^+)}}}=\frac{\det\nolimits^{\frac{1}{2}}{\Delta_0}}{\sqrt{\det\nolimits^{(T)}{(L_--i0^+)}}}=q^{-\frac{1}{48}}\bar{q}^{-\frac{1}{48}}\prod_{n=1}^\infty \frac{1}{1-q^{n}},
\end{equation}
cf.\ the determinant of the Dirac operator in \cite{GUILLARMOU20102464}. The infinite product is holomorphic in $q$. For the opposite choice of sign of the Chern-Simons action, we have instead $L_-\rightarrow (L_-+i\varepsilon)$ hence $q^{-\frac{1}{48}}\bar{q}^{-\frac{1}{48}}\prod_{n=1}^\infty \frac{1}{1-\bar{q}^{n}}$. 

\subsection{The Gravitational Chern-Simons Term}
The path integral is formally topologically invariant and metric-independent before gauge-fixing. By topological invariance we mean invariance under \textit{local} metric variations that vanish at the infinity, so that in particular they do
not change the boundary complex structure $q$. While the absolute value of the ratio of determinants $(\det{\Delta_0}/\sqrt{|\det{L_-}|})$, aka the Ray-Singer torsion, is a topological invariant \cite{RAY1971145}, the eta-invariant $\eta(L_-)$ is not. Such metric dependence was introduced when we deformed the action by an $i\varepsilon$-prescription to the steepest descent path. To restore topological invariance, refs.~\cite{Witten:1988hf}\cite{BarNatan:1991rn} observed that the following combination is a topological invariant by the Atiyah-Patodi-Singer Index Theorem \cite{Atiyah:1975jf,Atiyah:1976jg,Atiyah:1980jh}
\begin{align}
-\frac{1}{4}\eta(L_-)+\frac{I[\omega]}{24\pi},
\end{align}
where 
\begin{align}
I[\omega]&=\frac{1}{4\pi}\int_M {\rm tr}[\omega d\omega+\frac{2}{3}\omega\wedge\omega\wedge\omega]
\end{align}
is the gravitational Chern-Simons action of the spin connection $\omega$. $I[\omega]$ should be added to the action as a counterterm to restore topological invariance. This action is diffeomorphism invariant, but depends on the framing of the manifold (i.e. it has a Lorentz anomaly).

To compute $I[\omega]$, choose a local basis $\{f_a\}$, $a=2,3,1$, of orthonormal sections of a rank-3 vector bundle $V$ associated to the oriented orthonormal frame bundle $F_{SO(3)}(M)$, such that $\delta(f_a,f_b)={\rm diag}(1,\ldots,1)$ where $\delta$ is the metric on $V$. On the other hand, given local coordinates $\{x^\mu\}$ on $M$ we have the coordinate frame $\{\partial_\mu\}$ of $T(M)$. The Dreibein $e$, assuming invertibility, is an isomorphism between the tangent bundle $T(M)$ and $V$: $e(\partial_\mu)\equiv e_\mu^a f_a$.

\begin{figure}[ht]
\begin{center}
\hspace*{1.6cm}\includegraphics[width=0.5\linewidth,keepaspectratio]{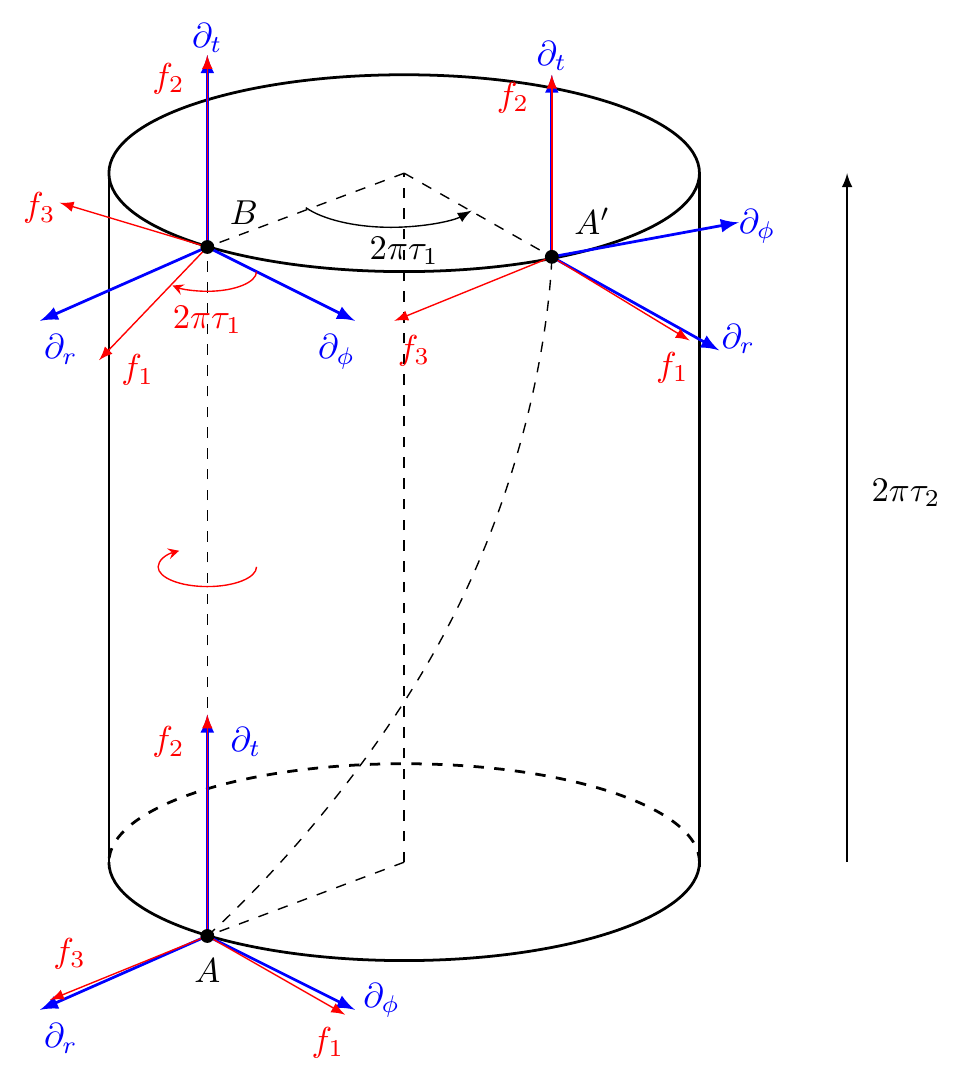}
\caption{Framing in thermal $AdS_3$. Blue arrows denote coordinate frames $\{\partial_\mu\}$ of $T(M)$ and red arrows denote local frames $\{f_a\}$ of $V$. The Dreibein is an isomorphism between them.}\label{fig:framing}
\end{center}
\end{figure}

In the present case, thermal $AdS_3$ is topologically a solid cylinder with top and bottom identified with a canonical twist; see \hyperref[fig:framing]{Figure \ref*{fig:framing}}. Suppose that at point $A$, the frame $\{\partial_\mu\}$ of $T(M)$ and $\{f_a\}$ of $V$ coincide, i.e. $e^a_\mu$ is diagonal. Under the torus identification with a twisting angle $2\pi\tau_1$, the frame of $V$ at point $A'$ (identified with $A$) is the same as that at $A$, while the frame $\{\partial_\mu\}$ of $T(M)$ is entirely determined by the coordinate system $\{x^\mu\}$. This means that along the oriented geodesic generated by $\partial_t$ from $A$ to $B$ on the top of the cylinder, $\{f_a\}$ is rotated by an angle of $-2\pi\tau_1$ about the $\partial_t$-direction. 

Without loss of generality, we take the twist to be uniform along $\partial_t$. The spin connections can be obtained from those without the twist $e^{(0)}$, $\omega^{(0)}$ by a rotation, 
\begin{align}
e^a_\mu&=h^a_{\;\;b}e^{(0)b}_\mu,\;e_a^\mu=(h^{-1})^b_{\;\;a}e^{(0)\mu}_b,\;\omega^a_{\mu \;b}=h^a_{\;\;c}\omega^c_{\mu \;d}(h^{-1})^d_{\;\;b}-\partial_\mu h^a_{\;\;c}(h^{-1})^c_{\;\;b},
\end{align}
where
\begin{align}
h^a_{\;\;b}&=\begin{pmatrix}
1 & 0 &0\\
0 & \cos{\left(\frac{\tau_1}{\tau_2}t \right)} & -\sin{\left(\frac{\tau_1}{\tau_2}t \right)}\\
0 & +\sin{\left(\frac{\tau_1}{\tau_2}t \right)} & \cos{\left(\frac{\tau_1}{\tau_2}t \right)}
\end{pmatrix}.
\end{align}
After a straightforward computation we have
\begin{align}
\omega_t&=\begin{pmatrix}
0 & r \cos{\left(\frac{\tau_1}{\tau_2}t \right)} & r \sin{\left(\frac{\tau_1}{\tau_2}t \right)}\\
-r \cos{\left(\frac{\tau_1}{\tau_2}t \right)} &0 & \frac{\tau_1}{\tau_2}\\
-r \sin{\left(\frac{\tau_1}{\tau_2}t \right)} & -\frac{\tau_1}{\tau_2} & 0
\end{pmatrix},\;\omega_r=0,\;\omega_\phi=\begin{pmatrix}
0&0&0\\0&0& -\sqrt{1+r^2}\\0&\sqrt{1+r^2}&0
\end{pmatrix}.
\end{align}
Note that both $\omega$ and $\omega^{(0)}$ are singular as one-forms: $\omega_\phi,\omega^{(0)}_\phi|_{r=0}\neq0$.

The gravitational Chern-Simons action is then
\begin{align}
I[\omega]&=\frac{1}{4\pi}\int_M {\rm tr}[\omega_t \partial_r \omega_\phi-\omega_\phi \partial_r \omega_t]=\frac{1}{4\pi}\left(2\frac{\tau_1}{\tau_2} \right)\int_{T^2}d^2x \left(\sqrt{1+r_\infty^2}-1\right).
\end{align}
Here the trace ${\rm tr}$ is taken over the local frame indices $a,b,\ldots$. We have introduced a cutoff $r_\infty$ to regularize the integral. To renormalize it we add a counterterm which only depends on data on the $r_\infty$ cutoff surface and does not depend on the bulk metric. First we parametrize the cutoff surface by $y^i=(t,\phi)$ and denote the Hodge dual and the induced metric on this surface as $\tilde{\star}$ and $\gamma_{ij}$ respectively. Next we define $\omega^a\equiv -\frac{1}{2}\epsilon^{abc}\omega_{bc}$. The pullbacks of the Dreibein $e$ and the spin connection $\omega^a$ on the cutoff surface, including a pullback in the $a=3$ direction of the local basis, are respectively
\begin{align}
\tilde{e}^a&\equiv \tilde{e}^a_i dy^i=\left( e^a_\mu \frac{\partial x^\mu}{\partial y^i}\right)dy^i,\;\tilde{\omega}^a\equiv \tilde{\omega}^a_{i} dy^i=\left( \omega^a_{\mu} \frac{\partial x^\mu}{\partial y^i}\right)dy^i,
\end{align}
where now $a=2,1$. We propose a 2D diffeomorphism-invariant counterterm, not necessarily a unique choice,
\begin{align}
I^{(\omega)}_{ct}[\gamma]&=\frac{1}{2\pi}\int_{r=r_\infty}\tilde{\omega}^a\wedge \tilde{\star}\tilde{e}_a=\frac{1}{2\pi}\int_{r=r_\infty} d^2x \sqrt{\gamma} \left(\gamma^{ij}\tilde{\omega}^a_{i}\tilde{e}_{ja}\right)=-\frac{1}{4\pi}\left( 2\frac{\tau_1}{\tau_2}\right)\int_{T^2}d^2x r_\infty,
\end{align}
where $i,j=t,\phi$. The renormalized gravitational Chern-Simons action is then
\begin{align}
I_{ren}[\omega]&=I[\omega]+I^{(\omega)}_{ct}[\gamma]\overset{r_\infty\rightarrow \infty}{=}-2\pi\tau_1.
\end{align}
The total counterterm to be added to the action to restore topological invariance is $\Delta I[\omega]\equiv-\frac{iI_{ren}[\omega]}{24}$, which contributes a pure phase to the path integral, 
\begin{align}
\exp{(-\Delta I[\omega])}=\exp{\left(+\frac{iI_{ren}[\omega]}{24}\right)}=\exp{\left(-\frac{2\pi i\tau_1}{24}\right)}=(\bar{q}/q)^{\frac{1}{48}}.\label{eq:gravCS}
\end{align} 

In the end, the renormalized topologically-invariant path integral at one-loop $O(k^0)$ reads
\begin{align}
Z_{U(1)}(k,\bar{A}=0)&=e^{-\Delta I[\omega]}\frac{\det{\Delta_0}}{\sqrt{\det{(L_--i0^+)}}}=(\bar{q}/q)^{\frac{1}{48}} (q\bar{q})^{-\frac{1}{48}}\prod_{n=1}^\infty \frac{1}{1-q^{n}}\\
&=q^{-\frac{1}{24}}\prod_{n=1}^\infty \frac{1}{1-q^{n}},
\end{align}
which is exactly the vacuum character of the $\widehat{\mathfrak{u}}(1)$ current algebra with central charge $c=1$, i.e. the partition function of a chiral boson. It is holomorphic in $q$, and also agrees with the WZW calculation~\eqref{eq:chiralboson}. Moreover, it is one-loop exact.

We are now ready to compare with the standard WZW method reviewed in \hyperref[sec:chWZW]{Appendix \ref*{sec:chWZW}}. Starting from the same action, both give the identical partition function of a chiral boson which is holomorphic in $q$. However, the holomorphicity in $q$ has completely different origins. In the standard WZW method, one must first compactify thermal $AdS_3$ into a solid torus $T$ with a boundary $\partial T$ to define a theory on the boundary. Holomorphicity then comes from imposing chiral boundary condition $A_{\bar{w}}|_{\partial T}=$const. In covariant gauge-fixing we instead take thermal $AdS_3$ non-compact without a boundary. Holomorphicity comes from the unique deformation to the steepest descent path when $k>0$, or, equivalently, from  the $i\varepsilon$-prescription together with the gravitational Chern-Simons counterterm which restores topological invariance.

\section{The \texorpdfstring{$\widehat{\mathfrak{su}}(2)_k$}{su(2)-k-hat} Vacuum Character with a Chemical Potential}\label{su2vac}
\subsection{Chern-Simons Action with a Holomorphic Current}\label{KLact}
Consider now the $SU(2)$ Chern-Simons path integral introduced in ref.~\cite{Kraus:2006nb},
\begin{align}
Z_{SU(2)}(k>0,A_{\bar{w}}|_{\partial M})=\int_{\mathcal{U}_{SU(2)}} DA\; \exp{(-I[A]) }=\int_{\mathcal{U}_{SU(2)}} DA\;\exp{\left(+ikW[A]-I_{bt}[A]\right) },\;k>0,\label{eq:PIchem}
\end{align}
where the boundary term $I_{bt}[A]$ of the action is
\begin{align}
I_{bt}[A]&=+\frac{k}{8\pi}\int_{\partial T}d^2x\sqrt{g}g^{\alpha\beta}{\rm Tr}[A_\alpha A_\beta]=\frac{k}{8\pi}\int_{\partial T}d^2x\delta^{\alpha\beta}{\rm Tr}[A_\alpha A_\beta],\label{eq:CSbt}
\end{align}
and $(\alpha,\beta)=(\phi,t)$, which allows for a boundary holomorphic current. The total action is
\begin{align}
I[A]&=-\frac{ik}{4\pi}\int_T {\rm Tr}[AdA+\frac{2}{3}A\wedge A\wedge A]+\frac{k}{8\pi}\int_{\partial M}\Tr[A_{\phi}^2+A_t^2].\label{eq:actionI}
\end{align}
Upon variation of $A$, ($\varepsilon^{x\phi t}=+1$)
\begin{align}
\delta I[A]=(\text{EOM})+\frac{k}{\pi}\int_{\partial M}d^2x {\rm Tr}[A_w\delta A_{\bar{w}}]\equiv(\text{EOM})+i\int_{\partial M}d^2x {\rm Tr}[J_w\delta A_{\bar{w}}].
\end{align}
The equation of motion is $F=0$ so any classical solution is a flat connection. An appropriate boundary condition is to fix $A_{\bar{w}}$. We impose\footnote{Our boundary condition of $A_{\bar{w}}$ is one-half that in \cite{Kraus:2006nb}, due to their definition of the current $J_0=2j^3_0$, which is an integer for all highest weight states.}
\begin{align}
\text{Boundary Condition: }&A_{\bar{w}}|_{\partial M}=\frac{i\chi}{2\tau_2} T^3,\;\chi\in\mathbb{C}.\label{BC1}
\end{align}
The overall sign of action \eqref{eq:actionI} is uniquely fixed by that of the boundary term. The choice of the latter can be justified using ref.~\cite{Kraus:2006nb} in two ways. First, in the BTZ background, by correctly identifying $A_{\bar{w}}$ with the chemical potential and $A_w$ with the holomorphic current in the canonical partition function, one gets the correct black hole entropy of a charged black hole. Second, the thermal $AdS_3$ background is the NS-NS vacuum in a supergravity theory, so it 
has conformal weights $L_0=\tilde{L}_0=0$. By setting $A_w=1$, one gets the maximally charged R vacuum in the holomorphic sector. The choice of the sign of the boundary term gives the correct positive shift in the boundary stress tensor such that $L_0=c/24$, in agreement with the prediction from spectral flow discussion in the dual CFT language. The action that gives a boundary anti-holomorphic current has the same boundary term with the same sign as in \eqref{eq:actionI}, but an opposite sign for the Chern-Simons action.

\subsection{Classical Solutions}\label{classsol}
Since we work in the background field gauge, we need to find all flat connections modulo gauge transformations that approach the identity at infinity. These are the classical saddle points around which we perform an expansion using a covariant gauge-fixing. Here we prove that all flat connections that respect the chiral boundary condition~\eqref{BC1} can be gauge-transformed to having $\bar{A}_r=0$ everywhere, i.e. that $\bar{A}_r=0$ is a good gauge \textit{given~\eqref{BC1}}.

Thermal $AdS_3$ has a solid-torus topology, and in this section we consider the case where no source is present. Hence a flat connection $\bar{A}$ has trivial holonomy around the contractible cycle parametrized by $\phi$, but it can have non-trivial holonomy around the non-contractible cycle parametrized by $t$,
\begin{align}
P\exp{\left(\oint d\phi \bar{A}_{\phi}\right)}=+\mathds{1},\;P\exp{\left(\oint dt \bar{A}_{t}\right)}=\exp{(\mu^a T^a)},\label{eq:expeom}
\end{align}
where $P$ means path-ordered. Since we allow for complex background, we have $\mu^a\in \mathbb{C}$. We write a flat connection as $\bar{A}=U^{-1}dU$ for some $U(r,w,\bar{w})$, and to factor out the holonomies, we decompose
\begin{align}
U(r,w,\bar{w})=V(r,w,\bar{w})e^{\mu^a T^a t}e^{2 N \phi T^a n^a},\;N\in\mathbb{Z},
\end{align}
where $|n|=1$ and $V(r,w,\bar{w})$ is doubly-periodic. Then
\begin{align}
\bar{A}=&(2 N  T^a n^a) d\phi+e^{-2 N \phi T^a n^a}(\mu^a T^a) e^{2 N \phi T^a n^a}dt\nonumber\\
&+e^{-2 N \phi T^a n^a}e^{-\mu^a T^a t}(V^{-1}dV)e^{\mu^a T^a t}e^{2 N \phi T^a n^a}.
\end{align}
The chiral boundary condition~\eqref{BC1} is solved by
\begin{align}
T^a n^a=T^3,\;\mu^a T^a=\left(\frac{\chi}{\tau_2}+2 i N\right) T^3\text{ and }V^{-1}\partial_{\bar{w}}V=0.
\end{align}
Since $V$ has no singularities, is doubly-periodic and holomorphic, it must be independent of $w,\bar{w}$: $V(r,w,\bar{w})=V(r)$. Moreover, 
\begin{align}
\bar{A}_r(r,\phi,t)=e^{-2 N \phi T^3}e^{-\left(\frac{\chi}{\tau_2}+2 i N\right) T^3t}(V^{-1}\partial_r V)e^{\left(\frac{\chi}{\tau_2}+2 i N\right) T^3 t}e^{2 N \phi T^3}
\end{align}
is doubly-periodic, which for generic $\chi\in\mathbb{C}$ implies that $V(r)$ is generated by $T^3$. Next, we decompose $V(r)=V(\infty)\hat{V}(r)$. Since we are considering complex solutions, $V(\infty)$ is some constant $SL(2,\mathbb{C})$ matrix that commutes with $T^3$ and that drops out of the expression $\bar{A}=U^{-1}dU$ because it is a redundancy of 
our parametrization~\cite{Elitzur:1989nr}. $\hat{V}(r)$ is a local gauge parameter with $\hat{V}(r\rightarrow \infty)=\mathds{1}$ 
so it can be removed by a gauge transformation that vanishes at infinity. This proves our assertion that given the chiral boundary condition~\eqref{BC1}, including the case $\chi=0$, we can always choose the representative flat connections to have $\bar{A}_r=0$. Furthermore, regularity requires $\bar{A}_\phi|_{r=0}=0$, so we have $\bar{A}_\phi=0$ everywhere. In the end, we have only one classical solution in this problem,
\begin{align}
\bar{A}=-\frac{i\chi}{2\tau_2} T^3dw+\frac{i\chi}{2\tau_2} T^3d\bar{w}=\frac{\chi}{\tau_2} T^3 dt.\label{eq:Abart}
\end{align}

\subsection{Twisted Periodicities and One-Loop Determinants}\label{sec:twisted}
We take $\chi$ real throughout the one-loop calculation and determine the unique steepest descent path (equivalent to the $(L_--i\varepsilon)$ deformation) as discussed in \hyperref[sec:gfcom]{Section \ref*{sec:gfcom}}. Only at the end 
of the calculation we continue $\chi$ to complex values assuming that the path varies continuously with $\chi\in\mathbb{C}$. This avoids the need to find the steepest descent path \cite{Witten:2010cx} for general complex background connections. 

We trade the flat background connection for twisted periodicity conditions of the fluctuations in thermal $AdS_3$. The covariant derivative $D_\mu$ acts on an $\mathfrak{su}(2)$-valued field $\psi=\psi^aT^a$ by
\begin{align}
D_\mu \psi&=\nabla_\mu \psi+[\bar{A},\psi],\;D_{t} \psi=\nabla_{t} \psi^{(3)}T^3+\left(\nabla_{t} \psi^{(+)}+\frac{i\chi}{\tau_2}\psi^{(+)}\right)T^++\left(\nabla_{t} \psi^{(-)}-\frac{i\chi}{\tau_2}\psi^{(-)}\right)T^-.
\end{align}
This implies that for $\psi^{(\pm)}$ doubly-periodic, defining $\psi^{(\pm)}=e^{\mp \frac{i\chi}{\tau_2} t}\tilde{\psi}^{(\pm)}$ we have 
\begin{align}
D_{\mu} \psi^{(\pm)}T^\pm=e^{\mp \frac{i\chi}{\tau_2} t}\nabla_{\mu}\tilde{\psi}^{(\pm)}T^\pm,
\end{align}
which implies twisted periodicities for $\tilde{\psi}$
\begin{align}
\tilde{\psi}^{(\pm)}(\phi+2\pi)&=\tilde{\psi}^{(\pm)}(\phi),\;\tilde{\psi}^{(\pm)}(w+2\pi\tau,\bar{w}+2\pi\bar{\tau})=e^{\pm 2\pi i \chi}\tilde{\psi}^{(\pm)}(w,\bar{w}).
\end{align}
The problem now becomes to find resonance modes with twisted periodicity conditions of the \textit{untwisted} Laplacians. Mode expanding we have
\begin{align}
\tilde{\psi}^{(\pm)}&=R(x)e^{-i\omega t+ik\phi}e^{\pm \frac{i\chi}{\tau_2} t}=R(x)e^{-i(\omega \mp \frac{\chi}{\tau_2})t+ik\phi},\;k\in \mathbb{Z},\;\omega=\frac{-n+k\tau_1}{\tau_2},\;n\in \mathbb{Z}.
\end{align}
That is, re-labeling $\omega'=\frac{-(n\pm \chi)+k\tau_1}{\tau_2}$, we shift $n\in \mathbb{Z}$ to $n'=n\pm \chi$. (Indeed, $n$ is the quantum number along the non-contractible cycle.)

\subsubsection*{Twisted Scalar Laplacian}
We compute the functional determinant for the action $\int \varphi^{(\mp)} (-D^2+m^2)\varphi^{(\pm)}=\int \tilde{\varphi}^{(\mp)} (-\nabla^2+m^2)\tilde{\varphi}^{(\pm)}$. The determinants for both $+$ and $-$ give the same answer, since $\Delta_0=-D^2$ is self-adjoint), and is found by shifting $n\rightarrow n\pm \chi$,
\begin{align}
&\quad\det\nolimits^{(\pm)}{(\Delta_0+s(s-2))}\nonumber\\
&\propto \prod_{\substack{k>0\\p\geq0\\n\in \mathbb{Z}}}%
    \begin{array}{c}
      \left( (n\pm\chi)^2+\left[\tau_2(s+2p+k)-ik\tau_1\right]^2 \right)\\
      \left( (n\pm\chi)^2+\left[\tau_2(s+2p+k)+ik\tau_1\right]^2 \right)
    \end{array}\times \prod\limits_{\substack{p\geq0\\n\in \mathbb{Z}}} \left( (n\pm\chi)^2+\tau_2^2(s+2p)^2 \right)\nonumber\\
&\propto \prod_{\substack{k>0\\p\geq0}}\frac{(1-e^{2\pi i \chi}q^{k}(q\bar{q})^{\frac{s}{2}+p})(1-e^{-2\pi i \chi}q^{+k}(q\bar{q})^{\frac{s}{2}+p})}{(1-e^{2\pi i \chi})^2} \frac{(1-e^{2\pi i \chi}\bar{q}^k(q\bar{q})^{\frac{s}{2}+p})(1-e^{-2\pi i \chi}\bar{q}^k(q\bar{q})^{\frac{s}{2}+p})}{(1-e^{2\pi i \chi})^2} \nonumber\\
&\quad\times \prod_{\substack{p\geq0}} \frac{(1-e^{2\pi i \chi}(q\bar{q})^{\frac{s}{2}+p})(1-e^{-2\pi i \chi}(q\bar{q})^{\frac{s}{2}+p})}{(1-e^{2\pi i \chi})^2}\nonumber\\
&=\prod_{l,l'=0}^\infty \left(1-e^{2\pi i \chi}q^{l+\frac{s}{2}}\bar{q}^{l'+\frac{s}{2}}\right)\left(1-e^{-2\pi i \chi}q^{l+\frac{s}{2}}\bar{q}^{l'+\frac{s}{2}}\right).\label{eq:twist1q1}
\end{align}
The infinite product of $\frac{1}{1-e^{2\pi i \chi}}$ factors regularized by the Riemann zeta function cancel exactly. Matching with the heat kernel result~\eqref{eq:hkscatw} to get the local terms, we have
\begin{equation}
\det\nolimits^{(\pm)}{\Delta_0}=q^{-\frac{1}{24}}\bar{q}^{-\frac{1}{24}}\prod_{l,l'=0}^\infty \left(1-e^{2\pi i \chi}q^{l+1}\bar{q}^{l'+1}\right)\left(1-e^{-2\pi i \chi}q^{l+1}\bar{q}^{l'+1}\right).
\end{equation}

\subsubsection{Twisted \texorpdfstring{$L_-$}{L}}
The calculation for the transverse part of the determinant of the twisted $L_-$ is done similarly. Note that since there is no shift in the angular quantum number $k$, the poles are simply those in \eqref{eq:transversepoles} with $\omega\rightarrow \omega'$. Thus we have
\begin{align}
&\quad \det\nolimits^{(T)(\pm)}{(L_--im)}\nonumber\\
&\propto \prod_{\substack{k>0\\p\geq0\\n\in \mathbb{Z}}}%
    \begin{array}{c}
      \left((n\pm\chi)^2+\left[\tau_2(2(p+1)+k+m)+ik\tau_1\right]^2\right)\\
      \left((n\pm\chi)^2+\left[\tau_2(2p+k+m)-ik\tau_1\right]^2\right)
    \end{array}\times\prod_{\substack{p\geq0\\n\in \mathbb{Z}}}\left((n\pm\chi)^2+\left[\tau_2(2(p+1)+m)\right]^2\right)\nonumber\\
&\propto\prod_{l,l'=0}^\infty \left(1-e^{2\pi i \chi}q^{l+\frac{m}{2}+1}\bar{q}^{l'+\frac{m}{2}}\right)\left(1-e^{-2\pi i \chi}q^{l+1+\frac{m}{2}}\bar{q}^{l'+\frac{m}{2}}\right).\label{eq:twist1q22}
\end{align}
So again we have that half of the product is shifted by $e^{2\pi i \chi}$ and half shifted by $e^{-2\pi i \chi}$. This agrees with the heat kernel approach using \cite{Giombi:2008vd} but only when $q=\bar{q}$, because of the first-order nature of $L_-$, as in the previous section:
\begin{equation}
\det\nolimits^{(T)\pm}{(L_--i0^+)}=(\text{an overall factor})\times \prod_{l,l'=0}^\infty \left(1-e^{2\pi i \chi}q^{l+1}\bar{q}^{l'}\right)\left(1-e^{-2\pi i \chi}q^{l+1}\bar{q}^{l'}\right).\label{eq:twist1q3}
\end{equation}
The undetermined local factor is a pure phase for $\chi\in\mathbb{R}$.

\subsection{The Eta Invariant \texorpdfstring{$\eta(\bar{A})$}{eta(A)}}\label{sec:levelshift}

We have determined $\det\nolimits^{(T)\pm}{(L_--i0^+)}$ up to an overall factor. Because we are considering a real background, $\bar{A}^\dagger= -\bar{A}$ (recall that the $T^a$ are anti-Hermitian), i.e. $\chi\in\mathbb{R}$, we still have
\begin{align}
\det\nolimits{^{(T)\pm}}{(L_--i0^+)}= \left(\det\nolimits{^{(T)\pm}}{(L_-+i0^+)}\right)^*,\;\bar{A}^\dagger= -\bar{A}.
\end{align}
Assuming again that there is no multiplicative anomaly, i.e.\ that the operator relation $(L_--i0^+)(L_-+i0^+)=\Delta_1$ implies $\det{(L_--i0^+)}\det{(L_-+i0^+)}=\det{(\Delta_1)}$,  
\begin{align}
\det\nolimits{^{(T)\pm}}{(\Delta_1)}=\det\nolimits{^{(T)\pm}}{(L_--i0^+)}\det\nolimits{^{(T)\pm}}{(L_-+i0^+)}.\label{eq:delta1A}
\end{align}
Together with~\eqref{eq:hkvec2} one deduces that the undetermined factor is indeed a pure phase. When $\bar{A}=0$ there is no such local phase, as discussed in the previous section.

Define the eta invariant $\eta(\bar{A})$ in the presence of a non-trivial connection $\bar{A}$ as in~\eqref{eq:etainv}. By considering the variation of $\eta(\bar{A})$ under a \textit{local} deformation of $\bar{A}$ that vanishes at the boundary, ref. \cite{BarNatan:1991rn}\cite{Witten:1988hf} found that $\eta(\bar{A})$ differs from $\eta(0)$ by a metric-independent term,
\begin{align}
\eta(\bar{A})=\eta(0)-\frac{4h}{\pi}\left(\frac{I[\bar{A}]}{-ik}\right)\text{ mod }2. \label{m1b}
\end{align}
Here $h$ is the dual Coxeter number; $h=N$ for $SU(N)$ and $h=-2$ for $SL(N,\mathbb{R})$. $I[\bar{A}]$ is the on-shell Chern-Simons action of $\bar{A}$, which is purely imaginary when $\bar{A}^\dagger= -\bar{A}$. This one-loop correction effectively shifts the level $k\rightarrow k+h$. One subtlety is that since the deformation is local, the $\bar{A}$ that appears in~\eqref{m1b} vanishes at infinity. On the other hand, our background $\bar{A}$ is non-vanishing at infinity by our choice of boundary condition. To obtain $\eta(\bar{A})$ in our case we should in principle consider a continuous deformations from $\bar{A}=0$ to our $\bar{A}$, which would be non-local. Instead, we simply assume that it shifts the level in the boundary terms in the same way $k\rightarrow k+2$. If that were not the case, the classical variational problem would not give chiral
boundary conditions at one loop.

\subsection{The One-Loop \texorpdfstring{$\widehat{\mathfrak{su}}(2)_k$}{su(2)-k-hat} Vacuum Character}\label{sec:su2vac}
The path integral at one-loop $O(k^0)$ is, after continuing to $\chi\in\mathbb{C}$,
\begin{equation}
Z_{SU(2)}(k,A_{\bar{w}}|_{\partial M}=i\chi/2\tau_2 T^3)=e^{+\frac{\pi (k+2)\chi^2}{4\tau_2}}q^{-\frac{3}{24}}\prod^\infty_{n=1}\frac{1}{1-q^n}\frac{1}{1-q^n e^{+2\pi i \chi}}\frac{1}{1-q^n e^{-2\pi i \chi}},\;\chi\in\mathbb{C}. \label{m2b}
\end{equation}
Here we have included three copies of the gravitational Chern-Simons counterterm~\eqref{eq:gravCS} to restore topological invariance and included the shift $k\rightarrow k+2$. There does not appear to be pathologies as we meromorphically continue to complex $\chi$. Eq.~\eqref{m2b} is identical to the one-loop WZW result~\eqref{eq:WZWsu2} that we computed by reducing exactly the same action~\eqref{eq:actionI} to a gauged WZW action (there the same continuation of $\chi$ was also done and we ignored the $k+2$ shift). The fact that we get a partition function holomorphic in $q$ agrees with our expectation, since the boundary condition we chose allows for a boundary holomorphic current, which implies that the theory describes the holomorphic sector of the $\widehat{\mathfrak{su}}(2)_k$ current algebra. In \hyperref[KLact]{Section \ref*{KLact}} we have argued that the overall sign of the action \eqref{eq:actionI} is uniquely fixed, and it gives us the deformation $(L_--i\varepsilon)$ equivalent to the steepest descent path. The result \eqref{m2b} is a non-trivial confirmation of the validity of our calculation.

The canonical Kac-Moody character $\mathcal{Z}_{SU(2)}(k,\chi)$ as a trace over an irreducible representation is known to be different from the path integral expression $Z_{SU(2)}(k,\chi)$ by exactly the tree-level path integral (with the shift $k+2$)\cite{Kraus:2006nb}. From this we recover the $\widehat{\mathfrak{su}}(2)_k$ vacuum character (see \eqref{eq:su2ch2}, which is valid in the strip $\Im{\chi}\in(-\tau_2,\tau_2)$ where it makes sense as geometric sums), at $O(k^0)$ where $c=3-\frac{6}{k+2}=3+O(1/k)$
\begin{align}
&\quad\mathcal{Z}_{SU(2)}(k,\chi,J=0)\equiv Tr_{0}\left[ q^{L_0-\frac{c}{24}}e^{2\pi i \chi j^3_0} \right]=q^{-\frac{c}{24}}\prod^\infty_{n=1}\frac{1}{1-q^n}\frac{1}{1-q^n e^{+2\pi i \chi}}\frac{1}{1-q^n e^{-2\pi i \chi}}.
\end{align}
The $O(1/k)$ corrections to the central charge which should resum to $c=\frac{k\text{dim}\mathfrak{g}}{k+2}$, a result we simply quote in this work, can in principle be computed perturbatively by considering the higher-loop Feynman diagrammatics, or, in the WZW approach, Wilson lines with endpoints on the boundary; see \cite{Besken:2018zro} and references therein.

\section{Wilson Loops and \texorpdfstring{$\widehat{\mathfrak{su}}(2)_k$}{su(2)-k-hat} Characters of Unitary Representations}\label{sec:wilson}
In this section we compute $\widehat{\mathfrak{su}}(2)_k$ characters of unitary representations by including a Wilson loop in the Chern-Simons theory. As shown in \hyperref[sec:sl2k]{Appendix \ref*{sec:sl2k}} and in particular in~\eqref{eq:su2ch2}, a unitary representation $J$ of $\widehat{\mathfrak{su}}(2)_k$ ($k\in\mathbb{Z}^+$) is labeled by the spin $J=0,1/2,1,\ldots,k/2$ of the $\mathfrak{su}(2)$ Lie algebra. The maximum value $J=k/2$ is a consequence of demanding no negative-norm states in the representation, which is of course necessary for unitarity. The Sugawara construction gives the highest weight of $L_0$ as $h=\frac{J(J+1)}{k+2}$. Except for the central charge, which has $O(1/k)$ corrections, the characters are one-loop $O(k^0)$ exact.

\subsection{Wilson Loop as a Quantum System}
Consider a Lie group $G$ and define the Wilson Loop of an irreducible representation $R$, which is a gauge-invariant and metric-independent observable, as
\begin{align}
W_R[C]\equiv Tr_R P\exp{\left(\oint_C A \right)}=Tr_R P\exp{\left(\oint_C A^a T^a_{(R)}\right)},
\end{align}
where $C$ is an oriented loop in the manifold and $P$ means path-ordered. The connection $A$ is expanded in the $\mathfrak{g}$ generators $T^a_{(R)}$ in the irreducible representation $R$. The Wilson loop operator is intrinsically quantum mechanical. To consistently apply a semi-classical treatment to the path integral we need to represent it by a local theory on $C$. This was suggested in \cite{Witten:1988hf} and thoroughly reviewed in \cite{Beasley:2009mb}. In the rest of this subsection, we quote the essential facts from \cite{Beasley:2009mb} before proceeding with the calculation.

Consider for simplicity a Lie group $G$ endowed with a Killing metric ($\Tr$) which provides an isomorphism between $\mathfrak{g}$ and its dual $\mathfrak{g}^*$, $\mathfrak{g}\cong \mathfrak{g}^*$. In this case a co-adjoint orbit embedded in $\mathfrak{g}^*$ can be identified with an adjoint orbit $\mathcal{O}_\lambda\subset\mathfrak{g}$, defined as follows. Define the adjoint action of $G$ on $\lambda\in\mathfrak{g}$ as $\lambda \rightarrow S\lambda S^{-1}$, $S\in G$. Let $\mathfrak{t}\subset\mathfrak{g}$ be the Cartan sub-algebra. Given $\alpha\in\mathfrak{t}$, the adjoint action of $G$ sweeps out an adjoint orbit $\mathcal{O}_\alpha\equiv \{S\alpha S^{-1}|S\in G\}$. $\mathcal{O}_\alpha$ is isomorphic to the quotient $G/G_\alpha$, where $G_\alpha$ is the subgroup of $G$ that leaves $\alpha$ invariant under the right adjoint action, 
$G/G_\alpha\equiv \{S\in G\}/\{ \sim \}$, $S\sim Sh^{-1},\;h\in G_\alpha$. The isomorphism is then given on  $S\in G$ by 
$S\rightarrow S\alpha S^{-1}$  and passes to the quotient because of the definition of $G_\alpha$. For a generic element $\alpha\in \mathfrak{t}$, $G_\alpha$ is isomorphic to the maximal torus generated by $\mathfrak{t}$. An adjoint orbit $\mathcal{O}_\alpha$ is a symplectic manifold. Define the pre-symplectic one-form $\Theta_\alpha$ on $G$ by $\Theta_\alpha=\Tr[\alpha S^{-1}dS]$, $S\in G$. The symplectic form is given by $\omega_\alpha=d\Theta_\alpha$.

The crucial ingredient is the Borel-Weil-Bott theorem which states the isomorphism between two Hilbert spaces: 1) an irreducible representation $R$ of $G$ with highest/lowest weight $\alpha$, and 2) the Hilbert space constructed from quantizing the space of holomorphic sections of the holomorphic line bundle over the co-adjoint orbit $\mathcal{O}_{\alpha}=G/G_\alpha$, where the line bundle has the symplectic form $\omega_\alpha$ of $\mathcal{O}_\alpha$ as its curvature two-form. We can describe such Hilbert space with a single-particle quantum system.  The particle moves in $\mathcal{O}_\alpha$ and its worldline is $C$, so the system is described by the map $U:C\rightarrow \mathcal{O}_\alpha$. The particle has a first-order, one-dimensional Chern-Simons action:
\begin{align}
I_{C}[U]=i\oint_C ds (\Theta_\alpha)_m \frac{dU^m}{ds}
\end{align}
where we have picked a coordinate system $\{u^m\}$ in $\mathcal{O}_\alpha$ and parametrized $C$ by $s$. Furthermore, we couple $U$ to the bulk connection $A$ in order to promote the global $G$-invariance to a gauge redundancy. To this end, define a $G$-valued function $S:C\rightarrow G$ by the relation $U(s)=S\alpha S^{-1}$, $s\in C$. Such relation determines $S$ from $U$ only up to a local right action $Sh$ of $G_\alpha$, $h:C\rightarrow G_\alpha$. From
$I_{C}[U]$ we have the gauge-invariant action
\begin{align}
I_{C}[S,A]=i\oint_Cds (\Theta_\alpha)_m \frac{d_AU^m}{ds}=i\oint_C\Tr[\alpha\cdot S^{-1}(d+A)S].\label{eq:WLaction0}
\end{align}
The indeterminacy in $S$ gives rise to a $2\pi\mathbb{Z}$ shift in the action.

In the end, the path integral equivalent to Chern-Simons with a Wilson loop $W_R[C]$ insertion is
\begin{align}
Z_{G}(R)&=\int_{\mathcal{U}_{G}} DA\int_{G}DS\; \exp{\left(-I[A]-I_{C}[S,A]\right)}.
\end{align}

\subsection{\texorpdfstring{$SU(2)$}{SU(2)} Wilson Loops}
Back to our case $G=SU(2)$, we consider an irreducible spin-$J>0$ representation, in which case the (co-)adjoint orbit is $\mathcal{O}_{J>0}=SU(2)/U(1)\cong S^2$, which is a sphere of radius $J$ ($\mathcal{O}_{J=0}=SU(2)/SU(2)$ is a point). Consider the oriented loop $C_0$ at the origin $r=0$, along the positive $t$-direction, with the canonical twist $(+2\pi\tau_1)$ determined by the boundary complex structure. The loop $C_0$ is parametrized by $s\in [0,2\pi \tau_2)$, $x(s)=(t=s,r=0,\phi=\left(\frac{\tau_1}{\tau_2}\right)s)$. Consider the path integral with a Wilson loop of an irreducible spin-$J>0$ $SU(2)$ representation,
\begin{align}
&\quad Z_{SU(2)}(k,A_{\bar{w}}|_{\partial M}=i\chi/2\tau_2 T^3,J)=\int_{\mathcal{U}_{SU(2)}} DA\int_{SU(2)}DS\; \exp{\left(-I[A]-I_{C_0}[S,A,J]\right)},\\
&\quad I_{C_0}[S,A,J]\equiv +i2J\int_0^{2\pi\tau_2}ds \left.\Tr[(A_t+\left(\frac{\tau_1}{\tau_2}\right)A_\phi)S^{-1}T^3 S+\dot{S}^{-1}T^3 S ]\right|_{r=0}.\label{eq:WLaction}
\end{align}
The factor of $2$ comes from our normalization of the generators. This action was constructed explicitly in \cite{Diakonov:1989fc}. Here $I[A]$ is the action~\eqref{eq:actionI} from the last section and we keep the dependence of $Z_{SU(2)}$ on $C_0$ implicit.

In order that the power counting in $k$ remain valid in the perturbative expansion, we must keep $O(k^0)=2J/k>0$ finite as we take $k\rightarrow \infty$. 

\subsubsection{Tree Level}
The equations of motion in terms of $U=S^{-1}T^3 S\in \mathcal{O}_{J>0}/J$ read
\begin{subequations}
\begin{empheq}{align}
\frac{-ik}{2\pi}F_{r\phi}+i(2JU)\frac{\delta(r)}{2\pi}&=0,\\
\frac{+ik}{2\pi}F_{rt}+i(2JU) \left(\frac{\tau_1}{\tau_2}\right)\frac{\delta(r)}{2\pi}&=0,\\
dU+[A,U]&=0.
\end{empheq}
\end{subequations}
We impose the same chiral boundary condition as before
\begin{align}
\text{Boundary Condition: }{\bar{A}}_{\bar{w}}|_{\partial M}&=\frac{i\chi}{2\tau_2}T^3.
\end{align}
Since the boundary condition fixes ${\bar{A}}_{\bar{w}}\propto T^3$, we have both $\bar{A},\bar{U}\propto T^3$ on-shell where $J\bar{U}\in\mathcal{O}_{J>0} $ is constant. Moreover, following the same argument as in \hyperref[classsol]{Section \ref*{classsol}}, we can always choose the representative background solutions to have $\bar{A}_r=0$ given the chiral boundary condition. In the end, there are two distinct classical solutions, 
\begin{enumerate}
\item
\begin{equation}
\begin{aligned}
U&=\bar{U}_+=+T^3\\
\bar{A}_+&:\;\begin{dcases}
\bar{A}_\phi&=+\left(\frac{2J}{k}\right)T^3 \theta(r)\\
\bar{A}_t&=\left(\frac{\chi}{\tau_2}+\frac{\tau}{\tau_2}\left(\frac{2J}{k}\right) \right)T^3-\left(\frac{2J}{k}\right)\left(\frac{\tau_1}{\tau_2}\right)T^3 \theta(r)
\end{dcases},\label{eq:wilsonclass1}
\end{aligned}
\end{equation}

\item
\begin{equation}
\begin{aligned}
U&=\bar{U}_-=-T^3\\
\bar{A}_-&:\;\begin{dcases}
\bar{A}_\phi&=-\left(\frac{2J}{k}\right)T^3 \theta(r)\\
\bar{A}_t&=\left(\frac{\chi}{\tau_2}-\frac{\tau}{\tau_2}\left(\frac{2J}{k}\right) \right)T^3+\left(\frac{2J}{k}\right)\left(\frac{\tau_1}{\tau_2}\right)T^3 \theta(r)
\end{dcases}.\label{eq:wilsonclass2}
\end{aligned}
\end{equation}
\end{enumerate}
We have a solution with $U=T^3$ sitting on the ``north" pole of $\mathcal{O}_J= J S^2$ and another one with $U=-T^3$ sitting on the ``south" pole. (Which one is north is purely conventional of course.) The step function $\theta(r)$ has $\theta(0)=0$, $\theta(r>0)=1$ and $\frac{d\theta(r)}{dr}=\delta(r)$. Moreover, since $\bar{A}_{\pm,\phi}|_{r=0}=0$ without any extra constant terms, the only singularities of $\bar{A}_{\pm}$ are indeed the ones sourced by the Wilson loop, given by the step function. The two sets of solutions are inequivalent because to get one from the other one needs to transform $\bar{A}$ by a $\phi$-dependent gauge parameter non-vanishing at the boundary and  singular on $C$.

The holonomies of $\bar{A}_{\pm}$ around the origin (where the loop sits) satisfy
\begin{align}
\exp{\left( \oint d\phi \bar{A}_\phi \right)}\sim\exp{\left(2\pi\left(\frac{ 2J}{k}\right)T^3\right)} ,
\end{align}
where $\sim$ is the equivalence relation defined by $SU(2)$ conjugations. 

The on-shell Wilson loop actions and Chern-Simons actions with boundary terms for $(\bar{U}_\pm,\bar{A}_\pm)$ are
\begin{align}
\quad I_{C_0}[\bar{U}_\pm,\bar{A}_\pm,J]&=-2(2\pi i\tau)\left( \frac{J^2}{k}\right)\mp 2\pi i \chi J,\;I[\bar{A}_\pm]=+2\pi i\tau\left(\frac{J^2}{k}\right)-\frac{\pi k\chi^2}{4\tau_2},
\end{align}
so the tree-level path integrals for the solutions $(\bar{U}_\pm,\bar{A}_\pm)$ are 
\begin{align}
Z_{SU(2),\pm}(k,A_{\bar{w}}|_{\partial M}=i\chi/2\tau_2 T^3,J)=e^{\frac{\pi k\chi^2}{4\tau_2}}q^{\frac{J^2}{k}}e^{\pm2\pi i \chi J}.\label{eq:treeWL}
\end{align}
The first term is the same factor that makes the path integral different from the canonical Kac-Moody character, as mentioned in the end of \hyperref[sec:su2vac]{Section \ref*{sec:su2vac}}. The second term, $q^{\frac{J^2}{k}}$, comes from both actions and already includes the phase due to the canonical twist of the loop discussed in \cite{Witten:1988hf}. The missing $q^{\frac{J}{k}}$ factor and the shift $k\rightarrow k+2$ compared to the expected quantity $q^{L_0}=q^{\frac{J(J+1)}{k+2}}$ come from $O(k^0)$ and higher-order corrections.

\subsubsection{One-Loop Level}
Consider the ``north pole" classical solution $\bar{A}_+$
\begin{align*}
\bar{A}_\phi&=\left(\frac{2J}{k}\right)T^3,\;\bar{A}_t=\frac{\chi}{\tau_2}T^3+i\left(\frac{2J}{k}\right)T^3.
\end{align*}
Despite that the background connection has a singular curvature at the origin, in computing functional determinants of the 
fluctuations, whose domains are fixed only by demanding self-adjointness, the fluctuations must still be regular at the origin. 
We have therefore consistently neglected the discontinuity of $\bar{A}$ at the origin as it does not appear to affect the one-loop 
calculation. 

The CFT unitarity bound, which follows from demanding no negative-norm states in the Hilbert space, is $0\leq 2J/k\leq 1$. For non-integer $2J/k$, one cannot construct a heat kernel as a twisted image sum because one would also have to change the periodicity of the fluctuations in $\phi$ by shifting the angular quantum number $k$, while the heat kernels we use are by construction periodic in $\phi$. When $2J/k=1$, we may ignore the shift $k\rightarrow k\pm 1$ (in the quantum number not in the level) and construct such a heat kernel, but when $\chi=0$ the background is not real ($\bar{A}_t=iT^3$), so we still have an undetermined local factor in $\det\nolimits^{(T)}(L_--i0^+)$ which is not a phase, cf.~\eqref{eq:twist1q3}.

We should make a few remarks about the analytic continuation here. The background connection is complex for generic $\chi$ due to the canonical twist in $C_0$. We again do a simple analytic continuation from a real background by defining $\chi'\equiv \chi+i\tau_2(2J/k)$ and take $\chi'\in\mathbb{R}$, i.e. $\Im{\chi}=-i\tau_2(2J/k)$. In this case we do have a steepest descent path given by $(L_--i\varepsilon)$. After performing the functional integral, we restore the $\chi'\equiv \chi+i\tau_2(2J/k)$ and meromorphically continue $\chi$ to the complex plane, assuming again that the steepest descent path varies continuously with it. 

In \hyperref[sec:WLapp]{Appendix \ref*{sec:WLapp}}, we have performed the resonance pole computation for $0<2J/k<1$ with $k,J>0$, shifting both quantum numbers $n$ and $k$ accordingly and checked for regularity of the modes at the origin. The ratio of determinants valid for $0<2J/k<1$ is~(see~\eqref{eq:C1rd})
\begin{align}
\frac{\det\nolimits^{\frac{1}{2}}{\Delta_0}}{\sqrt{\det\nolimits^{(T)}{L_-}}}\propto(q\bar{q})^{-\frac{3}{48}}\prod_{n=1}^\infty\frac{1}{1-q^n}\prod_{n=0}^\infty\frac{1}{1-e^{-2\pi i\chi}q^n}\prod_{n=1}^\infty\frac{1}{1-e^{2\pi i\chi}q^n}.\label{eq:chWL0}
\end{align}
Note that $2J/k$ drops out of the infinite products and there is a shift in the middle infinite product. It is an equality up to local terms that should only shift the level $k$ and $J$. 

Pulling out the usual $\exp{\left(\pi k\chi^2/4\tau_2\right)}$ from the tree-level path integral, the one-loop, $O(k^0)$, steepest descent path integral of $(\bar{U}_+,\bar{A}_+)$ for $0<2J/k< 1$ (ignoring the shifts in $k$ and $J$) is
\begin{align}
&\quad e^{-\frac{\pi k\chi^2}{4\tau_2}}Z_{SU(2),+}(k,A_{\bar{w}}|_{\partial M}=i\chi/2\tau_2 T^3,J)\nonumber\\
&=q^{\frac{J^2}{k}-\frac{3}{24}}e^{2\pi i \chi J}\prod_{n=1}^\infty\frac{1}{1-q^n}\prod_{n=0}^\infty\frac{1}{1-e^{-2\pi i\chi}q^n}\prod_{n=1}^\infty\frac{1}{1-e^{2\pi i\chi}q^n}\\
&=q^{\frac{J^2}{k}-\frac{3}{24}}\left(\frac{e^{2\pi i \chi J}}{1-e^{-2\pi i \chi}}\right)\prod_{n=1}^\infty\frac{1}{1-q^n}\prod_{n=1}^\infty\frac{1}{1-e^{-2\pi i\chi}q^n}\prod_{n=1}^\infty\frac{1}{1-e^{2\pi i\chi}q^n}.\label{eq:chWL}
\end{align}

We do the same calculation in \hyperref[sec:WLapp]{Appendix \ref*{sec:WLapp}} for the ``south pole" solution $(\bar{U}_-,\bar{A}_-)$; from~eq. \eqref{eq:C1rdneg} we have
\begin{align}
&\quad e^{-\frac{\pi k\chi^2}{4\tau_2}}Z_{SU(2),-}(k,A_{\bar{w}}|_{\partial M}=i\chi/2\tau_2 T^3,J)\nonumber\\
&=q^{\frac{J^2}{k}-\frac{3}{24}}\left(\frac{e^{-2\pi i \chi J}}{1-e^{+2\pi i \chi}}\right)\prod_{n=1}^\infty\frac{1}{1-q^n}\prod_{n=1}^\infty\frac{1}{1-e^{-2\pi i\chi}q^n}\prod_{n=1}^\infty\frac{1}{1-e^{2\pi i\chi}q^n}\\
&=-q^{\frac{J^2}{k}-\frac{3}{24}}\left(\frac{e^{-2\pi i \chi (J+1)}}{1-e^{-2\pi i \chi}}\right)\prod_{n=1}^\infty\frac{1}{1-q^n}\prod_{n=1}^\infty\frac{1}{1-e^{-2\pi i\chi}q^n}\prod_{n=1}^\infty\frac{1}{1-e^{2\pi i\chi}q^n}.\label{eq:chWL2}
\end{align}
The total $SU(2)$ Chern-Simons path integral with a Wilson loop is then the sum of the steepest descent path integrals~\eqref{eq:chWL} and~\eqref{eq:chWL2} of both saddle points $(\bar{U}_\pm,\bar{A}_\pm)$, 
\begin{align}
&\quad e^{-\frac{\pi k\chi^2}{4\tau_2}}Z_{SU(2)}(k,A_{\bar{w}}|_{\partial M}=i\chi/2\tau_2 T^3,J)\\
&=e^{-\frac{\pi k\chi^2}{4\tau_2}}\left(Z_{SU(2),+}(k,A_{\bar{w}}|_{\partial M}=i\chi/2\tau_2 T^3,J)+Z_{SU(2),-}(k,A_{\bar{w}}|_{\partial M}=i\chi/2\tau_2 T^3,J)\right)\nonumber\\
&=q^{\frac{J^2}{k}-\frac{3}{24}}\left(\sum_{\alpha=0}^{2J}e^{2\pi i \chi (J-\alpha)}\right)\prod_{n=1}^\infty\frac{1}{1-q^n}\prod_{n=1}^\infty\frac{1}{1-e^{-2\pi i\chi}q^n}\prod_{n=1}^\infty\frac{1}{1-e^{2\pi i\chi}q^n}.\label{eq:chWL3}
\end{align}
This is in fact the $\widehat{\mathfrak{su}}(2)_k$ character~\eqref{eq:su2ch2} for any $0<2J/k<1$ at $O(k^0)$, so indeed both saddles points contribute to the path integral. This also confirms that our simple analytic continuation of $\chi$ is valid even in the presence of a Wilson loop. In eq.~\eqref{eq:su2ch} of \hyperref[sec:su2k]{Appendix \ref*{sec:su2k}}, we decompose~\eqref{eq:su2ch2} into a sum of two terms $\mathcal{Z}_{SU(2)+}$ and $\mathcal{Z}_{SU(2)-}$. The $Z_{SU(2),\pm}$ compute the contributions to the character given by $\mathcal{Z}_{SU(2),\pm}$. The undetermined local terms at $O(k^0)$ simply shift $k\rightarrow k+2$ and $J\rightarrow J+1/2$ in the exponent of $q$. 

For the special case $2J/k=1$, which saturates the unitarity bound, the ratio of determinants for $\bar{A}_+$ (for $\bar{A}_-$ replace $\chi\rightarrow -\chi$) is~(see eq.~\eqref{eq:C2rd}):
\begin{align}
&\quad\left.\frac{\det\nolimits^{\frac{1}{2}}{\Delta_0}}{\sqrt{\det\nolimits^{(T)}{L_-}}} \right|_{\bar{A}_+,\; 2J/k=1}\propto(q\bar{q})^{-\frac{3}{48}}\prod_{n=1}^\infty\frac{1}{1-q^n}\prod_{n=0}^\infty\frac{1}{1-e^{-2\pi i\chi}q^n}\prod_{n=2}^\infty\frac{1}{1-e^{2\pi i\chi}q^n}.
\end{align}
which has an extra shift in the last infinite product, as it can be verified with the heat kernel computation (for $q=\bar{q}$). 

\section{The \texorpdfstring{$\widehat{\mathfrak{sl}}(2,\mathbb{R})_{k'}$}{sl(2,R)-k-hat} Characters from Chern-Simons Theory}\label{sl2rcs}
So far we have computed characters of Kac-Moody algebras based on compact groups: $U(1)$ and $SU(2)$. 
Motivated by gravity, we discuss how to (and how not to) obtain holomorphic $\widehat{\mathfrak{sl}}(2,\mathbb{R})_{k'}$ characters out of Chern-Simons Theory.

\subsection{Analytic Continuation of Chern-Simons Theory I}\label{sec:anacont1}
\subsubsection*{Continuing The Level $k$}
Consider the following Chern-Simons path integrals for the compact group $H$,
\begin{align}
Z_{H}(k)&=\int_{\mathcal{U}_{H}} DA\exp{\left(+ikW[A]\right)},\quad\tilde{Z}_{H}(k)=\int_{\mathcal{U}_{H}} DA\exp{\left(-ikW[A]\right)}.
\end{align}
\underline{We define them for a level $k>0$}, which is possibly quantized. For simplicity consider a real background connection. We wish to analytically continue $Z_{H}(k)$ from $k$ to $t\in\mathbb{C}$, following ref. \cite{Witten:2010cx}. To this end we complexify $A\in \mathcal{U}_H$ to $\mathcal{A}\in\mathcal{U}_{H_{\mathbb{C}}}$, and define the analytic continuation as
\begin{align}
Z_{H}(t)&=\int_{\mathcal{C}} DA\exp{\left(+itW[\mathcal{A}]\right)}
\end{align}
for some path $\mathcal{C}$ satisfying the matching condition $Z_{H}(t)=Z_{H}(k+h)$ at $t=k$. The shift by the dual Coxeter number $h$ of $H$ comes from the Jacobian between the integration measures $D\mathcal{A}$ and 
$DA$~\cite{Witten:2010cx}:
\begin{align}
DA=D\mathcal{A}\exp{(ihW[\mathcal{A}])}.
\end{align}
We will at times keep this shift in $h$ as well as that of the spin $J$ implicit in the analytic continuation; they can be obtained from one-loop corrections.

Two questions arise: 1) How does the steepest descent path vary wrt. $t$ in the analytic continuation $Z_H(t)$, and 2) is it true that $Z_{H}(t=-k)=\tilde{Z}_H(k)$? As we will see, the answer to the latter is in the negative since we have chosen a principal branch $\rm{Arg}(z)\in(-\pi,\pi]$ for the square root.

\begin{figure}[ht]
\begin{center}
\includegraphics[width=0.45\linewidth,keepaspectratio] {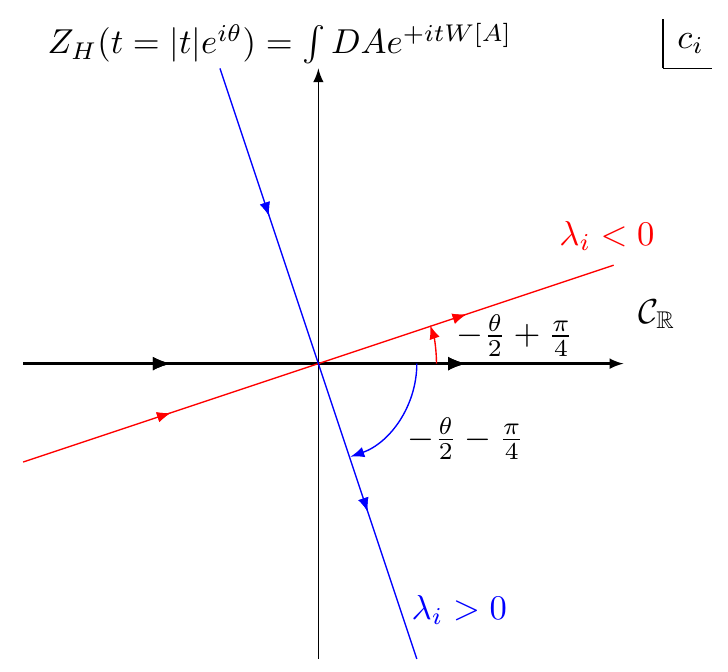} 
\caption{Deformation of Steepest Descent Paths of $Z_H(t)$ wrt. $t$,~\eqref{eq:ZHtdef}.}\label{fig:SD2}
\end{center}
\end{figure}

To answer the first question, first let $t=|k|\exp{(i\theta)}$, $\theta\in(-\pi,\pi]$. Following \hyperref[sec:SDH]{Section \ref*{sec:SDH}}, we gauge-fix and expand $H=\sum_i c_i H_i$, $c_i\in\mathbb{R}$, in terms of the orthonormal eigenfunctions of $L_-H_i=\lambda_i H_i$; the eigenvalue problem is independent of $t$. Discarding the positive delta-function norms of $\{H_i\}$, the path integral $Z_H(t)$, ignoring the ghost determinants is (recall $(\cdot,\cdot)=-\int_M \Tr[\cdot \star\cdot]$)
\begin{align}
Z_{H}(t)=\int DH\exp{\left(-\left(\frac{t}{2\pi}\right)\frac{i}{2} (H,L_-H)\right)}=\prod_i \int^\infty_{-\infty} dc_i\exp{\left(-\frac{i}{2}e^{i\theta} \lambda_i c^2_i\right)}.\label{eq:ZHt}
\end{align}
In arriving at the last equality we have again canonically normalized $H$ by a positive number, $H\rightarrow H/\sqrt{2\pi|k|}$. For $t\in \mathbb{C}$ the Gaussian integral in $c_i$ along the real line $\mathcal{C}={\mathbb{R}}$ is no longer purely oscillatory, and in fact diverges at $\pm \infty$ when $\lambda_i \sin\theta >0$, rendering the path integral ill-defined. Also, a direct application of the $i\varepsilon$ deformation to $L_-$ does not yield a steepest descent path. Therefore, we must change the original path of integration $\mathcal{C}={\mathbb{R}}$ in order that the path integral converge. To do this, for each eigenvalue $\lambda_i$, we rotate the contour in the $c_i$-plane by $(-\theta/2)$ from the real line \textit{independently of }$\text{sign}(\lambda_i)$ so that the integral is once again purely oscillatory. In other words, we define $c_i=\exp{(-i\theta/2)}c_i'$ and integrate over $c'_i\in\mathbb{R}$. After that we can do the $(-\text{sign}(\lambda_i)\pi/4)$ rotations or equivalently the $-i\varepsilon$ deformation as before; see \hyperref[fig:SD2]{Fig.\ref*{fig:SD2}}:
\begin{align}
Z_{H}(t)&\rightarrow\prod_i e^{-\frac{i\theta}{2}}\int^\infty_{-\infty} dc'_i\exp{\left(-\frac{i}{2} \lambda_i {c'}^2_i \right)}\\
&\rightarrow\prod_ie^{-\frac{i\theta}{2}}\int^\infty_{-\infty} dc'_i\exp{\left(-\frac{i}{2} (\lambda_i-i\varepsilon) {c'}^2_i \right)}\label{eq:ZHtdef}\\
&=\prod_ie^{-\frac{i\theta}{2}}\frac{1}{\sqrt{\varepsilon+i\lambda_i}}=\prod_ie^{-\frac{i\theta}{2}}\frac{1}{\sqrt{|\lambda_i|}}\exp{\left(-\frac{i\pi}{4}\text{sign}(\lambda_i) \right)}.
\end{align}
In this way we obtained a smooth deformation of the steepest descent path wrt. $t$ in the principal branch and hence an analytic continuation of $Z_H(t)$ from $t=k>0$. In particular, for $Z_H(t)$ we always have the deformation $(L_--i\varepsilon)$ for all $t\in\mathbb{C}$. In fact at one loop the only difference from the $t=k>0$ case is the infinite product of the phase $\exp{(-i\theta/2)}$ coming from the Jacobians in rotating the contours. This phase is independent of $\text{sign}(\lambda_i)$ therefore does not interfere with the eta-invariant; we can consistently discard it.

\begin{figure}[ht]
\makebox[\linewidth][c]{%
 \subfigure[$Z_{H}(t=|k|\exp{(i\pi)})=\int DA \exp{(+itW[A])}$.]{
\includegraphics[width=0.45\linewidth,keepaspectratio] {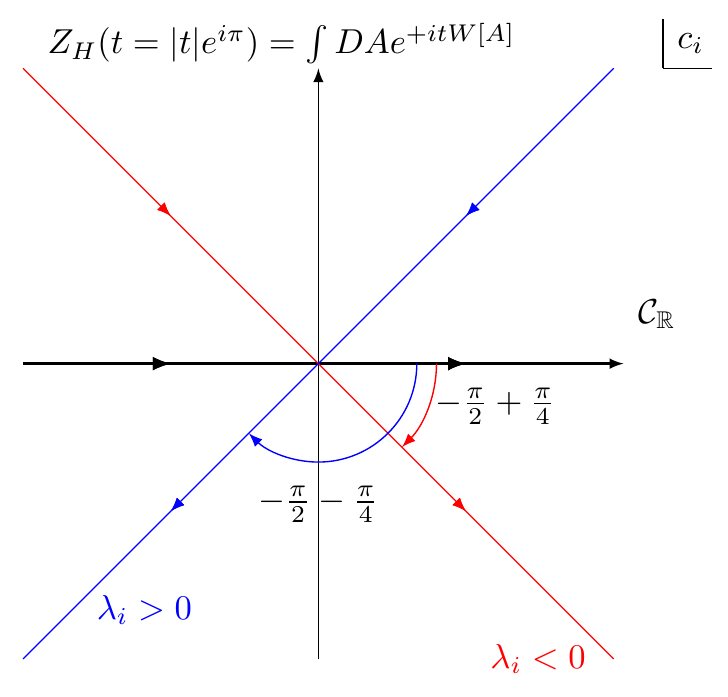} \label{fig:SD3a}}%
 \subfigure[$\tilde{Z}_{H}(k>0)=\int DA \exp{(-ikW[A])}$.]{
\includegraphics[width=0.51\linewidth,keepaspectratio] {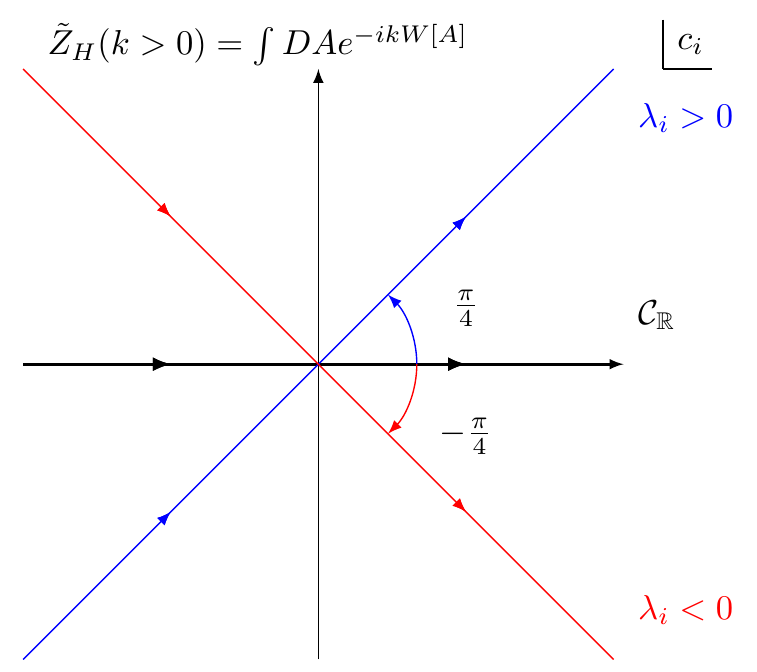} \label{fig:SD3b}}%
 }
\caption{Steepest Descent Paths of $Z_{H}(t=|k|e^{i\pi})$ VS. $\tilde{Z}_{H}(k>0)$.}\label{fig:SD3}
\end{figure}

That at the special value $t=-k$, $Z_H$ looks the same as $\tilde{Z}_H(k>0)$ is a disguise. The steepest descent path of $\tilde{Z}_H(k>0)$ implies that the Gaussian integration contour is rotated by an angle $(+\text{sign}(\lambda_i)\pi/4)$. In the above analytic continuation of $Z_H(t)$, since we have chosen the principal branch $\rm{Arg}(z)\in(-\pi,\pi]$, we take $t=|k|\exp{(+i\pi)}$. Then for $\lambda_i<0$, the contour is rotated by an angle $(-\pi/2+\pi/4)=-\pi/4$, the same as in $\tilde{Z}_H(k)$, but for $\lambda_i>0$ we have a rotation of $(-\pi/2-\pi/4)=(+\pi/4-\pi)$, off by exactly $(-\pi)$ compared to that in $\tilde{Z}_H(k)$; see \hyperref[fig:SD3]{Fig.\ref*{fig:SD3}}. As far as each individual Gaussian integral is concerned, a steepest descent contour is defined up to $(\pm \pi)$ rotations so both are equally good. However, these $\exp{(-i\pi)}$ phases appear only for $\lambda_i>0$ and the regularized sum together with the original $\exp{(-i\text{sign}(\lambda_i)\pi/4)}$ is not equal to $\exp{(+i\text{sign}(\lambda_i)\pi/4)}$. Alternatively, we can ask what it takes to get $\tilde{Z}_H(k)$ from $Z_H(k)$. To do this, one has to take $t=|k|\exp{(i\pi)}$ for $\lambda_i<0$ but $t=|k|\exp{(-i\pi)}$ for $\lambda_i>0$. But this involves two branches of the square root, not 
simply an analytic continuation; therefore, in doing so in a $\text{sign}(\lambda_i)$-dependent manner we are actually defining a different path integral: $\tilde{Z}_H(k)$. To conclude, the analytically continued $Z_{H}(t=-k)$ is not, and cannot be $\tilde{Z}_H(k)$. Furthermore, depending on whether we analytically continue $Z_H(k)$ from $k>0$ or $k<0$ (i.e. $\tilde{Z}_H(k>0)$), we get two distinct continuations for each independent direction of the Lie algebra $\mathfrak{h}$ of $H$.

\subsubsection*{Continuing The Spin $J$}
Consider the $SU(2)$ path integral including the spin-$J$ Wilson loop action, evaluated around say the ``north pole" solution $(\bar{U}_+,\bar{A}_+)$,
\begin{align}
Z_{SU(2),+}(k,J)=\int_{\mathcal{U}_{SU(2)}} DA\; \exp{\left(+ikW[A]-I_{C_0}[U,A,J]\right)}.
\end{align}
For the purpose of what follows, we will only need to know the analytic continuation $(k>0,J>0)\rightarrow (t=|k|\exp{(i\pi)},R=|J|\exp{(i\pi)})$. To this end, let us consider the case $(t=|t|\exp{(i\theta)},R=|R|\exp{(i\theta)})$, i.e. continuing $k$ and $J$ to complex $t$ and $R$ with the same phase. Then there is no relative phase between the two actions and the classical solutions,
 hence the one-loop computation is the same as before. Together with the continuation of $k$ just discussed, such continuation effectively simply replaces the $k$ and $J$ in the final answer by their continuation. The same applies to $Z_{SU(2),-}$.

\subsection{\texorpdfstring{$\widehat{\mathfrak{sl}}(2,\mathbb{R})_{k'}$}{sl(2,R)-k-hat} Characters from Analytic Continuation of Chern-Simons Theory}\label{sec:sl2ra}
From equation~\eqref{eq:sl2ch3} of \hyperref[sec:sl2k]{Appendix \ref*{sec:sl2k}}, we know that the (non-unitary) $\widehat{\mathfrak{sl}}(2,\mathbb{R})_{k'}$ Kac-Moody characters of the vacuum and discrete representations, holomorphic in $q$ by definition, are the same as the ones obtained by formally replacing $k=-k'$ and $J=-l$ or $J=+l'$ in the $\widehat{\mathfrak{su}}(2)_k$ characters. However, the $\mathcal{Z}_{SU(2)}(k,\chi,J)$ are only defined for positive integer $k$, while $k'$ can be an arbitrary positive real number. The correct way to execute this operation is first to continue analytically $\mathcal{Z}_{SU(2)}(k,\chi,J=0)$ and $\mathcal{Z}_{SU(2),\pm}(k,\chi,J)$ to $k\rightarrow t-2\in\mathbb{C}$ and $J\rightarrow R\in\mathbb{C}$. Then the precise statement is:
\begin{align}
\mathcal{Z}_{SL(2,\mathbb{R})}(k',\chi,l=0)&=\mathcal{Z}_{SU(2)}(t=-k'+2,\chi,J=0)\\
\mathcal{Z}_{SL(2,\mathbb{R})}(k',\chi,l\in\mathbb{Z}^+/2)&=\mathcal{Z}_{SU(2),-}(t=-k'+2,\chi,R=-l)\\
\mathcal{Z}_{SL(2,\mathbb{R})}(k',\chi,l'\in\mathbb{Z}^-/2)&=\mathcal{Z}_{SU(2),+}(t=-k'+2,\chi,R=l').
\end{align}

These relations suggest the corresponding relations in Chern-Simons theory from an analytic continuation of $SU(2)$ Chern-Simons Theory with $k>0$ and spin $0\leq J\leq k/2$. Explicitly, we need the analytic continuation $(k,J)\rightarrow (-k,l'\in\mathbb{Z}^-)$ and $(-k,-l\in\mathbb{Z}^-)$. As we have just discussed in the last subsection, in such continuation the $i\varepsilon$ prescription stays the same, meaning that starting from $Z_H(k>0)$, which gives $\widehat{\mathfrak{su}}(2)_{k}$ characters holomorphic in $q$, we get $\widehat{\mathfrak{sl}}(2,\mathbb{R})_{k'}$ characters also holomorphic in $q$. Furthermore, we only need to continue $k$ and $J$ with the same phase, $(k,J)\rightarrow(t=|t|\exp{(i\theta)},R=|R|\exp{(i\theta)})$, in which case the perturbative calculations also remain the same. In the end we simply have to take the final answers of $Z_{SU(2),\pm}(k,\chi,J)$ that we have computed and replace respectively $(k,J)\rightarrow (-k,l'\in\mathbb{Z}^-)$ and $(-k,-l\in\mathbb{Z}^-)$ to obtain the $\widehat{\mathfrak{sl}}(2,\mathbb{R})_{k'}$ characters. At this point, one should not be surprised by the fact that an $SU(2)$ path integral computes an $\widehat{\mathfrak{sl}}(2,\mathbb{R})_{k'}$ character after an analytic continuation; after all, the steepest descent path integral $Z_{H',\mathcal{J}_\sigma}$ from each saddle point is independent of the choice of the real form $H'$ of $SU(2)_{\mathbb{C}}=SL(2,\mathbb{C})$. 

\subsection{Covariant Gauge-Fixing of \texorpdfstring{$SL(2,\mathbb{R})$}{SL(2,R)} Chern-Simons Theory}
We have just shown that the analytic continuation of $SU(2)$ Chern-Simons Theory gives $\widehat{\mathfrak{sl}}(2,\mathbb{R})_{k'}$ characters holomorphic in $q$. But can we get them directly from $SL(2,\mathbb{R})$ Chern-Simons Theory using a covariant gauge-fixing? The answer is again no. First of all, the $L^2$-norm~\eqref{eq:L2norm} we used for compact groups is not positive-definite for non-compact groups such as $SL(2,\mathbb{R})$. A consistent gauge-fixing procedure was devised in \cite{BarNatan:1991rn} as follows.

Consider a non-compact gauge group $H'$ and its complexification $G=H'_\mathbb{C}$. As before we use a basis $\{T^a\}$ for $\mathfrak{g}$ such that any $\mathfrak{g}$-valued field $\psi=\sum_a \psi^a(x)T^a$ has $\psi^a(x)$ real when $\psi\in\mathfrak{h}$, where $\mathfrak{h}$ generates the maximal compact subgroup $H$ of $G$. Define an operation $T$ on an adjoint bundle by
\begin{align}
T\left(\psi^a(x)T^a\right)\equiv \psi^{a*}(x)T^a.
\end{align}
In real geometry, $T$ commutes with the metric i.e. the Hodge-star $\star$. The $L^2$-norm of two $\mathfrak{g}$-valued $p$-forms $u$ and $v$ is defined as
\begin{align}
(u,v)_{T}=-\int_M d\mu \Tr[u\star Tv].\label{eq:BWnorm}
\end{align}
This is a legitimate positive-definite inner product for all $\mathfrak{g}$-valued adjoint bundles, which is therefore 
also true when restricted to $\mathfrak{h}'$-valued adjoint bundles. When restricted to $\mathfrak{h}$ the $T$-operation has no effect so one recovers~\eqref{eq:L2norm}. Consider the $H'$ Chern-Simons theory of a connection $A$. Expand it around a flat background $\bar{A}$. In terms of the fluctuation $A'$ (we drop the prime from now on) one can use the following background-gauge gauge-fixing term
\begin{align}
V_{T}=\frac{k}{2\pi}\int_M\Tr[\bar{c}\star D \star TA],
\end{align}
for which the total gauge-fixed action for $Z_{H'}=\int_{\mathcal{U}_{H'}} DA\exp{\left(+ikW[A]\right)}$ is
\begin{align}
&\quad I[A]-\delta_B V_{T}=I[\bar{A}]+\frac{i}{2}(H',\hat{L}_- H')_{T}+(\bar{c},\hat{\Delta}_0 c)_{T}.\label{eq:Igf2}
\end{align}
Once again we have canonically normalized the fields with positive constants $\sim 1/\sqrt{k}$, and $\varphi'\equiv T\varphi$. We have defined $\hat{\Delta}_0\equiv -TD^\mu TD_\mu$ and 
\begin{align}
H'=\begin{pmatrix}
A\\ \varphi'
\end{pmatrix},\;\hat{L}_-=\begin{pmatrix}
T\frac{\varepsilon^{\mu\nu\rho}}{\sqrt{g}}D_\mu & -D^\mu T \\
D^\rho T &0
\end{pmatrix}.
\end{align}
$\hat{\Delta}_0$ and $\hat{L}_-$ are both self-adjoint wrt. the new $T$-norm~\eqref{eq:BWnorm}.

Now consider $\bar{A}=0$, in which case $T$ commutes with $D=d$ and so $\hat{L}_-=L_-T$. Denote respectively with $Q_0$ and $Q_\perp$ the restrictions of $L_-$ (no hat) on the compact $K$ and non-compact $P$ directions of the adjoint bundle. $\hat{L}_-$ decomposes into a direct sum: $\hat{L}_-=Q_0\oplus (-Q_\perp)$. Decompose $H'$ into compact and non-compact components, ${H'}^K$ and ${H'}^P$. The eigenvalue equation of $\hat{L}_-=L_-T$ reads
\begin{align}
\hat{L}_-{H'}^K_i&=L_-{H'}^K_i=\lambda_i{H'}^K_i,\;\hat{L}_-{H'}^P_i=-L_-{H'}^P_i=-\lambda_i{H'}^P_i.
\end{align}
Normalize $({H'}^K_i,{H'}^K_j)_{T}=({H'}^P_i,{H'}^P_j)_{T}=\delta_{ij}$ and $({H'}^K_i,{H'}^P_j)_{T}=0$, and expand $H'=\sum_i c_i{H'}^K_i+\sum_i \tilde{c}_i{H'}^P_i$, with $c_i$ and $\tilde{c}_i$ real. The Chern-Simons path integral containing $H'$ reads
\begin{align}
&\quad Z_{H'}(k)\sim\int DH' \exp{\left(\frac{i}{2}\int \Tr [H'T\hat{L}_-H']\right)}\nonumber\\
&=\int DH' \exp{\left(-\frac{i}{2}\int -\Tr[(c_i{H'}^K_i+\tilde{c}_i{H'}^P_i)T\left\{\lambda_j(c_j{H'}^K_j-\tilde{c}_j{H'}^P_j)\right\}]\right)}\nonumber\\
&=\prod_i\int^\infty_{-\infty} dc_i\int^\infty_{-\infty} d\tilde{c}_i \exp{\left(-\frac{i}{2}\lambda_i(c^2_i-\tilde{c}^2_i)\right)}=\prod_i\int^\infty_{-\infty} dc_i \exp{\left(-\frac{i}{2}\lambda_ic^2_i\right)} \int^\infty_{-\infty} d\tilde{c}_i \exp{\left(+\frac{i}{2}\lambda_i\tilde{c}^2_i\right)}.\label{eq:NCGaussian}
\end{align}
The path integral is now expressed in terms of the eigenvalues $\lambda_i$ of $L_-$ (no hat). In~\eqref{eq:NCGaussian} the $c_{i}$ and $\tilde{c}_i$ Gaussian integrands have opposite signs. Compared to the Gaussian integrals~\eqref{eq:ZHt} of $Z_{H}(k>0)$ (before continuation) for a compact group $H$, the $T$-operation in \cite{BarNatan:1991rn} amounts to assigning $Z(k>0)$ and $\tilde{Z}(k>0)$ respectively to the compact and non-compact directions of $H'$, thereby arriving also at the main result of \cite{BarNatan:1991rn} that the gravitational contribution to the eta-invariant is $(\dim{K}-\dim{P})\eta_{grav}$. This however means that the resulting partition function cannot be holomorphic in $q$ (unless $H'$ is compact), therefore when $H'=SL(2,\mathbb{R})$, $Z_{H'}(k)$ is not a (non-unitary) $\widehat{\mathfrak{sl}}(2,\mathbb{R})_{k}$ character holomorphic in $q$. This is not a contradiction or inconsistency: the two definitions of the partition function for a non-compact group involve two choices of analytic continuation of Chern-Simons theory, which are distinct as we have argued in \hyperref[sec:anacont1]{Section \ref*{sec:anacont1}}. This seems to suggest that the gauge-fixing in \cite{BarNatan:1991rn} involving the $T$-operation is equivalent to imposing chiral boundary conditions on the compact components of $A$ but anti-chiral conditions on the non-compact components in the Hamiltonian reduction to a chiral WZW model, at least when the background is trivial, cf.\ Appendix E in~\cite{Mikhaylov:2014aoa}. This is in contrast to imposing chiral boundary condition on all components from which one obtains a holomorphic character. 

\section{Gravity and the Virasoro Characters}\label{sec:grav}
\subsection{Analytic Continuation of Chern-Simons Theory II}
We review here some facts about the analytic continuation of Chern-Simons Theory \cite{Witten:2010cx}, which are relevant to the computation of the 
gravity partition function. Consider the path integral of a connection $\mathcal{A}$ in $G=H_\mathbb{C}$
\begin{align}
Z_{G}(t,t^*)&=\int_{\mathcal{U}_G} D\mathcal{A}\exp{\left( itW[\mathcal{A}]+ it^* W[\mathcal{A}^*] \right)}\equiv \int_{\mathcal{U}_G} D\mathcal{A}\exp{\left( i\mathcal{I}[\mathcal{A},\mathcal{A}^*]\right)}, \label{m3b}
\end{align}
where $2t=l+is$, $l\in\mathbb{Z}$ is quantized by demanding that the path integral be invariant under a gauge transformation not connected to the identity, and $s\in\mathbb{R}$ such that $\mathcal{I}$ is real and that the path integral is (oscillatory) convergent. This is not a contour integral since the action is not holomorphic in $\mathcal{A}$. However, we analytically continue this integral following \cite{Witten:2010cx}\cite{Witten:1989ip} by defining a new path integral over two independent $\mathfrak{g}$ connections, $\mathcal{A}$ and $\tilde{\mathcal{A}}$,
\begin{align}
Z'_G(t,\tilde{t})\equiv \int_\mathcal{C} D\mathcal{A}D\tilde{\mathcal{A}}\exp{\left( itW[\mathcal{A}]+ i\tilde{t}W[\tilde{\mathcal{A}}] \right)},
\end{align}
where the contour $\mathcal{C}$ is middle-dimensional in $\mathcal{U}_G$ and is defined such that $Z'_G(t,\tilde{t}=t^*)=Z_{G}(t,t^*)$. This analytic continuation is equivalent to a complexification of $G$, isomorphic to $G\times G$. The standard procedure is to deform and decompose the contour $\mathcal{C}$ into a piecewise sum of steepest descent paths of the contributing critical points. The action of $Z'_G$ is a sum of the Chern-Simons actions of $\mathcal{A}$ and $\tilde{\mathcal{A}}$, so the critical points of $Z'_G$ factorize as $\bar{\mathcal{A}}_\sigma\times \bar{\tilde{\mathcal{A}}}_\tau$, where $\bar{\mathcal{A}}_\sigma$ and $\bar{\tilde{\mathcal{A}}}_\tau$ are flat connections. The steepest descent path from each critical point also factorizes into the one from $\bar{\mathcal{A}}_\sigma$ and that from $\bar{\tilde{\mathcal{A}}}_\tau$ respectively: $\mathcal{J}_{\sigma,\tau}=\mathcal{J}_{\sigma}\times \tilde{\mathcal{J}}_\tau$. A similar factorization holds for the steepest ascent paths. Then $Z'_G(t,\tilde{t})$ can be expressed as
\begin{align}
Z'_G(t,\tilde{t})&=\sum_{\sigma,\tau}\mathfrak{n}_{\sigma,\tau} \int_{\mathcal{J}_\sigma}D\mathcal{A}\exp{\left(\vphantom{\tilde{A}} itW[\mathcal{A}]\right)}\times \int_{\tilde{\mathcal{J}}_\tau}D\tilde{\mathcal{A}}\exp{\left( i\tilde{t}W[\tilde{\mathcal{A}}]\right)}\\
&=\sum_{\sigma,\tau}\mathfrak{n}_{\sigma,\tau} \int_{\mathcal{J}_\sigma}DA\exp{\left(\vphantom{\tilde{A}} i(t-h)W[\mathcal{A}]\right)}\times \int_{\tilde{\mathcal{J}}_\tau}D\tilde{A}\exp{\left( i(\tilde{t}-h)W[\tilde{\mathcal{A}}]\right)}\\
&\equiv\sum_{\sigma,\tau}\mathfrak{n}_{\sigma,\tau}Z_{H',\sigma}(t-h)Z_{H',\tau}(\tilde{t}-h)\label{eq:anacont}
\end{align}
where $\mathfrak{n}_{\sigma,\tau}$ is the net intersection number of $\mathcal{C}$ with the oriented steepest ascent path from each critical point. In the first line, the steepest descent path $\mathcal{J}_\sigma$ in $\mathcal{U}_G$ is uniquely determined by the Morse function-- taken to be the real part of the action-- and depends only on the critical point and the parameters characterizing the path integral. In the third line we denote the integral as an analytic continuation from that of some real form $H'$ of $G$. As emphasized multiple times $\mathcal{J}_\sigma$ hence $Z_{H',\sigma}(t)$ are independent of the choice of $H'$. $Z_{H',\sigma}(t)$ is the analytic continuation of the steepest descent path integral of $H'$
\begin{align}
Z_{H',\sigma}(k_1)\equiv \int_{\mathcal{J}_\sigma}DA\exp{\left(ik_1W[A] \right)}
\end{align}
initially defined for some $k_1$. As we have seen in \hyperref[sec:anacont1]{Section \ref*{sec:anacont1}} (where we chose $H'=H$ compact), the choice $k_1>0$ or $k_1<0$ from which the analytic continuation is performed gives two distinct results (for each independent direction of the Lie algebra), due to the square-root branch cut and the infinite number of eigenvalues in $\mathbb{R}$ of $L_-$. Such choice is an ambiguity not fixed at this stage, which is presumably specified by the problem under consideration.

\subsection{Gravity as Chern-Simons Theory and the Brown-Henneaux Conditions}
In the first-order formalism, the dynamical variables of Einstein gravity are the Dreibein $e$ and the spin connections $\omega$. In three dimensions with a negative cosmological constant $\Lambda=-1$, the Euclidean Einstein-Hilbert action reads ($Z\sim e^{-I}$)
\begin{align}
I_{EH}[e,\omega]&=-\frac{1}{16\pi G_N}\int_{M}\sqrt{g}(R-2\Lambda)=-\frac{1}{16\pi G_N}\int_M \epsilon_{abc}\left(e^a\wedge R^{bc}+\frac{1}{3} e^a\wedge e^b\wedge e^c\right),
\end{align}
where $R^{ab}[\omega]=d\omega^{ab}+\omega^{ad}\wedge \omega_{d}^{\;\;b}$ is the curvature two-form of the spin connections $\omega$ and $G_N$ is the Newton constant. As shown in \cite{Banados:1998ys} (see also~\cite{Witten:2018lgb}), for a canonical ensemble fixing the (conformal) boundary complex structure $\tau$, i.e. the angular potential $\tau_1$ and the inverse temperature $\tau_2$, the appropriate boundary term added to the Einstein-Hilbert action is \textit{half} the Gibbons-Hawking term. The total gravity action is therefore
\begin{align}
I_{g}[e,\omega]&=-\frac{1}{16\pi G_N}\left(\int_{M}\sqrt{g}(R-2\Lambda)+\int_{\partial M}\sqrt{h}K\right),\label{gravaction}
\end{align}
where $h$ and $K$ are respectively the induced metric and the extrinsic curvature on the (conformal) boundary. The signs chosen here in the path integral give the classical Bekenstein-Hawking entropy with the correct sign.

Einstein gravity in three dimensions can be recast as a Chern-Simons theory \cite{Achucarro:1987vz}\cite{Witten:1988hc}. In the present case, i.e. Euclidean asymptotically $AdS_3$ manifolds, define an $\mathfrak{sl}(2,\mathbb{C})$ connection $\mathcal{A}_{g}$ of the non-compact gauge group $G=SL(2,\mathbb{C})$ in terms of $e$ and $\omega$ as:
\begin{align}
\mathcal{A}_{g}&=(\omega^a+ie^a)T^a,\;-\mathcal{A}_{g}^\dagger=(\omega^a-ie^a)T^a,
\end{align}
where $\omega^a=-\frac{1}{2}\epsilon^{abc}\omega_{bc}$ as before and we have used the same generators as in the $SU(2)$ case. 

Taking $\tilde{\mathcal{A}}_g=-\mathcal{A}_{g}^\dagger$ as an independent field, the connections $\mathcal{A}_{g}$ and $\tilde{\mathcal{A}}_{g}$ corresponding to asymptotically Euclidean $AdS_3$  metrics \cite{Brown:1986nw} in terms of the Fefferman-Graham radial coordinate $r'$ (for pure $AdS_3$, $r'=\frac{r}{2}+\frac{\sqrt{1+r^2}}{2}$), assume the form \cite{Coussaert:1995zp}
\begin{align}
\mathcal{A}_{g}\overset{r\rightarrow \infty}{\rightarrow}\begin{pmatrix}
\frac{dr'}{2r'} & \frac{L(w,\bar{w})}{r'}dw\\
r' dw & -\frac{dr'}{2r'}
\end{pmatrix},\;\tilde{\mathcal{A}}_{g}\overset{r\rightarrow \infty}{\rightarrow}\begin{pmatrix}
-\frac{dr'}{2r'} & -r'd\bar{w}\\
-\frac{\bar{L}(w,\bar{w})}{r'}d\bar{w} & \frac{dr'}{2r'}
\end{pmatrix}.
\end{align}
The leading radial dependence can be gauged away by a (singular) $SL(2,\mathbb{C})$ matrix not in $SU(2)$, under which the Chern-Simons actions and the integration measures are invariant:
\begin{align}
b&=\diag{(\sqrt{r'},1/\sqrt{r'})}\notin SU(2),\\
\begin{split}
\mathcal{A}_{g}\rightarrow \mathcal{B}&=b\mathcal{A}_{g}b^{-1}-db b^{-1}\\
&=\begin{pmatrix}
0& L(w,\bar{w})\\
1 & 0
\end{pmatrix}dw
\end{split}\quad,
\begin{split}
\tilde{\mathcal{A}}_{g}\rightarrow \tilde{\mathcal{B}}&=b^{-1}\tilde{\mathcal{A}}_{g}b-db^{-1} b\\
&=\begin{pmatrix}
0& -1\\
-\bar{L}(w,\bar{w}) & 0
\end{pmatrix}d\bar{w}.
\end{split}\label{eq:Bfield}
\end{align}
In the Hamiltonian formalism, the Brown-Henneaux conditions, which are classical asymptotic conditions imposed at the tree-level, in terms of the continued Chern-Simons connections ($\mathcal{B}$ and $\tilde{\mathcal{B}}$) can be defined in three steps (see \cite{Coussaert:1995zp} where they work in Lorentzian signature). First, one imposes the chiral boundary conditions, which are compatible with the Chern-Simons actions with no boundary terms, 
\begin{align}
\text{Boundary Conditions: }\mathcal{B}_{\bar{w}}|_{r'\rightarrow \infty}=0&,\;\tilde{\mathcal{B}}_{w}|_{r'\rightarrow \infty}=0.\label{eq:BHBC}
\intertext{This \textit{formally} reduces the Chern-Simons Theories to two 
chiral WZW models with opposite chiralities. Second, one imposes boundary first-class constraints}
\text{First-Class Constraints: }\mathcal{B}_w^{(+)}|_{r'\rightarrow \infty}\overset{!}{=}1&,\;\tilde{\mathcal{B}}_{\bar{w}}^{(-)}|_{r'\rightarrow \infty}\overset{!}{=}-1,\label{eq:BHcons}
\intertext{where the superscripts $(+)$ and $(-)$ denote lower- and upper- triangular components respectively. They generate gauge transformations \textit{with support at the boundary}, which are gauge-fixed by:}
\text{Gauge Fixing: } \mathcal{B}_w^{(3)}|_{r'\rightarrow \infty}\overset{!}{=}0&,\;\tilde{\mathcal{B}}_{\bar{w}}^{(3)}|_{r'\rightarrow \infty}\overset{!}{=}0.\label{eq:BHgf}
\end{align}
The last two steps \textit{formally} reduce the two chiral WZW models into two Virasoro algebras. We emphasize that the gauge redundancies here are tree-level and have support at the boundary; they are independent of the bulk gauge redundancies that we have considered at one loop, since the latter only include proper gauge transformations vanishing at the boundary. We will perform the last two steps~\eqref{eq:BHcons} and~\eqref{eq:BHgf} in the path integral language in \hyperref[sec:BOO]{Section \ref*{sec:BOO}}.

These procedures were also discussed recently in \cite{Cotler:2018zff} in Lorentzian signature where, using the standard Hamiltonian reduction \cite{Elitzur:1989nr} and path integral quantization of co-adjoint orbits, the holomorphic Virasoro characters were reproduced. 

\subsection{Defining the Analytic Continuation for Gravity}
The gravity action $I_{g}$ for the canonical ensemble~\eqref{gravaction} can be rewritten as
\begin{align}
I_{g}[\mathcal{A}_g]&=ik_{g}W[\mathcal{A}_g]-ik_{g}W[-\mathcal{A}_g^\dagger]=ik_{g}W[\mathcal{A}_g]-ik_{g}W[\mathcal{A}^*_g]\label{gravact}
\end{align}
with no boundary terms, where the $k_{g}=l/4G_N>0$ is related classically to the central charge $c$ by $c=6k_g=3l/2G_N$. 
We recall that because we defined $X^*\equiv X^*_a T^a$ and the generators are anti-Hermitian,  $-\mathcal{A}^\dagger_g$ and $\mathcal{A}^*_g$ are equal. The gravity partition function for real geometry is then
\begin{align}
Z_{g}(k_g,\tau)&=\int De \int D\omega \exp{\left(-I_{g}[e,\omega] \right)}=\int D\mathcal{A}_{g}\exp{\left(-ik_{g}W[\mathcal{A}_g]+ik_{g}W[\mathcal{A}^*_g] \right)},\label{eq:gravZ}
\end{align}
where the subscript $g$ refers to gravity. Notice that the integrand is real, so it is not of the general form given in~\eqref{m3b}, where the integrand is a pure phase. This observation motivates us to define eq.~\eqref{eq:gravZ} by its analytic continuation (keeping the dependence on $\tau$ implicit)
\begin{align}
Z'_{SL(2,\mathbb{C})}(k_g)&\equiv \int_{\mathcal{C}} D\mathcal{A}_gD\tilde{\mathcal{A}}_g\exp{\left( -ik_g W[\mathcal{A}_g]+ik_g W[\tilde{\mathcal{A}_g}]\right)}\\
&=\sum_{\sigma,\tau}\mathfrak{n}_{\sigma,\tau} \int_{\mathcal{J}_\sigma}D\mathcal{A}_g\exp{\left( -ik_gW[\mathcal{A}_g]\right)}\times \int_{\tilde{\mathcal{J}}_\tau}D\tilde{\mathcal{A}}_g\exp{\left( ik_gW[\tilde{\mathcal{A}}_g]\right)},\label{eq:Zcont}\\
&\equiv\sum_{\sigma,\tau}\mathfrak{n}_{\sigma,\tau}Z_{SU(2),\sigma}(t=-k_g)\tilde{Z}_{SU(2),\tau}(\tilde{t}=-k_g) ,
\end{align}
where $\tilde{\mathcal{A}}_g$ is now independent of $\mathcal{A}_g$. The contour $\mathcal{C}$ is some middle-dimensional path in $\mathcal{U}_{SL(2,\mathbb{C})}\times \mathcal{U}_{SL(2,\mathbb{C})}$. We do not attempt to answer what the contour $\mathcal{C}$ is, or what $\mathfrak{n}_{\sigma,\tau}$ is for each critical point, cf. \cite{Maloney:2007ud}. We only attempt to compute the contribution of thermal $AdS_3$ to this path integral by choosing $H'=SU(2)$ compact.

In this problem, the Euclidean gravity partition function $Z_{g}(k_g,\tau)$ is real when evaluated on a real metric such as thermal $AdS_3$. The same must hold \textit{by definition} for its analytic continuation $Z'_{SL(2,\mathbb{C})}(k_g)$, so if we define $Z_{SU(2),\sigma}(t)$ as an analytic continuation from $Z_{SU(2)}(k>0)$, $\tilde{Z}_{SU(2),\tau}(\tilde{t})$ must be defined as an analytic continuation from $\tilde{Z}_{SU(2)}(k>0)$ such that they have the opposite $i\varepsilon$ prescriptions and are complex conjugate of each other on a real geometry. 

The question remains as to  why we continued $Z_{SU(2)}(k>0)$ instead of $\tilde{Z}_{SU(2)}(k>0)$. We do not have an a priori argument for choosing one over the other, nor we have a mathematical argument to rule out either one. In existing
literature (see e.g. \cite{Maloney:2007ud}\cite{Cotler:2018zff}) one typically starts with the first-order formalism in Lorentzian gravity, which is described by two independent $SL(2,\mathbb{R})$ Chern-Simons Theories. Upon imposing Brown-Henneaux constraints, its quantization results in a Hilbert space of two vacuum orbits $\widehat{\text{Diff}}(S^1)/PSL(2,\mathbb{R})$ of the Virasoro group, describing left and right movers. The partition function defined as a trace over the Hilbert space is then computed accordingly. It is only from the computation in Lorentzian gravity in the first-order formalism that we expect to get holomorphic $\widehat{\mathfrak{sl}}(2,\mathbb{R})_{k'}$ characters reduced to holomorphic Virasoro by imposing the Brown-Henneaux conditions. We can only take it as a \textit{definition}. Since the Brown-Henneaux conditions contain $\mathcal{A}_{g\bar{w}}|_{\partial M}=0$, we should take in~\eqref{eq:anacont} $Z_{SU(2)}(t)$ as a continuation of $Z_{SU(2)}(k>0)$ which computes $\widehat{\mathfrak{su}}(2)_{k}$ characters holomorphic in $q$, from $t=k>0$ to $t=k_g \exp{(i\pi)}$. Similarly, we should take $\tilde{Z}_{SU(2)}(\tilde{t})$ as a continuation of $\tilde{Z}_{SU(2)}(k>0)$ (i.e. $Z_{SU(2)}(k<0)$) which computes $\widehat{\mathfrak{su}}(2)_{k}$ characters holomorphic in $\bar{q}$, from $t=k>0$ to $t=k_g \exp{(i\pi)}$.

This resonates with our argument in \hyperref[sec:CSWZW]{Section \ref*{sec:CSWZW}} in the WZW language on the ill-definedness of Euclidean $AdS_3$ gravity and the necessity of analytic continuation. The Euclidean gravity path integral was also computed in the second-order formalism in \cite{Giombi:2008vd}. There one has the usual peculiar analytic continuation, namely the Wick rotation of the conformal mode as imposed in \cite{Gibbons:1978ac}. Such rotation only flips the sign of the negative eigenvalues of a scalar Laplacian of the metric and is otherwise inconsequential. It appears that the 
analytic continuations needed for Euclidean gravity in the first-order formalism are more delicate.

\subsection{\texorpdfstring{$\widehat{\mathfrak{sl}}(2,\mathbb{R})_{k_g}$}{sl(2,R)-k-hat} Characters from Gravity}
\subsubsection*{Classical Solutions and Holonomy in Genus-One Topology}
On the equations of motion, $L=L(w)$ and $\bar{L}=\bar{L}(\bar{w})$ are (anti-)holomorphic and there are no sub-leading terms in $r'$ in either $\bar{\mathcal{A}}_g$ or $\bar{\tilde{\mathcal{A}}}_g$. Moreover, in genus one topology, $L$ and $\bar{L}$ are constants, being (anti-)holomorphic, doubly-periodic and non-singular. From now on we focus on the thermal $AdS_3$ case, $L=\bar{L}=-\frac{1}{4}$, i.e.
\begin{align}
\bar{\mathcal{A}}_g&=\begin{pmatrix}
\frac{dr'}{2r'} & 0\\ 0&-\frac{dr'}{2r'}
\end{pmatrix}+\begin{pmatrix}
0 & -\frac{1}{4r'}\\ r'&0
\end{pmatrix}(d\phi+idt)\\
\bar{\tilde{\mathcal{A}}}_g&=\begin{pmatrix}
-\frac{dr'}{2r'} & 0\\ 0&\frac{dr'}{2r'}
\end{pmatrix}+\begin{pmatrix}
0 & -r'\\ \frac{1}{4r'}&0
\end{pmatrix}(d\phi-idt).
\end{align}

The holonomy of $\bar{\mathcal{A}_g}$ around a contractible cycle parametrized by $\phi$ is
\begin{alignat}{3}
H_\phi&=\exp{\left( \oint d\phi \bar{\mathcal{A}_g}\right)}&&\sim\exp{\left( 2\pi T^3\right)}=\begin{pmatrix}
e^{-i\pi}&0\\0&e^{i\pi}
\end{pmatrix}=-\mathds{1}.\label{eq:minus1hol}\\
\intertext{Meanwhile around a non-contractible cycle, including a canonical twist, the holonomy is}
H_\tau&=\exp{\left(\;\; \oint_w \bar{\mathcal{A}_g}\;\;\right)}&&\sim\exp{\left( 2\pi \tau T^3\right)}=\begin{pmatrix}
e^{-\pi i\tau}&0\\0&e^{\pi i\tau}
\end{pmatrix}=\begin{pmatrix}
q^{-\frac{1}{2}}&0\\0&q^{\frac{1}{2}}
\end{pmatrix},
\end{alignat}
where $\sim$ is the equivalence relation by an $SL(2,\mathbb{C})$ conjugation. The $-\mathds{1}$ holonomy and $\bar{\mathcal{A}}_{g\phi} (r\rightarrow 0)\neq 0$ indicate that $\bar{\mathcal{A}}_g$ is singular as a one-form in the complex Lie group $SL(2,\mathbb{C})$, although it is regular in $SL(2,\mathbb{C})/\mathbb{Z}_2$ (identifying $\pm S\in SL(2,\mathbb{C})$) and corresponds to a regular metric on thermal $AdS_3$. This singular nature in $SL(2,\mathbb{C})$ was noted in \cite{Banados:1998ta}, where it was suggested that if one insists on working in $SL(2,\mathbb{C})$ instead of $SL(2,\mathbb{C})/\mathbb{Z}_2$, one should source the $-\mathds{1}$ holonomy by a Wilson loop located at the origin at the expense of introducing a discontinuity into $\mathcal{A}_g$ and $\tilde{\mathcal{A}}_g$ at the origin, as we will do now.

\subsubsection*{The Path Integral with a Wilson Loop}
Let us take the $\mathcal{A}_g$ sector. Since $\mathcal{A}_g$ diverges in $r$ near infinity, we consider its $b$-transform $\mathcal{B}$,
\begin{align}
\bar{\mathcal{B}}=b\bar{\mathcal{A}}_g b^{-1}-dbb^{-1}=\begin{pmatrix}
0&-\frac{1}{4}\\1 &0
\end{pmatrix}dw,
\end{align}
which is justified by the fact that in the Hamiltonian approach, the Brown-Henneaux conditions are imposed on $\mathcal{B}$, 
which is regular in $r$ near infinity. Next, we further re-scale and rotate it by a constant matrix $V$ into
\begin{align}
\bar{\mathcal{B}}\rightarrow \bar{\mathcal{A}}&=V\bar{\mathcal{B}}V^{-1}-dVV^{-1}=T^3(d\phi+idt),\;V=\begin{pmatrix}
e^{\frac{i\pi}{4}}& \frac{1}{2}e^{-\frac{i\pi}{4}}\\
-e^{\frac{i\pi}{4}} & \frac{1}{2}e^{-\frac{i\pi}{4}}
\end{pmatrix}
\end{align} 
and work with $\bar{\mathcal{A}}$. This constant $V$-transformation leaves the action and the integration measure invariant.

Since we can identify $\bar{\mathcal{A}}$ with $\bar{A}_+$ with $\chi=0$ in \hyperref[sec:wilson]{Section \ref*{sec:wilson}} after introducing a discontinuity to $\bar{\mathcal{A}}$ at $r=0$, we take $Z_{SU(2),+}(k>0,J>0)$ with the same Wilson loop $C_0$ at the origin with a canonical twist. The loop ought to source the $-\mathds{1}$ holonomy, so we need to take $J=k/2$, corresponding to saturating the unitarity bound of the $\widehat{\mathfrak{su}}(2)_k$ current algebra. We could have transformed $\bar{\mathcal{A}}_g$ into $\bar{\mathcal{A}}=-T^3dw$, introduced a discontinuity at $r=0$ and identified it with $\bar{A}_-$ with $\chi=0$, leading to using $Z_{SU(2),-}$. However,  since $\chi=0$, the choice of $Z_{SU(2),\pm}$ is not important. Using~\eqref{eq:gravyay}, the total one-loop $O(k^0)$ path integral is then 
\begin{align}
Z_{SU(2)}(k\in\mathbb{Z}^+,J=k/2)&=q^{\frac{(k/2)(k/2+1)}{k+2}-\frac{3}{24}}\prod_{n=1}^\infty\frac{1}{1-q^n}\prod_{n=0}^\infty\frac{1}{1-q^n}\prod_{n=2}^\infty\frac{1}{1-q^n},\label{eq:sl2rgrav0}
\end{align}
where compared to the $0<J<k/2$ case there is an extra shift in the last product. We have included one-loop shifts in $k$ and $J$ in~\eqref{eq:sl2rgrav0} (which we assume to exist without having computed them). Note that the Brown-Henneaux boundary condition~\eqref{eq:BHBC} implies that the chemical potential $\chi$ vanishes. In this situation where a Wilson loop is present, there exists a bosonic zero mode $\varphi_0$. To take care of the divergence that it generates, we factor it out by inserting $\delta(\varphi_0)$ in the Chern-Simons path integral and integrate over the non-zero modes. The corresponding Jacobian is equal to one so the resulting path integral $Z'_{SU(2)}$ is simply~\eqref{eq:sl2rgrav0} without the $n=0$ term,
\begin{align}
Z'_{SU(2)}(k\in\mathbb{Z}^+,J=k/2)&=q^{\frac{(k/2)(k/2+1)}{k+2}-\frac{3}{24}}\prod_{n=1}^\infty\frac{1}{1-q^n}\prod_{n=1}^\infty\frac{1}{1-q^n}\prod_{n=2}^\infty\frac{1}{1-q^n}.\label{eq:sl2rgrav0no0}
\end{align}
Under the analytic continuation of the gravity partition function prescribed in the last subsection, eq.~\eqref{eq:sl2rgrav0no0} is then continued into a holomorphic $\widehat{\mathfrak{sl}}(2,\mathbb{R})_{k_g}$ character as in \hyperref[sec:sl2ra]{Section \ref*{sec:sl2ra}},
\begin{align}
\left.Z'_{SU(2)}(k\in\mathbb{Z}^+,J=k/2)\right|_{\substack{k\rightarrow t=-k_g\\J\rightarrow R= -k_g/2}}&=q^{-\frac{k_g}{4}-\frac{3}{24}}\prod_{n=1}^\infty\frac{1}{1-q^n}\prod_{n=1}^\infty\frac{1}{1-q^n}\prod_{n=2}^\infty\frac{1}{1-q^n},\label{eq:sl2rgrav}
\end{align}
under which the classical solution remains the same. \eqref{eq:sl2rgrav} is of course holomorphic in $q$. A similar computation is made for the other (anti-holomorphic) sector. Indeed $q^{-\frac{k_g}{4}}$ is the holomorphic part of the on-shell gravity partition function $(q\bar{q})^{-\frac{k_g}{4}}$ for thermal $AdS_3$. As we will see in the next subsection, after imposing the remaining Brown-Henneaux conditions in the path integral formalism, the bosonic zero mode will  cancel against a fermionic zero mode.

\subsection{Brown-Henneaux Conditions in the Path Integral Formalism}\label{sec:BOO}
So far we only imposed the boundary conditions~\eqref{eq:BHBC}. To recover the Virasoro character it still remains to implement the tree-level boundary constraints~\eqref{eq:BHcons} and the gauge-fixing~\eqref{eq:BHgf} in the path integral formalism.

The reduction from the $SL(2,\mathbb{R})$ current algebra to the Virasoro algebra was carried out in ref.~\cite{Bershadsky:1989mf} using the WZW language in Lorentzian signature. In that case, since the connections are real, the constraint \eqref{eq:BHcons} for each sector in Lorentzian signature, i.e. $\mathcal{B}^{(\pm)}_{\mp}|_{\partial M}\overset{!}{=}\pm 1$ for the right- and left-moving sector respectively, is imposed by inserting into the path integral a delta function $Z_{BH,L}\equiv\int D\lambda \exp{(i\int_{\partial M}\lambda (\mathcal{B}^{(\pm)}_{\mp}\mp 1))}$, where $\lambda$ is a real auxiliary scalar field and the subscript $L$ denotes Lorentzian. This is possible since both $\lambda$ and the constraint are real. For each sector, the $SL(2,\mathbb{R})$ Chern-Simons path integral 
combined with the delta function $Z_{BH,L}$  possesses a boundary gauge redundancy, once an appropriate transformation rule is assigned to the real auxiliary scalar $\lambda$. For the right-moving sector, this redundancy consists of $SL(2,\mathbb{R})$ gauge transformations $U_L$ that belong to the Borel subgroup of $SL(2,\mathbb{R})$ generated by the strictly upper-triangular matrix and are non-vanishing at infinity,
\begin{align}
U_L=\exp{\left(f(x^+,x^-)t^- \right)}=\begin{pmatrix}
1&f(x^+,x^-)\\0&1
\end{pmatrix},\;t^-\equiv\begin{pmatrix}0&1\\0&0
\end{pmatrix}.
\end{align}
For the left-moving sector, the relevant Borel subgroup is instead generated by the strictly lower-triangular matrix. In the Hamiltonian formalism, this is gauge-fixed precisely by the Brown-Henneaux condition \eqref{eq:BHgf}. Alternatively, the gauge-fixing can be done in the BRST formalism in the path integral, as in \cite{Bershadsky:1989mf}.

In Euclidean signature, since $\mathcal{B}$ is a complex $\mathfrak{sl}(2,\mathbb{C})$ connection, we expect the analog of the above procedures to require some analytic continuation, which we now determine. We start with the Euclidean gravity action \eqref{gravact} and path integral \eqref{eq:gravZ} written in terms of $\mathcal{B}$, before the analytic continuation that treats $\mathcal{B}^*$ as a field independent of $\mathcal{B}$,
\begin{align}
Z_{g}=\int D\mathcal{B}\exp{\left(-I_g[\mathcal{B}] \right)}\;,\; I_g[\mathcal{B}]&=ik_{g}W[\mathcal{B}]-ik_{g}W[\mathcal{B}^*].\label{gravact2}
\end{align}
The Brown-Henneaux boundary constraint~\eqref{eq:BHcons} is imposed in the path integral formalism by inserting a delta function $Z_{BH}$ into $Z_{g}$ in \eqref{gravact2}, 
\begin{align}
Z_{BH}\equiv \int D\lambda \exp{\left(-I_{BH}[\mathcal{B},\lambda] \right)}\; ,\;I_{BH}[\mathcal{B},\lambda]=\frac{ik_g}{2\pi}\int_{\partial M}\left(\lambda(\mathcal{B}_w^{(+)}-1)+\lambda^*((\mathcal{B}_{w}^{(+)})^*-1)\right),\label{IBH}
\end{align}
where $\lambda$ is a \textit{complex} auxiliary scalar field. Since $I_{BH}$ is purely imaginary, it imposes the constraints $\Re{(\mathcal{B}_w^{(+)})-1\overset{!}{=}0}$ and $\Im{(\mathcal{B}_w^{(+)})\overset{!}{=}0}$ upon integrating out $\lambda$. 

Next, we need a gauge redundancy similar to that of the two copies of the Borel subgroup of $SL(2,\mathbb{R})$ in the Lorentzian case. To find it, consider a gauge transformation $U$ non-vanishing at infinity belonging to the Borel subgroup of $SL(2,\mathbb{C})$, 
\begin{align}
U=\exp{\left(f(w,\bar{w})t^- \right)}=\begin{pmatrix}
1&f(w,\bar{w})\\0&1
\end{pmatrix}.
\end{align}
Under such a gauge transformation, since $\mathcal{B}\rightarrow U\mathcal{B}U^{-1}-dUU^{-1}$, \eqref{gravact2} transforms as
\begin{align}
\Delta(I_{g}[\mathcal{B}])&=\frac{k_g}{2\pi}\int_{\partial M} \left(\mathcal{B}^{(+)}_w\partial_{\bar{w}}f(w,\bar{w})+(\mathcal{B}^{(+)}_w)^*\partial_{w}f^*(w,\bar{w}) \right). \label{deltaIg}
\end{align}
Here we have used the boundary condition \eqref{eq:BHBC} before the analytic continuation, $\mathcal{B}_{\bar{w}}|_{\partial M}=0$, and the fact that $U$ contributes a vanishing topological term to $\Delta(I_g[\mathcal{B}])$.

We would like to assign a transformation rule to $\lambda$ that makes $Z_gZ_{BH}$ invariant under the Borel subgroup of $SL(2,\mathbb{C})$ (or at least some subgroup therein). However, we immediately run into a problem: $\Delta(I_g[\mathcal{B}])$ in \eqref{deltaIg} is real since $I_g[\mathcal{B}]$ is purely real, while $I_{BH}$ in \eqref{IBH} is purely imaginary. Therefore, to find a non-trivial invariance we must perform an analytic continuation that treats $\lambda$ and $\tilde{\lambda}=\lambda^*$ as independent, and continue $Z_{BH}$ into an integral $Z'_{BH}$ over $\lambda$ and $\tilde{\lambda}$ along the contour $\tilde{\lambda}=\lambda^*$,
\begin{align}
Z'_{BH}&\equiv \int_{\tilde{\lambda}=\lambda^*} D\lambda D\tilde{\lambda} \exp{\left(-I_{BH}[\mathcal{B},\lambda,\tilde{\lambda}] \right)}\; ,\;I_{BH}[\mathcal{B},\lambda,\tilde{\lambda}]=\frac{ik_g}{2\pi}\int_{\partial M}\left(\lambda(\mathcal{B}_w^{(+)}-1)+\tilde{\lambda}((\mathcal{B}_w^{(+)})^*-1)\right).\label{IBH2}
\end{align}
Then $Z_gZ'_{BH}$ does have the desired gauge redundancy by assigning the transformation rules
\begin{align}
\lambda\rightarrow \lambda'=\lambda+i\partial_{\bar{w}}f(w,\bar{w}),\;\tilde{\lambda}\rightarrow \tilde{\lambda}'=\tilde{\lambda}+i\partial_{w}f^*(w,\bar{w}).
\end{align}
This transformation deforms the path of integration, but the result of integration over any contour of the form $\tilde{\lambda}=\lambda^* + g(w,\bar{w})$ does not depend on the  arbitrary function $g(w,\bar{w})$, since the integration over $\lambda$ still results in a delta function. We take a step further by treating $\mathcal{B}$ and $\tilde{\mathcal{B}}=\mathcal{B}^*$ as independent. The path integral $Z_g$ is now continued into the term $Z'_{SL(2,\mathbb{C})}$ in \eqref{eq:Zcont}, that integrates over some middle-dimensional contour $\mathcal{C}$ in the $(\mathcal{B},\tilde{\mathcal{B}})$ field space. Next we introduce the ghosts $b_\partial$, $c_\partial$, $\tilde{b}_\partial$ and $\tilde{c}_\partial$, and the auxiliary complex scalars $\varphi_\partial$ and $\tilde{\varphi}_\partial$, where $b_\partial$, $c_\partial$ and $\varphi_\partial$ are independent of their tilde counterparts. The subscript $\partial$ distinguishes them from their \underline{totally unrelated} one-loop bulk counterparts. We \textit{define} the BRST transformations as follows, ($Dc_\partial=dc_\partial t^-+[\mathcal{B},t^-]c_\partial$ and $D\tilde{c}_\partial=d\tilde{c}_\partial t^++[\tilde{\mathcal{B}},t^+]\tilde{c}_\partial$),
\begin{equation}
\begin{aligned}
\delta_{B\partial} \mathcal{B}&=iDc_{\partial}&\;&&\; & \;&,\;&&\delta_{B\partial} \tilde{\mathcal{B}}&=iD\tilde{c}_\partial&,\\
\delta_{B\partial} \lambda&=\partial_{\bar{w}}c_{\partial}&,\;&&\delta_{B\partial}c_{\partial}&=0&,\;&&\delta_{B\partial} \tilde{\lambda}&=\partial_{w}\tilde{c}_{\partial}&,\;&&\delta_{B\partial}\tilde{c}_{\partial}&=0,\\
\delta_{B\partial} b_\partial&=\varphi_\partial&,\;&&\delta_{B\partial}\varphi_{\partial}&=0&,\;&&\delta_{B\partial} \tilde{b}_\partial&=\tilde{\varphi}_{\partial}&,\;&&\delta_{B\partial}\tilde{\varphi}_{\partial}&=0.
\end{aligned}\label{BRST2}
\end{equation}
With such definition we have essentially separated the Borel subgroup of $SL(2,\mathbb{C})$ into holomorphic and anti-holomorphic parts. To complete the gauge-fixing, we insert the following gauge-fixing path integral $Z'_{gf}$, which is \textit{defined} to be a contour integral along $\tilde{\varphi}_\partial=\varphi^*_\partial$, with a BRST-exact action,
\begin{align}
Z'_{gf}&\equiv \int_{\tilde{\varphi}_\partial=\varphi^*_\partial}D\varphi_\partial D\tilde{\varphi}_\partial\exp{\left(-I_{gf}[\lambda,\varphi_{\partial},b_{\partial},c_{\partial},\tilde{\lambda},\tilde{\varphi}_{\partial},\tilde{b}_{\partial},\tilde{c}_{\partial}] \right)},\\
I_{gf}&=\delta_{B\partial}\left(\frac{ik_g}{2\pi} \int_{\partial M}\left(\lambda b_{\partial}+\tilde{\lambda}\tilde{b}_\partial\right)\right)=\frac{ik_g}{2\pi}\int_{\partial M}\left(b_{\partial }\partial_{\bar{w}}c_{\partial}+\tilde{b}_\partial\partial_w \tilde{c}_\partial+\lambda \varphi_{\partial}+\tilde{\lambda}\tilde{\varphi}_\partial\right).
\end{align}
The total boundary-gauge-fixed path integral $Z'_{SL(2,\mathbb{C})} Z'_{BH} Z'_{gf}$ contains
\begin{align}
\int_{\tilde{\varphi}_\partial=\varphi^*_\partial}D\varphi_\partial D\tilde{\varphi}_\partial \int_{\tilde{\lambda}=\lambda^*} D\lambda D\tilde{\lambda}\exp{\left[ -\frac{ik_g}{2\pi}\int_{\partial M}\left(\lambda(\mathcal{B}_w^{(+)}-1)-\tilde{\lambda}((\tilde{\mathcal{B}}_{\bar{w}}^{(-)})+1)+\lambda \varphi_{\partial}+\tilde{\lambda}\tilde{\varphi}_\partial\right)\right]}.
\end{align}
The contour $\tilde{\varphi}_\partial=\varphi^*_\partial$, which is part of the definition of $Z'_{gf}$, together with the contour $\tilde{\lambda}=\lambda^*$ imposes $\lambda=0$ and $\tilde{\lambda}=\lambda^*=0$. After that we have
\begin{align}
Z'_{SL(2,\mathbb{C})}Z'_{BH}Z'_{gf}=\int_\mathcal{C} D\mathcal{B}D\tilde{\mathcal{B}}Dc_{\partial}Db_{\partial}D\tilde{c}_{\partial}D\tilde{b}_{\partial}\exp{\left(-I_g[\mathcal{B},\tilde{\mathcal{B}}]-\frac{ik_g}{2\pi}\int_{\partial M}b_{\partial }\partial_{\bar{w}}c_{\partial}-\frac{ik_g}{2\pi}\int_{\partial M}\tilde{b}_{\partial }\partial_{w}\tilde{c}_{\partial}\right)}.
\end{align}					
The two independent ghosts systems here are what one would expect from the reduction of the two independent $SL(2,\mathbb{R})$ Chern-Simons theories in Lorentzian signature. Next we consider the two independent $bc$- and $\tilde{b}\tilde{c}$-ghost path integrals on the boundary torus 
\begin{align}
Z_{bc\partial}\equiv\int Dc_{\partial}Db_{\partial}\exp{\left(-\frac{ik_g}{2\pi}\int_{\partial M}b_{\partial }\partial_{\bar{w}}c_{\partial}\right) } ,\;
Z_{\tilde{b}\tilde{c}\partial}\equiv\int D\tilde{c}_{\partial}D\tilde{b}_{\partial}\exp{\left(-\frac{ik_g}{2\pi}\int_{\partial M}\tilde{b}_\partial \partial_{w}\tilde{c}_{\partial}\right) }\label{eq:boundbc0}.
\end{align}
Each of these $bc$-systems has two fermionic zero modes, $b_\partial=b_0$, $c_\partial=c_0$, where $b_0,c_0=$const. To factor them out we insert
into the path integrals the fermionic delta function $\delta(b_0)\delta(c_0)\equiv b_0c_0$, i.e. consider
\begin{align}
Z'_{bc\partial}\equiv\int Dc_{\partial}Db_{\partial}b_0 c_0\exp{\left(-\frac{ik_g}{2\pi}\int_{\partial M}b_{\partial }\partial_{\bar{w}}c_{\partial}\right) },\;Z'_{\tilde{b}\tilde{c}\partial}\equiv\int D\tilde{c}_{\partial} D\tilde{b}_{\partial}\tilde{b}_{0}\tilde{c}_{0}\exp{\left(-\frac{ik_g}{2\pi}\int_{\partial M}\tilde{b}_\partial \partial_{w}\tilde{c}_{\partial}\right) }.
\end{align}
They are easily computed using Riemann and Hurwitz zeta function regularizations (as it was also done in \cite{Axelrod:1989xt}). For $Z'_{bc\partial}$ we have
\begin{align}
&\quad Z'_{bc\partial}=\det\nolimits{'}{(\partial_{\bar{w}})}=q^{\frac{1}{12}}\prod^\infty_{n=1}\left(1-q^n \right)^2,\label{eq:boundbc}
\end{align}
where $\det\nolimits{'}{(\partial_{\bar{w}})}$ is the functional determinant of $\partial_{\bar{w}}$ without the fermionic 
zero modes. Combined with the $\widehat{\mathfrak{sl}}(2,\mathbb{R})$ character~\eqref{eq:sl2rgrav} (where
we discarded the bosonic zero mode) we have, for thermal $AdS_3$ (we write here the functional integral over $\mathcal{A}_g$)
\begin{align}
Z'_g(q)&=\left(q^{\frac{1}{12}}\prod^\infty_{n=1}\left(1-q^n \right)^2\right)\times \left(q^{-\frac{k_g}{4}}q^{-\frac{3}{24}}\prod_{n=1}^\infty\frac{1}{1-q^n}\frac{1}{1-q^{n+1}}\frac{1}{1-q^{n}}\right)\\
&=q^{-\frac{k_g}{4}-\frac{3}{24}+\frac{2}{24}} \prod_{n=2}^\infty\frac{1}{1-q^n}. \label{mmm6}
\end{align} 
The exponent $-k_g/4$ already contains the shifts in $k$ and $J$ (which we only quote without explicitly computing it). If we take the weight to be $h=0$, then the central charge at one-loop is $c=6k_g+3-2$. This is at $O(k^0)$ the same as the Virasoro central charge $c_{vir}$ given in 
ref.~\cite{Bershadsky:1989mf} Eq. 40,
\begin{align}
c_{vir}=-6k+\left(3-\frac{6}{k+2} \right)-2,
\end{align}
when we identify $k=-k_g$. Here the factor $-2$ comes from the same boundary $bc$-system while the middle term can be 
identified with our factor of $3$ that  comes from the one-loop gravitationally-renormalized volume factor. In the end, our analytic continuation gives us the holomorphic Virasoro vacuum character. Multiplying eq.~\eqref{mmm6} 
with the functional integral
over $\tilde{\mathcal{A}}_g$, we recover the gravity partition function of thermal $AdS_3$. To get a Virasoro character corresponding to a non-vacuum co-adjoint orbit $\widehat{\text{Diff}}(S^1)/U(1)$ of highest weight $0< h<c_{vir}/24$, one should instead consider an $SU(2)$ Wilson loop within the $\widehat{\mathfrak{su}}(2)_k$ unitarity bound, i.e. $0<2J/k<1$, and
  follow the same calculation as in \hyperref[sec:wilson]{Section \ref*{sec:wilson}}.

\subsection*{Acknowledgements}
We thank P. Kraus and A. Maloney for useful discussions and especially B. Oblak for kindly sharing his unpublished notes
 with us. We gratefully acknowledge the hospitality of the UCLA Bhaumik Institute, where part of this work was done. M.P. is supported in part by NSF grant PHY-1620039. C.Y. is supported by a James Arthur Graduate Award.

\begin{appendix}
\section{Evaluation of the Chiral WZW Partition Functions}\label{sec:chWZW}
We compute the $SU(2)$ Chern-Simons path integral at one-loop using the standard WZW method on thermal $AdS_3$. It is important to point out that thermal $AdS_3$ is a non-compact manifold without boundary. In order to make sense of the reduction of Chern-Simons theory to a chiral WZW model on a boundary, it is necessary and always understood that we compactify the manifold into a solid torus with a boundary, $T=D\times S^1$.

Consider the same $SU(2)$ Chern-Simons path integral on $T$ as in eq.~\eqref{eq:PIchem},
\begin{align}
Z_{SU(2)}(k>0,A_{\bar{w}}|_{\partial T})=\int_{\mathcal{U}_{SU(2)}} DA\; \exp{(-I[A]) }=\int_{\mathcal{U}_{SU(2)}} DA\;\exp{\left(+ikW[A]-I_{bt}[A]\right) },\;k>0
\end{align}
such that the total action is~\eqref{eq:actionI}
\begin{align}
I[A]&=-\frac{ik}{4\pi}\int_T {\rm Tr}[AdA+\frac{2}{3}A\wedge A\wedge A]+\frac{k}{8\pi}\int_{\partial T}Tr[A_{\phi}^2+A_t^2]\label{eq:actionI2}.
\end{align}
We use the same decomposition used in \hyperref[sec:CSWZW]{Section \ref*{sec:CSWZW}} for $A=A_0+\hat{A}$, where $A_0=A_tdt$ is the component along the $S^1$ direction and $\hat{A}$ is that on the disk $D$. After integrating out $A_t$ in the interior of $T$, which imposes the constraint $\hat{F}=0$, the action $I[A]$ is reduced to the chiral WZW action with $\hat{A}=U^{-1}\hat{d}U$,
\begin{align}
I[U,A_{\bar{w}}|_{\partial T}]&=-\frac{ik}{4\pi}\left( \int_{\partial T} {\rm Tr}[U^{-1}\partial_{\phi}UU^{-1}\partial_{t}U ]-\frac{1}{3}\int_T {\rm Tr}[(U^{-1}dU)^3]+\int_{\partial T} {\rm Tr}[-A_tA_{\phi} +\frac{i}{2}A_\phi^2+\frac{i}{2}A_t^2]\right).
\end{align}
Since the variation of $I[A]$ wrt. $A$ is ($\varepsilon^{x\phi t}=+1$)
\begin{align}
\delta I[A]=(\text{EOM})+\frac{k}{\pi}\int_{\partial T}d^2x {\rm Tr}[A_w\delta A_{\bar{w}}]\equiv(\text{EOM})+i\int_{\partial T}d^2x {\rm Tr}[J_w\delta A_{\bar{w}}],
\end{align}
we fix $A_{\bar{w}}|_{\partial T}=\bar{A}_{\bar{w}}$ as a boundary condition, which allows us to substitute in the boundary term
\begin{align}
A_t|_{\partial T}=-2i \bar{A}_{\bar{w}}+iA_\phi|_{\partial T}=-2i\bar{A}_{\bar{w}}+iU^{-1}\partial_{\phi}U.
\end{align}
Furthermore, we consider the special case $\bar{A}_{\bar{w}}=\frac{i}{2}h^{-1}(t)\partial_t h(t)$ and define $\tilde{U}\equiv h(t)Uh^{-1}(t)$. Hence we have
\begin{align}
I[U,A_{\bar{w}}|_{\partial T}]&= -\frac{ik}{4\pi}\left(-2i\int_{\partial T} {\rm Tr}[\bar{A}_{\bar{w}}^2]-2i\int_{\partial T} {\rm Tr}[U^{-1}\partial_{\phi}U(U^{-1}\partial_{\bar{w}}U-2\bar{A}_{\bar{w}}) ]-\frac{1}{3}\int_T {\rm Tr}[(U^{-1}dU)^3]\right)\label{eq:KLI32}\\
&=-\frac{ik}{4\pi}\left( -2i\int_{\partial T} {\rm Tr}[\bar{A}_{\bar{w}}^2]-2i\int_{\partial T} {\rm Tr}[\tilde{U}^{-1}\partial_{\phi}\tilde{U}\tilde{U}^{-1}\partial_{\bar{w}}\tilde{U} ]-\frac{1}{3}\int_T {\rm Tr}[(\tilde{U}^{-1}d\tilde{U})^3] \right).
\end{align}
This shows that up to the first term it is in fact a gauged chiral WZW action. By setting
\begin{align}
h(t)&=\begin{pmatrix}
e^{-\frac{i\chi}{2\tau_2}t}&0\\
0& e^{+\frac{i\chi}{2\tau_2}t}
\end{pmatrix},\;\chi\in\mathbb{C},
\end{align}
we have $A_{\bar{w}}=\bar{A}_{\bar{w}}=\frac{i\chi}{2\tau_2} T^3$ everywhere. This in particular is in agreement with the boundary condition we used in \eqref{BC1}. 

Now we choose $U(r,\phi,t)=U(\phi,t)$ with the parametrization 
\begin{align}
U=\exp{\left( f_3(\phi,t) T^3+f_+(\phi,t) T^++f_-(\phi,t) T^- \right)},
\end{align}
where $f_{3,\pm}(\phi,t)$ are doubly-periodic, $f_3$ is real and $f_\pm=(f_\mp)^*$ since $SU(2)$ is a compact gauge group. Substituting $\tilde{U}=h(t)Uh^{-1}(t)$, the path integral becomes
\begin{align}
&\quad Z_{SU(2)}(k>0,A_{\bar{w}}|_{\partial T})=e^{+\frac{\pi k\chi^2}{4\tau_2}}\int_{U\in SU(2)} DU \exp{\left(+\frac{k}{2\pi}\int_{\partial T} {\rm Tr}[U^{-1}\partial_{\phi}U(U^{-1}\partial_{\bar{w}}U-2\bar{A}_{\bar{w}})]\right)}\\
&=e^{+\frac{\pi k\chi^2}{4\tau_2}}\int_{U\in SU(2)} DU \exp{\left(+\frac{k}{2\pi}\int_{\partial T} {\rm Tr}[\tilde{U}^{-1}\partial_{\phi}\tilde{U}\tilde{U}^{-1}\partial_{\bar{w}}\tilde{U}]\right)}\\
&=e^{+\frac{\pi k\chi^2}{4\tau_2}}\int_{U\in SU(2)} DU \exp{\left(-\frac{k}{4\pi}\int_{\partial T}\partial_\phi f_3 \partial_{\bar{w}} f_3+\partial_\phi\tilde{f}_+\partial_{\bar{w}} \tilde{f}_-+\partial_\phi\tilde{f}_-\partial_{\bar{w}} \tilde{f}_++\ldots\right)}\\
&=e^{+\frac{\pi k\chi^2}{4\tau_2}}\int_{U\in SU(2)} DU \exp{\left(-\frac{1}{2}\int_{\partial T}\partial_\phi f_3 \partial_{\bar{w}} f_3+\partial_\phi\tilde{f}_+\partial_{\bar{w}} \tilde{f}_-+\partial_\phi\tilde{f}_-\partial_{\bar{w}} \tilde{f}_+ +o(k^0)\right)}\\
&\equiv e^{+\frac{\pi k\chi^2}{4\tau_2}}\det\nolimits^{-\frac{1}{2}}{D}.
\end{align}
In the third line we have defined $\tilde{f}_{\pm}\equiv \exp{(\pm \frac{i\chi}{\tau_2} t)}f_\pm$. As in the main text, we take $\chi\in\mathbb{R}$ throughout the calculation such that $\tilde{f}_\pm=(\tilde{f}_\mp)^*$, and analytically continue $\chi$ to complex values in the end. In the second to last line we have canonically normalized $f_{3,\pm}\rightarrow \sqrt{2\pi/k} f_{3,\pm}$ and expanded up to $O(k^0)$. Note that since the real part of the exponent at $O(k^0)$ is negative-definite because the Killing metric $\Tr$ is negative-definite, the path integral converges; see also \cite{Mikhaylov:2014aoa}.

Consider first the eigenfunctions with $\tilde{f}_\pm=0$. Fourier expand
\begin{align}
f_3(\phi,t)&=e^{im\phi-i\omega t},\;\omega=\frac{-n+m\tau_1}{\tau_2},\;n\in\mathbb{Z},\;m\in\mathbb{Z}\backslash \{0\}.
\end{align}
We have discarded the zero modes $m=0$ which correspond to the shift redundancy of the quadratic action under $f_{3,\pm}\rightarrow f_{3,\pm}+\alpha(t)$. This step can be done more carefully by a BRST gauge-fixing procedure, as in~\cite{bob}, or by computing the Jacobian of the change of coordinates that factors out the zero mode. The eigenvalues are
\begin{align}
\lambda_{m,n}^{(3)}&=-\frac{1}{2}im\left(im+i\left(-i\omega \right) \right)=\frac{im}{2\tau_2}(n-m\tau),
\end{align}
with which the one-loop determinant $\det{D^{(3)}}$ for the $(3)$-component $f_3$ is
\begin{align}
&\quad\det{D^{(3)}}=\exp{(\tr\ln{D^{(3)}})}=\exp{\left(\sum_{\substack{m\neq 0\\n\in\mathbb{Z}}}\ln{\left(\frac{im}{2\tau_2}\right)}+\sum_{\substack{m\neq 0\\n\in\mathbb{Z}}}\ln{\left( n-m\tau\right)}\right)}\propto q^{\frac{1}{12}}\prod_{n=1}^\infty (1-q^n)^2.
\end{align}
Here the trace of the logarithm of the eigenvalues has been evaluated with Riemann and Hurwitz zeta-function regularizations and we have discarded overall numerical constants. The $(3)$-component contribution to the path integral $Z$ at $O(k^0)$ is thus
\begin{align}
Z^{(3)}=q^{-\frac{1}{24}}\prod_{n=1}^\infty \frac{1}{1-q^n}\label{eq:chiralboson} .
\end{align}
This is of course the (exact) partition function of a chiral boson. This is also the partition function that one obtains from the $U(1)$ Chern-Simons Theory $Z_{U(1)}(k)$.

The other set of eigenfunctions have $f_3=0$ and $f_{\pm}=(f_{\mp})^*$. Consider first the term $\tilde{f}_-\partial_\phi \partial_{\bar{w}}\tilde{f}_+$ in the action. In Fourier modes, the eigenfunctions read
\begin{align}
\tilde{f}_+&=e^{\frac{i\chi}{\tau_2}t}e^{im\phi-i\omega t}=e^{im\phi-i\left(\omega-\frac{\chi}{\tau_2}\right) t},\;\omega=\frac{-n+m\tau_1}{\tau_2},\;n\in\mathbb{Z},\;m\in\mathbb{Z}\backslash \{0\}.
\end{align}
We simply have a shift in $n\rightarrow n+\chi$. The eigenvalues are
\begin{align}
\lambda_{m,n}^{(+)}&=\frac{im}{2\tau_2}((n+\chi)-m\tau).
\end{align}
Doing the same calculation of the one-loop determinant as we did for $D^{(3)}$, we find that $\det{D^{(+)}}$ is 
\begin{align}
\det{D^{(+)}}=\exp{(\tr\ln{D^{(+)}})}=q^{\frac{1}{12}}\prod_{n=1}^\infty (1-q^ne^{2\pi i \chi})(1-q^ne^{-2\pi i \chi}).
\end{align}
Since this expression is symmetric in $\chi$, $\det{D^{(-)}}$ involving the term $\tilde{f}_+\partial_\phi \partial_{\bar{w}}\tilde{f}_-$ gives the same answer. Therefore, the total path integral at one-loop $O(k^0)$ is
\begin{align}
Z(k,\chi)&=e^{+\frac{\pi k\chi^2}{4\tau_2}}q^{-\frac{3}{24}}\prod_{n=1}^\infty \frac{1}{1-q^n}\frac{1}{1-q^ne^{2\pi i \chi}}\frac{1}{1-q^ne^{-2\pi i \chi}}.\label{eq:WZWsu2}
\end{align}
Discarding the exponential, this is the canonical vacuum Kac-Moody character $\mathcal{Z}_{SU(2)}(k,\chi)$ 
as in eq.~\eqref{eq:su2ch2}, computed to $O(k^0)$, which is known to be different from the path integral expression $Z_{SU(2)}(k,\chi)$ by exactly the discarded exponential \cite{Kraus:2006nb}. This also suggests that the continuation from $\chi\in\mathbb{R}$ to $\mathbb{C}$ is valid.

It is clear that the holomorphicity in $q$ of the path integral comes entirely from the chiral boundary condition fixing $\bar{A}_{\bar{w}}|_{\partial T}$ on the boundary torus of the compact manifold $T$.

\section{Characters in Current Algebra}\label{kmalg}
We review basic facts about Kac-Moody algebra, following \cite{Blumenhagen:2009zz}, \cite{DiFrancesco:1997nk} and \cite{Dixon:1989cg}.
\subsection{\texorpdfstring{$\widehat{\mathfrak{su}}(2)_k$}{su(2)-k-hat}}\label{sec:su2k}
For $\widehat{\mathfrak{su}}(2)_k$ with the components of the current $j^3$ and $j^\pm=j^1\pm i j^2$, the Kac-Moody algebra reads
\begin{align}
[j^3_m,j^3_n]=\frac{mk}{2}\delta_{m+n,0},\;[j^3_m,j^\pm_n]=\pm j^\pm_{m+n},\;[j^+_m,j^-_n]=km \delta_{m+n,0}+2j^3_{m+n},
\end{align}
with all other commutators vanishing. In this subsection we use the convention that a compact Lie group has Lie algebra $[J^a,J^b]=if^{abc}J^c$ with the structure constants $f^{abc}$ real. The zero modes $j^{3,\pm}_0$ span an $\mathfrak{su}(2)$ Lie algebra. The stress-energy tensor is obtained by the Sugawara construction, which makes all currents have weight one.
The central charge for $\widehat{\mathfrak{su}}(2)_k$ is 
\begin{align}
c=\frac{k \dim{\mathfrak{g}}}{k+C_{\mathfrak{g}}}=\frac{3k}{k+2}=3-\frac{6}{k+2},\;
\end{align}
A highest-weight state $|J\rangle$ of $\widehat{\mathfrak{su}}(2)_k$ is that of the zero-mode $\mathfrak{su}(2)$ Lie algebra, defined by
\begin{align}
j^3_0|J\rangle&=J|J\rangle,\;2J\in \mathbb{N}\cup \{0\}\\
j^{+}_{0}|J\rangle&=0,\;j^{3,\pm}_{n}|J\rangle=0,\;n>0.
\end{align}
$J$ labels the $\mathfrak{su}(2)$ representation. By acting with $j^-_0$ on $|J\rangle$, one obtains descendants of $\mathfrak{su}(2)$ spanning an irreducible $\mathfrak{su}(2)$ representation,
\begin{align}
|J_\alpha\rangle&\equiv(j^-_{0})^\alpha|J\rangle,\;j^3_0 |J_\alpha\rangle=\left( J- \alpha \right)|J_\alpha\rangle,\;\alpha=0,1,\ldots,2J.
\end{align}
For each such $\mathfrak{su}(2)$ descendant, one obtains $\widehat{\mathfrak{su}}(2)_k$ descendants by acting on it with all the nonzero negative modes:
\begin{align}
\prod_{l>0}(j_{-l}^3)^{l_i}\prod_{m>0}(j_{-m}^+)^{m_i}\prod_{n>0}(j_{-n}^-)^{n_i}|J_\alpha\rangle,
\end{align}
where the products in $l,m,n$ start from one, and $l_i,m_i,n_i$ are non-negative integers. Together, these states span an irreducible $\widehat{\mathfrak{su}}(2)_k$ representation $R$ labeled by $J$.

All in all, we have, for $n>0$, the following relations in a highest-weight representation
\begin{align}
L_0|J_\alpha\rangle&=\frac{J(J+1)}{k+2}|J_\alpha\rangle,\\
L_0j^{3,\pm}_{-n}|J_\alpha\rangle&=\left(\frac{J(J+1)}{k+2}+n\right)j^{3,\pm}_{-n}|J_\alpha\rangle,\\
j^3_0j^3_{-n}|J_\alpha\rangle&=(J-\alpha)j^3_{-n}|J_\alpha\rangle,\\
j^3_0 (j^\pm_{-n})^m|J_\alpha\rangle&=\left((J-\alpha)\pm m \right)(j^\pm_{-n})^m|J_\alpha\rangle,\\
0\leq 2J&\leq k,\;k\in\mathbb{Z}^+.
\end{align}
The last relation comes from unitarity, i.e. from demanding that no negative-norm states exists in the spectrum.

Next, define the character with a fixed chemical potential $\chi\in \mathbb{C}$ along $j^3_0$ in the representation $J$ and holomorphic in $q$:
\begin{align}
\mathcal{Z}_{SU(2)}(k,\chi,J)\equiv Tr_{J}\left[ q^{L_0-\frac{c}{24}}e^{2\pi i \chi j^3_0} \right],\;k\in\mathbb{Z}^+,\;J=0,1,\ldots k/2,
\end{align}
where the trace is the sum over the representation $J$ of $\widehat{\mathfrak{su}}(2)_k$. A simple computation yields
\begin{align}
&\quad \mathcal{Z}_{SU(2)}(k,\chi,J)\nonumber\\
&=q^{\frac{J(J+1)}{k+2}-\frac{c}{24}}\left(\sum_{\alpha=0}^{2J} e^{2\pi i \chi \left(J-\alpha\right)}\right)\sum_{l_i\in \mathbb{N}\cup \{0\}}\sum_{m_i\in \mathbb{N}\cup \{0\}}\sum_{n_i\in \mathbb{N}\cup \{0\}}q^{\sum_l l l_i+\sum_m m m_i+\sum_n n n_i}e^{2\pi i \chi (\sum_m m_i-\sum_n n_i)}\nonumber\\
&=q^{\frac{J(J+1)}{k+2}-\frac{c}{24}}\left(\sum_{\alpha=0}^{2J} e^{2\pi i \chi \left(J-\alpha\right)}\right)\prod^\infty_{n=1}\frac{1}{1-q^n}\frac{1}{1-q^n e^{+2\pi i \chi}}\frac{1}{1-q^n e^{-2\pi i \chi}},\;k\in\mathbb{Z}^+.\label{eq:su2ch2}
\end{align}
The sum in $\alpha$ is over the spin-$J$ representation of the $\mathfrak{su}(2)$ Lie algebra of the zero-mode and the infinite products from the negative modes $j_{-n}^{3,\pm}$ are the same for all $\mathfrak{su}(2)$ representations. The last line coming from geometric sums makes sense only when $\Im{\chi}\in(-\tau_2,\tau_2)$. The character can be decomposed into two parts, according to~\eqref{eq:chWL3},
\begin{align}
\mathcal{Z}_{SU(2)}(k,\chi,J)&=q^{\frac{J(J+1)}{k+2}-\frac{c}{24}}\left(\frac{e^{2\pi i \chi J}-e^{-2\pi i \chi (J+1)}}{1-e^{-2\pi i \chi}}\right)\prod^\infty_{n=1}\frac{1}{1-q^n}\frac{1}{1-q^n e^{+2\pi i \chi}}\frac{1}{1-q^n e^{-2\pi i \chi}}\nonumber\\
&=q^{\frac{J(J+1)}{k+2}-\frac{c}{24}}\left(\frac{e^{2\pi i \chi J}}{1-e^{-2\pi i \chi}}+\frac{e^{-2\pi i \chi J}}{1-e^{+2\pi i \chi}}\right)\prod^\infty_{n=1}\frac{1}{1-q^n}\frac{1}{1-q^n e^{+2\pi i \chi}}\frac{1}{1-q^n e^{-2\pi i \chi}}\nonumber\\
&\equiv \mathcal{Z}_{SU(2),+}(k,\chi,J)+\mathcal{Z}_{SU(2),-}(k,\chi,J).\label{eq:su2ch}
\end{align}
The subscripts $\pm$ in the last line correspond to the classical solutions $\bar{A}_\pm$ in the path integral,~\eqref{eq:chWL} and~\eqref{eq:chWL2}.

\subsection{\texorpdfstring{$\widehat{\mathfrak{sl}}(2,\mathbb{R})_{k'}$}{sl(2,R)-k-hat}}\label{sec:sl2k}
$SL(2,\mathbb{R})$ is a non-compact group, and we take the maximal compact subgroup $U(1)$ as the $3$-direction. The Lie algebra reads
\begin{align}
[J^a,J^b]&=i\epsilon^{abc}\eta_{cd}J^d,\;\epsilon^{123}=+1,\;\eta_{cd}=\diag{(1,1,-1)}.
\end{align}
Define $J^\pm=J^1\pm i J^2$, the Lie algebra becomes
\begin{align}
[J^3,J^\pm]&=\pm J^\pm,\;[J^+,J^-]=-2J^3.
\end{align}
The Kac-Moody algebra reads ($k'>0$)
\begin{align}
[j^a_m,j^b_n]&=i\epsilon^{abc}\eta_{cd}j^d_{m+n}+\frac{k'}{2}m\eta^{ab}\delta_{m+n,0},\;\text{i.e.}\\
[j^3_m,j^3_n]&=-\frac{1}{2}k'm\delta_{m+n,0},\\
[j^3_m,j^\pm_n]&=\pm j^\pm_{m+n},\\
[j^+_m,j^-_n]&=k'm\delta_{m+n,0}-2j^3_{m+n}.
\end{align}
The zero modes $j^{3,\pm}_0$ span an $\mathfrak{sl}(2,\mathbb{R})$ Lie algebra.

The Sugawara stress-energy tensor is again constructed such that all $j^a$ are of weight one. In particular, 
\begin{align}
L_0&=\frac{1}{k'-2}\left( \frac{1}{2}(j^+_0j^-_0+j^-_0j^+_0)-(j^3_0)^2+\sum_{m=1}^\infty (j^+_{-m}j^-_m+j^-_{-m}j^+_m-2j^3_{-m}j^3_m) \right),
\end{align}
where we used that the dual Coxeter number of $\mathfrak{sl}(2,\mathbb{R})_{k'}$ is $C_{\mathfrak{g}}=-2$.

The central charge for $\widehat{\mathfrak{sl}}(2,\mathbb{R})_{k'}$ is
\begin{align}
c'=\frac{k'\dim{\mathfrak{g}}}{k'+C_{\mathfrak{g}}}=\frac{3k'}{k'-2}=3+\frac{6}{k'-2}.
\end{align}

Since $SL(2,\mathbb{R})$ is non-compact, unitary representations of the $\mathfrak{sl}(2,\mathbb{R})$ Lie algebra are all necessarily infinite-dimensional, with the exception of the trivial representation. Each is labeled by a number $l$ and the eigenvalue of $L_0$ of the highest (or lowest) weight state,
\begin{align}
L_0=\frac{J^2}{k'-2},\;J^2\equiv \frac{1}{2}(j^+_0j^-_0+j^-_0j^+_0)-(j^3_0)^2.
\end{align}  
In a unitary irreducible representation of the Lie Group $SL(2,\mathbb{R})$ which is the double cover of $SO(2,1)$ spanned by positive-norm states, $j^3_0$ takes integer or half-integer eigenvalues. For our purpose, we only consider the trivial and discrete representations. 
\begin{itemize}
\item Trivial Representation $|0\rangle$: It has $J^2=0$ and $j^\pm_0|0\rangle=0$.
\item Discrete Series $\mathfrak{D}^+_l$, $2l\in \mathbb{Z}^+$: It starts with a \textit{lowest}-weight state $|l\rangle$, defined by
\begin{align}
j^3_0|l\rangle&=l|l\rangle,\;j^-_0|l\rangle=0,\;J^2|l\rangle=-l(l-1)|l\rangle.
\end{align}
The descendants are obtained by acting on it with $j^+_0$
\begin{equation}
\begin{aligned}
|l_\alpha\rangle&\equiv (j^+_0)^\alpha|l\rangle,\;\alpha=0,1,\ldots,\\
j^3_0|l_\alpha\rangle&=(l+\alpha)|l_\alpha\rangle,\\
J^2|l_\alpha\rangle&=-l(l-1)|l_\alpha\rangle.
\end{aligned}
\end{equation}
\item Discrete Series $\mathfrak{D}^-_{l'}$, $2l'\in \mathbb{Z}^-$: It starts with a \textit{highest}-weight state $|l'\rangle$, defined by
\begin{align}
j^3_0|l'\rangle&=l'|l'\rangle,\;j^+_0|l'\rangle=0,\;J^2|l'\rangle=-l'(l'+1)|l'\rangle.
\end{align}
The descendants are obtained by acting on it with $j^-_0$
\begin{equation}
\begin{aligned}
|l'_\alpha\rangle&\equiv (j^-_0)^\alpha|l'\rangle,\;\alpha=0,1,\ldots
\\j^3_0|l'_\alpha\rangle&=(l'-\alpha)|l'_\alpha\rangle,
\\J^2|l'_\alpha\rangle&=-l'(l'+1)|l'_\alpha\rangle.
\end{aligned}
\end{equation}
\end{itemize}

With a unitary representation of $SL(2,\mathbb{R})$, one can construct an irreducible representation of $\widehat{\mathfrak{sl}}(2,\mathbb{R})_{k'}$, spanned by the descendants
\begin{align}
\prod_{l>0}(j_{-l}^3)^{l_i}\prod_{m>0}(j_{-m}^+)^{m_i}\prod_{n>0}(j_{-n}^-)^{n_i}|l_\alpha\rangle,
\end{align}
where $l_i,m_i,n_i$ are non-negative integers. They satisfy the raising and lowering relations as in $\widehat{\mathfrak{su}}(2)_k$, but without the condition $0\leq 2J\leq k$ which came from unitarity; all representations of $\widehat{\mathfrak{su}}(2)_k$ contain negative-norm states thus they are all non-unitary. 

For each of the representations defined above, we compute the following normalized character, to which all states contribute with a weight one, regardless of their norms:
\begin{align}
\mathcal{Z}_{SL(2,\mathbb{R})}(k',\chi,l)\equiv Tr_{l}\left[ q^{L_0-\frac{c'}{24}}e^{2\pi i \chi j^3_0} \right],\;k'>0,\;2l\in\mathbb{Z},\;\chi\in\mathbb{C}.
\end{align}
For the trivial representation $l=0$,
\begin{align}
\mathcal{Z}_{SL(2,\mathbb{R})}(k',\chi,0)&=q^{-\frac{c'}{24}}\prod_{n=1}^\infty \frac{1}{1-q^n}\frac{1}{1-q^n e^{+2\pi i \chi}}\frac{1}{1-q^n e^{-2\pi i \chi}}.\label{eq:sl2ch0}
\end{align}
For $\mathfrak{D}^+_l$, $2l\in \mathbb{Z}^+$, 
\begin{align}
&\quad\mathcal{Z}_{SL(2,\mathbb{R})}(k',\chi,l\in\mathbb{Z}^+/2)\nonumber\\
&=q^{\frac{-l(l-1)}{k'-2}-\frac{c'}{24}}\left( \frac{e^{2\pi i \chi l}}{1-e^{+2\pi i \chi}}\right)\prod_{n=1}^\infty \frac{1}{1-q^n}\frac{1}{1-q^n e^{+2\pi i \chi}}\frac{1}{1-q^n e^{-2\pi i \chi}},\;\Im{\chi}\in(0,\tau_2),\label{eq:sl2ch1}
\end{align}
where the second term comes from summing over the infinite-dimensional $SL(2,\mathbb{R})$ representation, which of course only makes sense when $\Im{\chi}>0$.
For $\mathfrak{D}^-_{l'}$, $2l'\in \mathbb{Z}^-$, 
\begin{align}
&\quad\mathcal{Z}_{SL(2,\mathbb{R})}(k',\chi,l'\in\mathbb{Z}^-/2)\nonumber\\
&=q^{\frac{-l'(l'+1)}{k'-2}-\frac{c'}{24}}\left( \frac{e^{2\pi i \chi l'}}{1-e^{-2\pi i \chi}}\right)\prod_{n=1}^\infty \frac{1}{1-q^n}\frac{1}{1-q^n e^{+2\pi i \chi}}\frac{1}{1-q^n e^{-2\pi i \chi}},\;\Im{\chi}\in(-\tau_2,0),\label{eq:sl2ch2}
\end{align}
where the second term only makes sense when $\Im{\chi}<0$.\\

The $\widehat{\mathfrak{sl}}(2,\mathbb{R})_{k'}$ characters~\eqref{eq:sl2ch0},~\eqref{eq:sl2ch1},~\eqref{eq:sl2ch2} in fact can be obtained respectively by a formal replacement of the level $k$ and spin $J$ in the $\mathcal{Z}_{SU(2),\pm}(k,\chi,J)$ in the $\widehat{\mathfrak{su}}(2)_k$ character~\eqref{eq:su2ch}:
\begin{equation}
\begin{aligned}
\mathcal{Z}_{SL(2,\mathbb{R})}(k',\chi,l=0)&=\left.\mathcal{Z}_{SU(2)}(k,\chi,J=0)\right|_{\substack{\Im{\chi}\in(-\tau_2,\tau_2)\\k=-k'}}\\
\mathcal{Z}_{SL(2,\mathbb{R})}(k',\chi,l\in\mathbb{Z}^+/2)&=\left.\mathcal{Z}_{SU(2),-}(k,\chi,J)\right|_{\substack{\Im{\chi}\in(0,\tau_2)\\J=-l,\;k=-k'}}\\
\mathcal{Z}_{SL(2,\mathbb{R})}(k',\chi,l'\in\mathbb{Z}^-/2)&=\left.\mathcal{Z}_{SU(2),+}(k,\chi,J)\right|_{\substack{\Im{\chi}\in(-\tau_2,0)\\J=+l',\;k=-k'}}.
\end{aligned}\label{eq:sl2ch3}
\end{equation}

\section{The Heat Kernel Method}\label{sec:heat}
We review the main results from \cite{Giombi:2008vd}, that we used to determine the local contributions to functional determinants and generalized to include a flat background connection $\bar{A}=\bar{A}_tdt$ trivial holonomy ($\pm\mathds{1}$) around a contractible cycle in the case $q=\bar{q}$.

In the upper-half space representation of the hyperbolic space $\mathbb{H}^3$, the Poincar\'e metric reads
\begin{equation}
ds^2=\frac{dy^2+dzd\bar{z}}{y^2},\;y>0\label{eq:Poinmet}
\end{equation}
where the coordinates in~\eqref{eq:Poinmet} are related to the global coordinates by 
\begin{align}
z=e^{iw}\cos{\theta},\;\bar{z}=e^{-i\bar{w}}&\cos{\theta},\;y=e^{\frac{i(w-\bar{w})}{2}}\sin{\theta},\;\sin{\theta}=\frac{1}{\sqrt{1+r^2}},\;\tan^2{\theta}=\frac{y^2}{z\bar{z}}.
\end{align}

Thermal $AdS_3$ is a quotient space $\mathbb{H}^3/\Gamma$ of the hyperbolic space $\mathbb{H}^3$ by a classical Schottky group $\Gamma$ generated by a single loxodromic element $\gamma$ under the identification $(z,\bar{z},y)\sim(qz,\bar{q}\bar{z},\sqrt{q\bar{q}}y)$, where $q=e^{2\pi i \tau}$, $|q|<1$.

\subsection{Background Flat Connection and The Twisted Laplace Operators}
The flat connection in thermal $AdS_3$ we used to compute the vacuum $\widehat{\mathfrak{su}}(2)_k$ character is $\bar{A}=\frac{\chi}{\tau_2}T^3dt$. Written in the Poincar\'e metric, it is
\begin{align}
\bar{A}_z=-\frac{1}{2}\frac{\bar{z}}{y^2+z\bar{z}}{\bar{A}}_t,\;{\bar{A}}_{\bar{z}}=-\frac{1}{z}\frac{z}{y^2+z\bar{z}}{\bar{A}}_t,\;{\bar{A}}_y=-\frac{y}{y^2+z\bar{z}}{\bar{A}}_t.
\end{align}
Note that $\bar{A}$ is divergenceless: $\nabla^\mu {\bar{A}}_\mu=0$. Because ${\bar{A}}$ is flat, it can be written as $\bar{A}=MdM^{-1}$, where
\begin{align}
M\equiv\begin{pmatrix}
(y^2+z\bar{z})^{-\frac{i\chi}{4\tau_2}} &0\\
0& (y^2+z\bar{z})^{\frac{i\chi}{4\tau_2}}
\end{pmatrix}\equiv \begin{pmatrix}
f(y,z,\bar{z})&0\\0&f^{-1}(y,z,\bar{z})
\end{pmatrix}.
\end{align}
Given an adjoint-valued field $\psi=\sum_a\psi^a T^a$, we have $D_\mu \psi^3T^3=\nabla_\mu \psi^3T^3$ and $D_\mu\psi^{\pm}T^\pm=(\nabla_\mu\psi^\pm \pm i\bar{A}^3_\mu \psi^\pm)T^\pm$. By defining 
\begin{align}
\psi^\pm&=(f(y,z,\bar{z}))^{\mp 2}\tilde{\psi}^\pm=(y^2+z\bar{z})^{\pm\frac{i\chi}{2\tau_2}}\tilde{\psi}^\pm,
\end{align}
we have
\begin{align}
D_\mu\psi^{\pm}T^\pm=(f(y,z,\bar{z}))^{\mp 2}\nabla_\mu \tilde{\psi}^\pm T^\pm.
\end{align}
Given that the original field $\psi^\pm$ is doubly-periodic under ($z\rightarrow e^{2\pi i}z$, $\bar{z}\rightarrow e^{-2\pi i}\bar{z}$), and $x\rightarrow \gamma x$, and given that $f(\gamma x)=(q\bar{q})^{-\frac{i\chi}{4\tau_2}}f(x)=e^{\frac{2\pi i\chi}{2}}f(x)$, the twisted periodicity of the tilde-fields is
\begin{align}
\tilde{\psi}^\pm (\gamma x)=e^{\pm 2\pi i\chi} \tilde{\psi}^\pm ( x).
\end{align}

\subsection{Twisted Heat Kernels in the Fundamental Domain}
\subsubsection*{Scalar}
The untwisted heat kernel $K^{(3)}_{\mathbb{H}^3/\Gamma}$ in the fundamental domain representing thermal $AdS_3$ is the sum over images of the heat kernel $K_{\mathbb{H}^3}$ in $\mathbb{H}^3$ (Eq. 4.4 in \cite{Giombi:2008vd}), 
\begin{align}
K^{(3)}_{\mathbb{H}^3/\Gamma}(t,x,x')&=\sum_{n\in \mathbb{Z}}K_{\mathbb{H}^3}(t,r(x,\gamma^n x')),\;K_{\mathbb{H}^3}(t,r(x,x'))=\frac{e^{-\frac{r^2}{4t}-t}}{(4\pi t)^{\frac{3}{2}}}\frac{r}{\sinh{r}}.\label{eq:Kscim}
\end{align}

Next, recall that the action of the ghosts $c$ and $\bar{c}$ in the gauge-fixed Chern-Simons action~\eqref{eq:Igf} is
\begin{align}
-\int_M d^3x \sqrt{g}\; {\rm Tr}[\bar{c}\Delta_0 c]=\int_M d^3x \sqrt{g}\; {\rm Tr}[\bar{c}D^\mu D_\mu c]=-\frac{1}{2}\int_Md^3x \sqrt{g}\; \left(\bar{c}^3 \nabla^2 c^3+\bar{c}^+ D^2 c^-+\bar{c}^- D^2 c^+\right).
\end{align}
Define the twisted tilde-fields $c^\pm=(f(y,z,\bar{z}))^{\mp 2}\tilde{c}^\pm,\;\bar{c}^\pm=(f(y,z,\bar{z}))^{\mp 2}\tilde{\bar{c}}^\pm$. The action becomes
\begin{align}
-\int_M d^3x \sqrt{g}\; {\rm Tr}[\bar{c}\Delta_0 c]=-\frac{1}{2}\int_M d^3x\sqrt{g}\; \left(\bar{c}^3 \nabla^2 c^3+\tilde{\bar{c}}^+ \nabla^2 \tilde{c}^-+\tilde{\bar{c}}^- \nabla^2 \tilde{c}^+\right).
\end{align}
We now need the heat kernels of the scalar Laplacian in the fundamental domain $\mathbb{H}^3/\Gamma$. They are obtained from the eigenfunctions with twisted periodicities:
\begin{align}
K^{(\pm)}_{\mathbb{H}^3/\Gamma}(t,x,x')=\sum_n e^{-\lambda_n t}\tilde{\varphi}_n^{\pm}(x)\tilde{\varphi}^{\mp}_n(x'),
\end{align}
which satisfy the heat equation with twisted periodicities,
\begin{empheq}[box=\empheqlbrace]{align}
(\partial_t-\nabla^2_x)K^{(\pm)}_{\mathbb{H}^3/\Gamma}(t,x,x')&=0,\\
K^{(\pm)}_{\mathbb{H}^3/\Gamma}(0,x,x')&=\delta(x,x'),\\
K^{(\pm)}_{\mathbb{H}^3/\Gamma}(t,\gamma x,x')&=e^{\pm 2\pi i \chi} K^{(\pm)}_{\mathbb{H}^3/\Gamma}(t, x,x'),\\
K^{(\pm)}_{\mathbb{H}^3/\Gamma}(t, x,\gamma x')&=e^{\mp 2\pi i \chi} K^{(\pm)}_{\mathbb{H}^3/\Gamma}(t, x,x').
\end{empheq}
The heat kernel is easily found by the method of images to be
\begin{align}
K^{(\pm)}_{\mathbb{H}^3/\Gamma}(t,x,x')&=\sum_{n\in \mathbb{Z}}e^{\pm 2\pi in \chi}K_{\mathbb{H}^3}(t,r(x,\gamma^n x')).
\end{align}

\subsubsection*{Vector}
For the twisted vector Laplacian of the vector fluctuation, 
\begin{align}
-\int_M \sqrt{g}\; {\rm Tr}[A_\mu g^{\mu \nu} \Delta_1 A_\nu],
\end{align}
we follow the same procedure. The untwisted heat kernel in the fundamental domain is
\begin{align}
K^{(3)}_{\mathbb{H}^3/\Gamma}{}_{\mu\nu'}(t,x,x')&=\sum_{n\in \mathbb{Z}} \frac{\partial(\gamma^n x)^{\rho'}}{\partial x^{\nu'}} K_{\mathbb{H}^3}{}_{\mu\rho'}(t,r(x,\gamma^n x')),
\end{align}
such that $K^{(3)}_{\mathbb{H}^3/\Gamma}{}_{\mu\nu'}(t,x,x') dx^\mu d{x'}^{\nu}$ is periodic under $x\rightarrow \gamma x$ and $x'\rightarrow \gamma x'$. The twisted kernels are then
\begin{align}
K^{(\pm)}_{\mathbb{H}^3/\Gamma}{}_{\mu\nu'}(t,x,x')&=\sum_{n\in \mathbb{Z}} e^{\pm 2\pi i n\chi}\frac{\partial(\gamma^n x)^{\rho'}}{\partial x^{\nu'}} K_{\mathbb{H}^3}{}_{\mu\rho'}(t,r(x,\gamma^n x')).
\end{align}

\subsection{Functional Determinants}
\subsubsection*{Scalar}
For the untwisted Laplacian, integrating over the fundamental domain and $t$ gives the logarithm of the determinant, which is made of a volume factor from the zero mode plus a sum over $n$ \textit{positive} integers (eq. 4.9 of \cite{Giombi:2008vd})
\begin{align}
\ln{\det{\Delta^{(3)}_0}}=-\left.\left(\frac{(1+m^2)^{3/2}}{6\pi}\right)\right|_{m=0}\text{Vol}(\mathbb{H}^3/\Gamma)-2\sum_{n=1}^\infty \frac{q^n \bar{q}^n}{n(1-q^n)(1-\bar{q}^n)}.
\end{align}
The volume is regularized by holographic renormalization (see \cite{Karch:2005ms}) as $\text{Vol}(\mathbb{H}^3/\Gamma)_{ren}=-\pi^2 \tau_2=\frac{i\pi^2}{2}(\tau-\bar{\tau})$. The determinant is then
\begin{align}
\det{\Delta^{(3)}_0}&=q^{-\frac{1}{24}}\bar{q}^{-\frac{1}{24}}\prod_{l,l'=0}^\infty \left(1-q^{l+1}\bar{q}^{l'+1}\right)^2.\label{eq:hksca}
\end{align}

Meanwhile, the computation of the determinants of twisted Laplacians with the heat kernels $K_{\mathbb{H}^3/\Gamma}^{(\pm)}(t,x,x')$ gives
\begin{align}
\ln{\det{\Delta^{(\pm)}_0}}&=-\frac{1}{6\pi}\text{Vol}(\mathbb{H}^3/\Gamma)-\sum_{n=1}^\infty \frac{q^n \bar{q}^n \left(e^{ 2\pi in \chi}+e^{-2\pi in \chi} \right)}{n(1-q^n)(1-\bar{q}^n)},\\\;\det{\Delta^{(\pm)}_0}&=q^{-\frac{1}{24}}\bar{q}^{-\frac{1}{24}}\prod_{l,l'=0}^\infty \left(1-e^{ 2\pi i \chi}q^{l+1}\bar{q}^{l'+1}\right)\left(1-e^{ -2\pi i \chi}q^{l+1}\bar{q}^{l'+1}\right).\label{eq:hkscatw}
\end{align}
Hence the local term, i.e. the volume factor, is the same for the untwisted determinants.

\subsubsection*{Vector}
For the transverse direction of the untwisted vector Laplacian, we have (Eq. 4.18 of \cite{Giombi:2008vd})
\begin{align}
\ln{\det\nolimits^{(T)}{\Delta_1^{(3)}}}&=-\left.\left(\frac{m^3-3m}{3\pi}\right)\right|_{m=0}\text{Vol}(\mathbb{H}^3/\Gamma)-2\sum_{n=1}\frac{q^n+\bar{q}^n}{n|1-q^n|^2},\;\\
\det\nolimits^{(T)}{\Delta_1^{(3)}}&=\prod_{l,l'=0}^\infty \left(1-q^{l+1}\bar{q}^{l'}\right)^2\left(1-q^{l}\bar{q}^{l'+1}\right)^2.\label{eq:hkvec}
\end{align}
The volume factor vanishes in the massless limit. This determinant in the massless limit is the product of $\det\nolimits^{(T)}{(L_--i\varepsilon)}$ and $\det\nolimits^{(T)}{(L_-+i\varepsilon)}$, and is equal to the square of $\det\nolimits^{(T)}{(L_--i\varepsilon)}$~\eqref{eq:transverse0} only when $q=\bar{q}$; see \hyperref[sec:4.3]{Section \ref*{sec:4.3}}. For the twisted determinants we have (cf. see Eq. 4.10 of \cite{Giombi:2008vd})
\begin{align}
\ln{\det\nolimits^{(T)}{\Delta_1^{(\pm)}}}&=-\sum_{n=1}^\infty \frac{(q^n+\bar{q}^n) \left(e^{ 2\pi in \chi}+e^{-2\pi in \chi} \right)}{n(1-q^n)(1-\bar{q}^n)}\\
\det\nolimits^{(T)}{\Delta_1^{(\pm)}}&=\prod_{l,l'=0}^\infty \left(1-e^{ 2\pi i \chi}q^{l+1}\bar{q}^{l'}\right)\left(1-e^{ -2\pi i \chi}q^{l+1}\bar{q}^{l'}\right)\left(1-e^{ 2\pi i \chi}q^{l}\bar{q}^{l'+1}\right)\left(1-e^{ -2\pi i \chi}q^{l}\bar{q}^{l'+1}\right).\label{eq:hkvec2}
\end{align}
The result agrees with the resonance pole calculation~\eqref{eq:twist1q22} when $q=\bar{q}$, up to an undetermined local phase.

\section{The Ratio of Determinants with a Wilson Loop}\label{sec:WLapp}
Here we compute the ratio of determinants~\eqref{eq:chWL0} in $SU(2)$ Chern-Simons theory in the presence of a Wilson loop. Working in the regime $2J/k=O(k^0)>0$, we take the classical solution $\bar{A}_+$ in~\eqref{eq:wilsonclass1} satisfying the boundary condition $A_{\bar{w}}|_{\partial M}=i\chi/2\tau_2 T^3$,
\begin{align}
\bar{A}_+&=\frac{\chi'}{\tau_2}T^3dt+\theta T^3d\phi,\;\theta\equiv 2J/k>0,
\end{align}
where $\chi'\equiv \chi+i\tau_2 \theta$. We take $\chi'\in \mathbb{R}$ throughout the computation and substitute $\chi'=\chi+i\tau_2(2J/k)$ in the end.

Following \hyperref[sec:twisted]{Section \ref*{sec:twisted}} we trade the flat background for fields with twisted periodicity conditions. Given an adjoint field $\psi=\psi^a T^a$, we define
\begin{align}
\psi^{(\pm)}&\equiv e^{\mp i(\frac{\chi'}{\tau_2}t+\theta \phi)}\tilde{\psi}^{(\pm)}\text{ such that } D_\mu\psi^{(\pm)}=e^{\mp i(\frac{\chi'}{\tau_2}t+\theta \phi)}\nabla_\mu \tilde{\psi}^{(\pm)}.
\end{align}
Next we expand the twisted field $\tilde{\psi}^{(\pm)}$ in modes
\begin{align}
\tilde{\psi}^{(\pm)}=R(x)e^{-i\omega t+ik\phi}e^{\pm i(\frac{\chi'}{\tau_2}t+\theta \phi)}=R(x)e^{-i(\omega\mp \frac{\chi'}{\tau_2})t+i(k\pm\theta)\phi}\equiv R(x)e^{-i\omega' t+ik'\phi},
\end{align}
where for $\tilde{\psi}^{(\pm)}$ respectively 
\begin{align}
\omega'&=\omega\mp\frac{\chi'}{\tau_2}=\frac{-(n\pm\chi')+k\tau_1}{\tau_2},\;k'=k\pm \theta.\label{eq:Ctwist}
\end{align}
Since both $\psi^{(+)}$ and $\psi^{(-)}$ are present, the sign of $\theta$ or $J$ is not important in this calculation. We consider separately the cases $0<2J/l<1$ and $2J/k=1$, respectively within the unitarity bound and saturating it.

\subsection{Case 1: \texorpdfstring{$0<2J/k<1$}{0<2J/k<1}}
\subsubsection*{Scalar}
The scattering pole solutions ($m^2=s(s-2)$) for general $\theta\in\mathbb{R}$, regular at the origin, read
\begin{align}
R(x)=(1-x)^\frac{s}{2} x^{\pm\frac{k'}{2}}F(\frac{s\pm k'+i\omega'}{2},\frac{s\pm k'-i\omega'}{2};1\pm k';x),\label{eq:Csc1}
\end{align}
where the plus and minus signs are for $k'\geq 0$ and $k'<0$ respectively.

For $\psi^{(+)}$ where $k'=k+\theta$ and $\omega'=\frac{-(n+\chi')+k\tau_1}{\tau_2}$, the determinant $\det\nolimits^{(+)}{(\Delta_0+s(s-2))}$ contains the following products from the scattering poles
\begin{itemize}
\item $k\geq -\theta$:
\begin{align}
&\quad \left[s+2p+(k+\theta)-i\left(\frac{-(n+\chi')+k\tau_1}{\tau_2}\right)\right]\left[s+2p+(k+\theta)+i\left(\frac{-(n+\chi')+k\tau_1}{\tau_2}\right)\right]\nonumber\\
&\propto (i(n+\chi')+[\tau_2(s+2p+\theta)-ik\tau])(-i(n+\chi')+[\tau_2(s+2p+\theta)+ik\bar{\tau}])
\end{align}
\item $k< -\theta$:
\begin{align}
&\quad \left[s+2p-(k+\theta)-i\left(\frac{-(n+\chi')+k\tau_1}{\tau_2}\right)\right]\left[s+2p-(k+\theta)+i\left(\frac{-(n+\chi')+k\tau_1}{\tau_2}\right)\right]\nonumber\\
&\propto (i(n+\chi')+[\tau_2(s+2p-\theta)-ik\bar{\tau}])(-i(n+\chi')+[\tau_2(s+2p-\theta)+ik\tau]).
\end{align}
\end{itemize}
For $\psi^{(-)}$ where $k'=k-\theta$ and $\omega'=\frac{-(n-\chi')+k\tau_1}{\tau_2}$, the determinant $\det\nolimits^{(-)}{(\Delta_0+s(s-2))}$ contains the following products from the scattering poles
\begin{itemize}
\item $k\geq +\theta$:
\begin{align}
&\quad \left[s+2p+(k-\theta)-i\left(\frac{-(n-\chi')+k\tau_1}{\tau_2}\right)\right]\left[s+2p+(k-\theta)+i\left(\frac{-(n-\chi')+k\tau_1}{\tau_2}\right)\right]\nonumber\\
&\propto (i(n-\chi')+[\tau_2(s+2p-\theta)-ik\tau])(-i(n-\chi')+[\tau_2(s+2p-\theta)+ik\bar{\tau}])
\end{align}
\item $k< +\theta$:
\begin{align}
&\quad \left[s+2p-(k-\theta)-i\left(\frac{-(n-\chi')+k\tau_1}{\tau_2}\right)\right]\left[s+2p-(k-\theta)+i\left(\frac{-(n-\chi')+k\tau_1}{\tau_2}\right)\right]\nonumber\\
&\propto (i(n-\chi')+[\tau_2(s+2p+\theta)-ik\bar{\tau}])(-i(n-\chi')+[\tau_2(s+2p+\theta)+ik\tau]).\label{eq:Csc5}
\end{align}
\end{itemize}
Then $\left(\det\nolimits^{(+)}{(\Delta_0+s(s-2))}\det\nolimits^{(-)}{(\Delta_0+s(s-2))}\right)=\prod_{k=0}^\infty f_k(s)$ for $0<\theta<1$ has the following contributions to $f_k(s)$ from the $k$-modes
\begin{itemize}
\item $k=0$:
\begin{align}
f_{k=0}(s)&\propto\prod_{\substack{p\geq 0\\n\in\mathbb{Z}}}(i(n+\chi')+\tau_2(s+2p+\theta))(-i(n+\chi')+\tau_2(s+2p+\theta))\nonumber\\
&\quad \times(i(n-\chi')+\tau_2(s+2p+\theta))(-i(n-\chi')+\tau_2(s+2p+\theta))\\
&\propto \prod_{p\geq 0}\left(1-(q\bar{q})^{\frac{s}{2}+p+\frac{\theta}{2}}e^{-2\pi i \chi'} \right)^2\left(1-(q\bar{q})^{\frac{s}{2}+p+\frac{\theta}{2}}e^{+2\pi i \chi'} \right)^2
\end{align}
\item $k\geq 1$:
\begin{align}
f_{k}(s)&\propto\prod_{\substack{p\geq 0\\n\in\mathbb{Z}}}(+i(n+\chi')+\tau_2(s+2p+\theta)-ik\tau)(-i(n+\chi')+\tau_2(s+2p+\theta)+ik\bar{\tau})\nonumber\\
&\quad \times(-i(n+\chi')+\tau_2(s+2p-\theta)-ik\tau)(+i(n+\chi')+\tau_2(s+2p-\theta)+ik\bar{\tau})\nonumber\\
&\quad \times(+i(n-\chi')+\tau_2(s+2p-\theta)-ik\tau)(-i(n-\chi')+\tau_2(s+2p-\theta)+ik\bar{\tau})\nonumber\\
&\quad \times(-i(n-\chi')+\tau_2(s+2p+\theta)-ik\tau)(+i(n-\chi')+\tau_2(s+2p+\theta)+ik\bar{\tau})\\
&\propto \prod_{p\geq 0}\left(1-(q\bar{q})^{\frac{s}{2}+p+\frac{\theta}{2}}e^{-2\pi i \chi'} q^k\right)^2\left(1-(q\bar{q})^{\frac{s}{2}+p-\frac{\theta}{2}}e^{+2\pi i \chi'} q^k\right)^2\nonumber\\
&\quad \times\left(1-(q\bar{q})^{\frac{s}{2}+p+\frac{\theta}{2}}e^{+2\pi i \chi'} \bar{q}^k\right)^2\left(1-(q\bar{q})^{\frac{s}{2}+p-\frac{\theta}{2}}e^{-2\pi i \chi'} \bar{q}^k\right)^2.
\end{align}
\end{itemize}
\subsubsection*{Transverse Vector}
The transverse scattering poles~\eqref{eq:transversepoles} of $(L_--im)$ for general $\theta\in\mathbb{R}$ for the twisted fields expanded in $k'$ and $\omega'$, are
\begin{subequations}
\begin{align}
k'= 0:&\begin{cases}
\frac{2+m_\star+k'+i\omega'}{2}&=-p,\;\text{or}\\
\frac{2+m_\star+k'-i\omega'}{2}&=-p
\end{cases}\label{eq:transversepolesb2}\\
k'>0:&\begin{cases}
\frac{2+m_\star+k'+i\omega'}{2}&=-p,\;\text{or}\\
\frac{m_\star+k'-i\omega'}{2}&=-p\qquad\quad,\quad\;p=0,1,\ldots.
\end{cases}\label{eq:transversepolesb}\\
k'<0:&\begin{cases}
\frac{m_\star-k'+i\omega'}{2}&=-p,\;\text{or}\\
\frac{2+m_\star-k'-i\omega'}{2}&=-p
\end{cases}\label{eq:transversepolesb3}
\end{align}
\end{subequations}

For the case $0<\theta<1$, within the unitarity bound, the shifted $k\pm\theta$ are non-integers so $k'\neq 0$ and we only need~\eqref{eq:transversepolesb} and~\eqref{eq:transversepolesb3}.

For $k'=k+\theta$ and $\omega'=\frac{-(n+\chi')+k\tau_1}{\tau_2}$, the determinant $\det\nolimits^{(T)(+)}{(L_--im)}$ contains the following products from the scattering poles
\begin{itemize}
\item $k> -\theta$:
\begin{align}
&\quad \left[m+2p+2+(k+\theta)+i\left(\frac{-(n+\chi')+k\tau_1}{\tau_2}\right)\right]\left[m+2p+(k+\theta)-i\left(\frac{-(n+\chi')+k\tau_1}{\tau_2}\right)\right]\nonumber\\
&\propto (-i(n+\chi')+[\tau_2(m+2p+2+\theta)+ik\bar{\tau}])(i(n+\chi')+[\tau_2(m+2p+\theta)-ik\tau])
\end{align}
\item $k< -\theta$:
\begin{align}
&\quad \left[m+2p-(k+\theta)+i\left(\frac{-(n+\chi')+k\tau_1}{\tau_2}\right)\right]\left[m+2p+2-(k+\theta)-i\left(\frac{-(n+\chi')+k\tau_1}{\tau_2}\right)\right]\nonumber\\
&\propto (-i(n+\chi')+[\tau_2(m+2p-\theta)+ik\tau])(i(n+\chi')+[\tau_2(m+2p+2-\theta)-ik\bar{\tau}]).
\end{align}
\end{itemize}

For $k'=k-\theta$ and $\omega'=\frac{-(n-\chi')+k\tau_1}{\tau_2}$, the determinant $\det\nolimits^{(T)(-)}{(L_--im)}$ contains the following products from the scattering poles
\begin{itemize}
\item $k> +\theta$:
\begin{align}
&\quad \left[m+2p+2+(k-\theta)+i\left(\frac{-(n-\chi')+k\tau_1}{\tau_2}\right)\right]\left[m+2p+(k-\theta)-i\left(\frac{-(n-\chi')+k\tau_1}{\tau_2}\right)\right]\nonumber\\
&\propto (-i(n-\chi')+[\tau_2(m+2p+2-\theta)+ik\bar{\tau}])(i(n-\chi')+[\tau_2(m+2p-\theta)-ik\tau])
\end{align}
\item $k< +\theta$:
\begin{align}
&\quad \left[m+2p-(k-\theta)+i\left(\frac{-(n-\chi')+k\tau_1}{\tau_2}\right)\right]\left[m+2p+2-(k-\theta)-i\left(\frac{-(n-\chi')+k\tau_1}{\tau_2}\right)\right]\nonumber\\
&\propto (-i(n-\chi')+[\tau_2(m+2p+\theta)+ik\tau])(i(n-\chi')+[\tau_2(m+2p+2+\theta)-ik\bar{\tau}]).
\end{align}
\end{itemize}
Then $\left(\det\nolimits^{(T)(+)}{(L_--im)}\det\nolimits^{(T)(-)}{(L_--im)}\right)=\prod_{k=0}^\infty g_k(m)$ has the following contributions to $g_k(m)$ from the $k$-modes
\begin{itemize}
\item $k=0$:
\begin{align}
g_{k=0}(m)&\propto \prod_{\substack{p\geq 0\\n\in\mathbb{Z}}}(-i(n+\chi')+[\tau_2(m+2p+2+\theta)])(i(n+\chi')+[\tau_2(m+2p+\theta)])\nonumber\\
&\quad\times(-i(n-\chi')+[\tau_2(m+2p+\theta)])(i(n-\chi')+[\tau_2(m+2p+2+\theta)])\nonumber\\
&\propto \prod_{p\geq 0}\left(1-(q\bar{q})^{\frac{m}{2}+p+1+\frac{\theta}{2}}e^{+2\pi i \chi'} \right)^2\left(1-(q\bar{q})^{\frac{m}{2}+p+\frac{\theta}{2}}e^{-2\pi i \chi'} \right)^2
\end{align}
\item $k\geq 1$:
\begin{align}
g_{k}(m)&\propto \prod_{\substack{p\geq 0\\n\in\mathbb{Z}}}(+i(n+\chi')+[\tau_2(m+2p+\theta)-ik\tau])(-i(n+\chi')+[\tau_2(m+2p+2+\theta)+ik\bar{\tau}])\nonumber\\
&\quad\times(-i(n+\chi')+[\tau_2(m+2p-\theta)-ik\tau])(+i(n+\chi')+[\tau_2(m+2p+2-\theta)+ik\bar{\tau}])\nonumber\\
&\quad \times(+i(n-\chi')+[\tau_2(m+2p-\theta)-ik\tau])(-i(n-\chi')+[\tau_2(m+2p+2-\theta)+ik\bar{\tau}])\nonumber\\
&\quad \times(-i(n-\chi')+[\tau_2(m+2p+\theta)-ik\tau])(+i(n-\chi')+[\tau_2(m+2p+2+\theta)+ik\bar{\tau}])\nonumber\\
&\propto \prod_{p\geq 0}\left(1-(q\bar{q})^{\frac{m}{2}+p+\frac{\theta}{2}}e^{-2\pi i \chi'} q^k\right)^2\left(1-(q\bar{q})^{\frac{m}{2}+p-\frac{\theta}{2}}e^{+2\pi i \chi'} q^k\right)^2\nonumber\\
&\quad \times\left(1-(q\bar{q})^{\frac{m}{2}+p+1+\frac{\theta}{2}}e^{+2\pi i \chi'} \bar{q}^k\right)^2\left(1-(q\bar{q})^{\frac{m}{2}+p+1-\frac{\theta}{2}}e^{-2\pi i \chi'} \bar{q}^k\right)^2.
\end{align}
\end{itemize}

\subsubsection*{The Ratio of Determinants}

After combining the determinants and plugging in the values $s=2$ and $m=0$, the $\bar{q}$ terms again all cancel exactly,
\begin{align}
&\quad\left.\left(\frac{\det\nolimits^{\frac{1}{2}}{\Delta_0}}{\sqrt{\det\nolimits^{(T)}{L_-}}} \right)^{(3)}\left(\left(\frac{\det\nolimits^{\frac{1}{2}}{\Delta_0}}{\sqrt{\det\nolimits^{(T)}{L_-}}} \right)^{(+)}\left(\frac{\det\nolimits^{\frac{1}{2}}{\Delta_0}}{\sqrt{\det\nolimits^{(T)}{L_-}}} \right)^{(-)}\right)\right|_{0<2J/k<1}\nonumber\\
&\propto \prod_{n=1}^\infty\frac{1}{1-q^n} \left(\frac{1}{1- (q\bar{q})^{\frac{\theta}{2}}e^{-2\pi i \chi'}}\prod_{k=1}^\infty\frac{1}{1- (q\bar{q})^{\frac{\theta}{2}}e^{-2\pi i \chi'}q^k}\frac{1}{1- (q\bar{q})^{-\frac{\theta}{2}}e^{+2\pi i \chi'}q^k}\right)\nonumber\\
&=\prod_{n=1}^\infty\frac{1}{1-q^n} \left(\prod_{k=0}^\infty\frac{1}{1- (q\bar{q})^{\frac{\theta}{2}}e^{-2\pi i \chi'}q^k}\prod_{k=1}^\infty\frac{1}{1- (q\bar{q})^{-\frac{\theta}{2}}e^{+2\pi i \chi'}q^k}\right)\nonumber\\
&=\prod_{n=1}^\infty\frac{1}{1-q^n} \prod_{k=0}^\infty\frac{1}{1-e^{-2\pi i \chi}q^k}\prod_{k=1}^\infty\frac{1}{1- e^{+2\pi i \chi}q^k}.\label{eq:C1rd}
\end{align}
In the last line we have substituted $\chi'=\chi+i\tau_2 \theta$ so $\exp{(2\pi i \chi')}=(q\bar{q})^{+\theta/2}\exp{(2\pi i \chi)}$. This recovers the contribution to $Z_{SU(2),+}$ by~\eqref{eq:chWL0}, in particular with a shift in the middle product. Instead of fixing $A_{\bar{w}}|_{\partial M}$ as boundary condition, if we had fixed $A_w|_{\partial M}\propto \chi T^3$, we would have ended up having $\chi'=\chi-i\tau_2 \theta$ and a ratio of determinants that mixes $q$ and $\bar{q}$ in the infinite products, thus not a sensible character.

For the other solution $\bar{A}_-$, notice that $\bar{A}$ and $-\bar{A}$ give the same ratio of determinants since both the $T^{(\pm)}$-components are included. Hence for $\bar{A}_-$,
\begin{align}
\bar{A}_-=\left(\frac{\chi}{\tau_2}-i\theta\right)T^3dt-\theta T^3d\phi,\;0<\theta<1,
\end{align}
the ratio of determinants~\eqref{eq:chWL2} contributing to $Z_{SU(2),-}$ is

\begin{align}
\left.\frac{\det\nolimits^{\frac{1}{2}}{\Delta_0}}{\sqrt{\det\nolimits^{(T)}{L_-}}}\right|_{\bar{A}_-,\;0<2J/k<1}&=\prod_{n=1}^\infty\frac{1}{1-q^n} \prod_{k=1}^\infty\frac{1}{1-e^{-2\pi i \chi}q^k}\prod_{k=0}^\infty\frac{1}{1- e^{+2\pi i \chi}q^k}.\label{eq:C1rdneg}
\end{align}

The shift appears in the last product.

\subsection{Case 2: \texorpdfstring{$2J/k=1$}{2J/k=1}, At The Unitarity Bound}
In the case $\theta=2J/k=1$ that saturates the unitarity bound, only the $k=1$ modes $f_{k=1}(s)$ and $g_{k=1}(m)$ that contribute to the determinants are different from the previous case (with $\theta=1$). For the scalar determinant, we have instead
\begin{align}
f_{k=1}(s)&\propto\prod_{\substack{p\geq 0\\n\in\mathbb{Z}}}(+i(n+\chi')+\tau_2(s+2p+1)-i\tau)(-i(n+\chi')+\tau_2(s+2p+1)+i\bar{\tau})\nonumber\\
&\quad \times(+i(n+\chi')+\tau_2(s+2p+1)+i\tau)(-i(n+\chi')+\tau_2(s+2p+1)-i\bar{\tau})\nonumber\\
&\quad \times(+i(n-\chi')+\tau_2(s+2p-1)-i\tau)(-i(n-\chi')+\tau_2(s+2p-1)+i\bar{\tau})\nonumber\\
&\quad \times(-i(n-\chi')+\tau_2(s+2p+1)-i\tau)(+i(n-\chi')+\tau_2(s+2p+1)+i\bar{\tau}),
\end{align}
while for $\det\nolimits^{(T)(\pm)}{(L_--im)}$, using~\eqref{eq:transversepolesb2},~\eqref{eq:transversepolesb} and~\eqref{eq:transversepolesb3} we have
\begin{align}
g_{k=1}(m)&\propto \prod_{\substack{p\geq 0\\n\in\mathbb{Z}}}(+i(n+\chi')+[\tau_2(m+2p+1)-i\tau])(-i(n+\chi')+[\tau_2(m+2p+3)+i\bar{\tau}])\nonumber\\
&\quad \times(+i(n+\chi')+[\tau_2(m+2p+3)+i\tau])(-i(n+\chi')+[\tau_2(m+2p+3)-i\bar{\tau}])\nonumber\\
&\quad \times(+i(n-\chi')+[\tau_2(m+2p+1)-i\tau])(-i(n-\chi')+[\tau_2(m+2p+1)+i\bar{\tau}])\nonumber\\
&\quad \times(-i(n-\chi')+[\tau_2(m+2p+1)-i\tau])(+i(n-\chi')+[\tau_2(m+2p+3)+i\bar{\tau}]).
\end{align}

\subsubsection*{The Ratio of Determinants}

The $k=1$ contribution to the ratio of determinants is
\begin{align}
\sqrt{\frac{f_{k=1}(s=2)}{g_{k=1}(m=0)}}\propto\sqrt{\prod_{n\in\mathbb{Z}}\frac{1}{i(n+\chi')+\tau_2-i\tau}\frac{1}{-i(n-\chi')+\tau_2-i\tau} }\propto \frac{1}{1-(q\bar{q})^{\frac{1}{2}}e^{-2\pi i \chi'}q}.
\end{align}
Including the $k=0$ and $k\geq 2$ modes we have
\begin{align}
\left.\frac{\det\nolimits^{\frac{1}{2}}{\Delta_0}}{\sqrt{\det\nolimits^{(T)}{L_-}}}\right|_{2J/k=1}&\propto \prod_{n=1}^\infty\frac{1}{1-q^n}\prod_{k=0}^\infty\frac{1}{1- (q\bar{q})^{\frac{1}{2}}e^{-2\pi i \chi'}q^k}\prod_{k=2}^\infty\frac{1}{1- (q\bar{q})^{-\frac{1}{2}}e^{+2\pi i \chi'}q^k}\\
&=\prod_{n=1}^\infty\frac{1}{1-q^n}\prod_{k=0}^\infty\frac{1}{1- e^{-2\pi i \chi}q^k}\prod_{k=2}^\infty\frac{1}{1- e^{+2\pi i \chi}q^k}.\label{eq:C2rd}
\end{align}
In the last line we have again substituted $\chi'=\chi+i\tau_2 \theta$ corresponding to the boundary condition $A_{\bar{w}}|_{\partial M}=i\chi/2\tau_2 T^3$. Compared to the $0<2J/k<1$ case,~\eqref{eq:C1rd}, the is an extra shift in the third product. 

In the special case where the chemical potential vanishes, $\chi=0$, we get
\begin{align}
\left.\frac{\det\nolimits^{\frac{1}{2}}{\Delta_0}}{\sqrt{\det\nolimits^{(T)}{L_-}}}\right|_{\substack{2J/k=1\\\chi=0}}\propto \prod_{n=1}^\infty\frac{1}{1-q^n} \prod_{k=0}^\infty\frac{1}{1- q^k}\prod_{k=2}^\infty\frac{1}{1- q^k},\label{eq:gravyay}
\end{align}
which has a bosonic zero mode that is factored out in~\eqref{eq:sl2rgrav0no0}. If we had plugged in $\chi=-2i\tau_2$ corresponding to fixing $A_{w}|_{\partial M}=0$, we would have found
\begin{align}
\left.\frac{\det\nolimits^{\frac{1}{2}}{\Delta_0}}{\sqrt{\det\nolimits^{(T)}{L_-}}}\right|_{\substack{2J/k=1\\\chi=-2i\tau_2}}\propto \prod_{n=1}^\infty\frac{1}{1-q^n} \prod_{k=1}^\infty\frac{1}{1- \bar{q}q^k}\frac{1}{1- \bar{q}^{-1}q^k},\label{eq:gravwrong}
\end{align}
which is not holomorphic in $q$.

\end{appendix}

\end{document}